\documentclass[a4paper,11pt]{article}

\usepackage{latexsym}
\usepackage{amsfonts} 
\usepackage[mathscr]{euscript}
\usepackage{amsmath,mathrsfs,bm,amssymb,color, ascmac}
\usepackage{graphicx}
\usepackage{cite}

\usepackage[all]{xy}

\title{\textsf{
Stability of ferromagnetism in many-electron systems
}}

\date{\empty}
\author{
Tadahiro Miyao\\ 
 {\it Department of Mathematics,}
{\it Hokkaido University,}\\
{\it Sapporo 060-0810, Japan}\\
E-mail:
 miyao@math.sci.hokudai.ac.jp
}

\newcommand{\h}{\mathfrak{H}}

\newcommand{\ex}{\mathrm{e}}
\newcommand{\Md}{M^{\dagger}}
\newcommand{\D}{\mathrm{dom}}
\newcommand{\R}{\mathrm{ran}}

\newcommand{\Fock}{\mathfrak{F}}

\newcommand{\dG}{d\Gamma}

\newcommand{\ran}{\mathrm{ran}}

\newcommand{\la}{\langle}
\newcommand{\ra}{\rangle}
\newcommand{\Tr}{\mathrm{Tr}}

\newcommand{\slim}{\mbox{$\mathrm{s}$-$\displaystyle\lim_{n\to\infty}$}}

\newcommand{\vep}{\varepsilon}
\newcommand{\BbbR}{\mathbb{R}}
\newcommand{\BbbN}{\mathbb{N}}
\newcommand{\BbbZ}{\mathbb{Z}}

\newcommand{\BbbC}{\mathbb{C}}
\newcommand{\vepsilon}{\varepsilon}
\newcommand{\vphi}{\varphi}

\newcommand{\Cone}{\mathfrak{P}}

\newcommand{\no}{\nonumber \\}

\newcommand{\bsigma}{{\boldsymbol \sigma}}

\newcommand{\Ne}{N_{\mathrm{el}}}

\newcommand{\ab}{{\boldsymbol a}}

\newcommand{\bl}{{\boldsymbol \rightsquigarrow}}
\newcommand{\bS}{{\boldsymbol S}}
\newcommand{\Bs}{\boldsymbol}
\newcommand{\HH}{\mathbb{H}_{\mathrm{HH}}[M]}
\newcommand{\bep}{{\boldsymbol \vepsilon}}
\newcommand{\Leads}{
\overset{\mathrm{\star}}{\leadsto}
}

\begin{document}

\newtheorem{define}{Definition}[section]
\newtheorem{Thm}[define]{Theorem}
\newtheorem{Not}[define]{Notation}
\newtheorem{Prop}[define]{Proposition}
\newtheorem{lemm}[define]{Lemma}
\newtheorem{rem}[define]{Remark}
\newtheorem{assum}{Condition}
\newtheorem{example}{Example}
\newtheorem{coro}[define]{Corollary}

\maketitle

\begin{abstract}
We construct a model-independent framework 
describing stabilities of ferromagnetism in  strongly correlated electron systems;
Our description relies on the operator theoretic correlation inequalities.
Within  the new framework, we reinterpret the Marshall-Lieb-Mattis theorem and Lieb\rq{}s theorem;
in addition,  from the new perspective, we prove that Lieb\rq{}s theorem still
holds true even if the electron-phonon and electron-photon interactions are taken into
account. 
We also examine  the Nagaoka-Thouless theorem and its stabilities.
These examples verify  the effectiveness of our new viewpoint.
\end{abstract}
\setcounter{tocdepth}{1}
\tableofcontents

\section{Introduction}
\setcounter{equation}{0}

\subsection{Overview}

Although magnetism has a long history, 
 its  mechanism  has been  mystery and are  actively studied even today.
Heisenberg was the  first to try to explain the mechanism in terms of the quantum 
mechanics \cite{Heisenberg}.
A modern approach to  magnetism, more precisely,  metalic ferromagnetism
was initiated by Gutzwiller, Kanamori and Hubbard \cite{Gutzwiller, Kanamori, Hubbard}.
They introduced  a very simplified model which is nowaday called the Hubbard model
to explain ferromagnetism. The Hubbard model on a finite lattice $\Lambda$ is given by 
\begin{align}
H_{\mathrm{H}}=\sum_{x,y\in \Lambda}\sum_{\sigma=\uparrow, \downarrow} t_{xy}
c_{x\sigma}^*c_{y\sigma}+\sum_{x, y\in \Lambda}\frac{U_{xy}}{2} (n_x-1)(n_y-1)\label{Hami}.
\end{align}
With regard to more precise definition, see Section \ref{Models}.
Despite the simplicity of the Hamiltonian, the Hubbard model involves the three
essential factors of many-electron systems: first, the 
electron itineracy which is described by the first term of 
the RHS of (\ref{Hami}); second, the electron-electron  Coulomb repulsion, the second 
term of the  RHS of (\ref{Hami}); finally, the Fermi statistics which is expressed by the fact that  
$H_{\mathrm{H}}$ acts on the fermionic Fock space.
We expect that ferromagnetism arises from an  exquisite interplay of these  factors.
To reveal  the interplay mathematically has been a long-standing  problem in the theory 
of ferromagnetism.
Note that if any one of the factors are lacking, ferromagnetism will  never appear,
which indicates  that the  mathematical  analysis of ferromagnetism is
essentially  nonperturbative. This aspect makes the problem  difficult.

Despite the extensive research regarding ferromagnetism, 
only few exact results are currently known.
Below, we explain two  rigorous results which are related with the present study.
\medskip

{\it Nagaoka-Thouless\rq{} theorem:}
In 1965, Nagaoka constructed  a first  rigorous example of the ferromagnetism \cite{Nagaoka}.
He proved that the ground state of the model exhibits ferromagnetism
when  one electron is fewer than half-filling 
and the Coulomb strength $U$ is infinitely large. 
We remark that Thouless also discussed the same mechanism in \cite{Thouless}.
\medskip

{\it Lieb\rq{}s theorem:}
In 1989, the ground state  of the Hubbard model on some bipartite lattices at
 half-filling is shown  to exhibit ferrimagnetism  by Lieb \cite{Lieb}.
His  method  called  the   spin-reflection positivity  originates from 
the axiomatic quantum field theory \cite{GJ, OS}.
In the Nagaoka-Thouless theorem,  the Coulomb strength is  assumed to be  infinitely large,
on the other hand,  such a restriction is unnecessary in Lieb's theorem. 
\medskip

Other than the  above,  flat-band ferromagnetism is attracting attention \cite{Mielke,Mielke2, MT, Tasaki22,Tasaki3},
but this is not our concerns in the present paper.     
 For a
review of the history and  rigorous results concerning the Hubbard model, 
we refer to \cite{Tasaki22}.

On the one hand, electrons always interact with phonons and  electromagnetic fields
 in actual metals;
on the other hand, ferromagnetism is experimentally  observed in
various metals and has a
wide range of uses in daily life.
Therefore, if  the Nagaoka-Thouless  and Lieb's theorems contain an essence of  
real ferromagnetism,
 these theorems  should be stable under the influence of 
the electron-phonon and electron-photon interactions. 
 Stabilities of magnetism have been studied by the author and obtained 
affirmative results; it is proved that the  Nagaoka-Thouless  theorem is  stable
 under the influence of the above-mentioned perturbations \cite{Miyao6};
in addition, Lieb's theorem is proved to be stable
 even if the electron-phonon interaction is taken into consideration \cite{Miyao5, Miyao7}.

The main purpose in the present paper is to reveal a mathematical framework describing 
the stability of ferromagnetism.  
As we will see, this is attained by introducing a new  concept of  stability class.
 The  stability classes are described by operator theoretic correlation inequalities established in  \cite{Miyao1, Miyao2, Miyao4, Miyao5, Miyao7}.
 The  main advantage of our approach is that we can clearly recognize a model-independent  structure
 behind stabilities of ferromagnetism  in many electron  systems.

In \cite{Miyao5, Miyao6, Miyao7}, we have already examined the stabilitiy problems.
 However, our approach  in the present study is quite different from those in \cite{Miyao5, Miyao6, Miyao7}, 
 and provide a unified viewpoint of stability of ferromagnetism.

\subsection{Summary of results}\label{SecResults}

\subsubsection{Positivity in Hilbert spaces}
To state our results, we have to introduce some terminologies concerning self-dual cones: 
Let $\Cone$ be a convex cone in the Hilbert space $\h$. We say that $\Cone$ is {\it self-dual} if 
$\Cone=\{\eta\in \h\, |\, \la \eta|\xi\ra\ge 0\ \forall \xi\in
\Cone \}.$ A vector $\eta\in \Cone$ is called {\it strictly positive
w.r.t. $\Cone$} whenever $\la \xi| \eta\ra>0$ for all $\xi\in
\Cone\backslash \{0\}$. We write this as $\eta>0 $
w.r.t. $\Cone$. In this way, once we fix a self-dual cone in $\h$, 
 the corresponding positivity can be  naturally  defined.
In general,  there could be infinitely many self-dual cones in $\h$,
which implies that we can introduce various  positivities in $\h$. 
 See Section \ref{GeneralThy} for further details.
  
\subsubsection{Basic definitions  in   many-electron systems}\label{BFMS}
In the present study,  we will examine many-electron systems.
To describe a brief summary of our results, we will give some basic definitions. 

Consider  electrons   in  a finite  lattice $\Lambda$.
The Hilbert space of the electrons is  given by 
\begin{align}
\mathfrak{E}=\bigoplus_{n\ge 0} \bigwedge^n (\ell^2(\Lambda)\oplus \ell^2(\Lambda)), \label{FermiSp}
\end{align}
 the fermionic Fock space over $\ell^2(\Lambda)\oplus \ell^2(\Lambda)$;
here, $\bigwedge^n \mathfrak{h}$ indicates the $n$-fold antisymmetric tensor product
of $\mathfrak{h}$.

The electron annihilation- and creation operators with spin $\sigma\in \{\uparrow, \downarrow\}$
 at $x\in \Lambda$ are denoted by $c_{x\sigma}$ and $c_{x\sigma}^*$, respectively. 
 Of course, $c_{x\sigma}$ and $c_{x\sigma}^*$ act on the Hilbert space $\mathfrak{E}$ and 
   satisfy the standard anticommutation relations:
\begin{align}
\{
c_{x\sigma}, c^*_{y\sigma\rq{}}
\}=\delta_{xy}\delta_{\sigma\sigma\rq{}},\ \ \ \{c_{x\sigma}, c_{y\sigma'}\}=0, 
\end{align}
where $\delta_{xy}$ is the Kronecker delta.

The electron number operator at $x\in \Lambda$ is given by $n_x=
n_{x\uparrow}+n_{x\downarrow}$
with $
n_{x\sigma}=
c_{x\sigma}^*c_{x\sigma}$.
The total electron number operator is  defined by 
\begin{align}
\Ne
=\sum_{x\in \Lambda} n_x.
\end{align}
The $n$-electron subspace of $\mathfrak{E}$  is given  by 
\begin{align}
\mathfrak{E}_{n}&=\ker(\Ne-n)=\bigwedge^n \big(\ell^2(\Lambda)\oplus \ell^2(\Lambda)\big). \label{Nsub}
\end{align}

The spin
 operators at $x\in \Lambda$ are defined by 
\begin{align} 
S_x^{(j)}=\frac{1}{2}\sum_{\sigma, \sigma'=\uparrow,
 \downarrow}c_{x\sigma}^* \Big(s^{(j)}\Big)_{\sigma \sigma'}
c_{x\sigma'},\ \  j=1,2,3,
\end{align} 
where $s^{(j)}\ (j=1,2,3)$ are  the Pauli matricies.
These 
satisfy the standard commutation relations:
\begin{align}
[S_x^{(j)}, S_y^{(k)}]=i\vepsilon_{jkl}\delta_{xy}S_x^{(l)},
\end{align}
where $\vepsilon_{jkl}$ is the Levi-Civita symbol.

The total  spin operators are defined by 
\begin{align}
S_{\mathrm{tot}}^{(j)}=\sum_{x\in \Lambda}S_{x}^{(j)},\ \ j=1,2,3. \label{TotalSpins}
\end{align}
In addition, we set
\begin{align}
S_{\mathrm{tot}}^2= \sum_{j=1}^3\Big(S_{\mathrm{tot}}^{(j)}\Big)^2
\end{align}
with eigenvalues $S(S+1)$. If $\vphi$ is an eigenvector of $S_{\mathrm{tot}}^2$ with 
$
S_{\mathrm{tot}}^2\vphi=S(S+1) \vphi
$, then we say that {\it $\vphi$ has total spin $S$.}

Now let us consider the $n$-electron subspace $\mathfrak{E}_n$.
For each $M\in \mathrm{spec}(S_{\mathrm{tot}}^{(3)})$, we define the  {\it $M$-subspace} of $\mathfrak{E}_n$ by $ \mathfrak{E}_n[M]=\ker(S_{\mathrm{tot}}^{(3)}-M) \cap \mathfrak{E}_n$.
Let $\mathfrak{H}$ be a Hilbert space containing $\mathfrak{E}$ as a subspace.
Thus, the total spin operators and the electron number operator can be viewed as operators acting on $\mathfrak{H}$, naturally.
  The $n$-electron subspace of $\mathfrak{H}$ is defined by $\mathfrak{H}_n=\ker(\Ne-n) \cap \mathfrak{H}$, and 
  the $M$-subspace of $\mathfrak{H}_n$ is defined by $\mathfrak{H}_n[M]=\ker(S_{\mathrm{tot}}^{(3)}-M) \cap \mathfrak{H}_n$.

In concrete applications examined in the present study, we will mainly  explore the case where $\h=\mathfrak{E}\otimes \mathfrak{X}$, where $\mathfrak{X}$ is some Hilbert space, e.g., the phonon-Fock space.  In this case, 
we can regard $\mathfrak{E}$ as a subspace of $\h$ in the following manner:
Given a normalized vector $x_0\in \mathfrak{X}$, let $\mathfrak{E}\otimes x_0$ be a closed subspace of 
$\mathfrak{E}\otimes \mathfrak{X}$ defined by $\mathfrak{E}\otimes x_0=\{\psi\otimes x_0\, |\, \psi\in \mathfrak{E}\}$. We readily confirm that a map $\tau: \mathfrak{E} \ni \psi\mapsto  \psi\otimes x_0\in \mathfrak{E}\otimes x_0$ is an isometry, which provides an identification $\mathfrak{E}$ with $\mathfrak{E}\otimes x_0$. Under this identification, $\mathfrak{E}$ can be viewed as a subspace of $\h$.\footnote{An  appropriate  choice of $x_0$ depends on the problem.}

For later use, we slightly extend   the definition of the $M$-subspace as follows. Let $Q$ be an orthogonal projection on $\mathfrak{E}_n$. We suppose the following condition:
\begin{flushleft}
{\bf (Q)} $Q$ commutes with the total spin operators $S_{\mathrm{tot}}^{(j)},\ j=1, 2, 3$.
\end{flushleft}
We call $Q\mathfrak{E}_n[M]$ the $M$-subspace as well.
Note that $Q$ is a projection onto a subspace where the problem becomes relatively easier to treat.
A typical example of $Q$ is the Gutzwiller projection given by (\ref{GutzP}), see also (\ref{DefQ}) as another example of $Q$.

Let $\mathbb{H}$ be the set of all Hilbert spaces which contain $\mathfrak{E}$ as a subspace.
We define a family of self-adjoint operators  by 
\begin{align*}
\mathscr{H}_n=\bigcup_{\mathfrak{H}\in \mathbb{H}}
\Big\{
H\, \Big|\, &\mbox{$H$ is a self-adjoint operator acting  on the $n$-electron subspace $\mathfrak{H}_n$, } \no
 & \mbox{ bounded from below,  and commutes  with $Q$ and  $S_{\mathrm{tot}}^{(j)},\ j=1, 2, 3$} 
\Big\}.
\end{align*}
Physically, $\mathscr{H}_n$ is the set of all Hamiltonians in which we are interested.
Note that even if $H$ is a self-adjoint operator on $Q\mathfrak{E}_n$ which commutes with $S_{\mathrm{tot}}^{(j)},\, j=1, 2, 3$, we can  naturally regard $H$ as an element in $\mathscr{H}_n$.\footnote{
More precisely, corresponding to the decomposition $\mathfrak{E}_n=\R(Q)\oplus \ker(Q)$, we have the natural identification $H\cong H\oplus 0$. We readily check that $H\oplus 0 \in \mathscr{H}_n$.
}
For a given $H\in \mathscr{H}_n$, let $\mathfrak{H}_H\in \mathbb{H}$ be a corresponding Hilbert space on which
$H$ acts.\footnote{ 
Needless to say, the number of electron in $\mathfrak{H}_H$ is equal to $n$: $\mathfrak{H}_H\subset \ker(\Ne-n)$.}
We set $H[M]=H\restriction \mathfrak{H}_H[M]$, the restriction of $H$ onto the $M$-subspace $\mathfrak{H}_H[M]$.

\subsubsection{Stability of ferromagnetism: A general form}
Our results  in Section \ref{ManyElUniTh} can be briefly summarized as follows.
\begin{Thm}[Summary of Section \ref{ManyElUniTh}]\label{SummaryMain}
Let $Q$ be   an  orthogonal projection on $\mathfrak{E}_n$ satisfying {\rm {\bf (Q)}}.
Let $H_*$ be a Hamiltonian acting on  $Q\mathfrak{E}_n$.
 We assume the 
 the following:
\begin{itemize}
\item[{\bf (U. 1)}] $H_*$ commutes with the total spin operators $S_{\mathrm{tot}}^{(j)},\ j=1,2,3$. In addition, the ground state of $H_*$ has total spin $S_*$ and, is unique apart from the trivial $(2S_*+1)$-degeneracy. 
\item[{\bf (U. 2)}] There exist an $M_*\in \mathrm{spec}(S_{\mathrm{tot}}^{(3)})$  with $|M_*| \le S_*$
and a self-dual cone $\Cone_*$ in the $M_*$-subspace $ Q\mathfrak{E}_n[M_*] $ such that the unique ground state $\psi_*$ of $H_*[M_*]$ fullfils $\psi_*>0$ w.r.t. $\Cone_*$.
\end{itemize}
Then there exists a family of Hamiltonians  $\mathscr{U}(H_*)\subset \mathscr{H}_n$  with the following properties:
\begin{itemize}

\item[{\rm (i)}] For every $H\in \mathscr{U}(H_*)$, the ground state of $H$ has the common total spin $S_*$, and is 
unique apart from the trivial $(2S_*+1)$-degeneracy.
\item[{\rm (ii)}] 
The cardinality of $\mathscr{U}(H_*)$ is greater than $\aleph_0$, the cardinality of the natural numbers.
In this sense, $\mathscr{U}(H_*)$ is rich enough. 
\item[{\rm (iii)}] For each $H\in \mathscr{U}(H_*)$, let $\mathfrak{H}_H$ be  the corresponding Hilbert space on which $H$ acts. Then there exists a self-dual cone $\Cone_H$ in the $M_*$-subspace $\mathfrak{H}_H[M_*]$ such that the unique ground state of $H[M_*]$ satisfies $\psi_H>0$ w.r.t. $\Cone_H$. Note that there could be another self-dual cone $\Cone'_H$ such that $\psi_H>0$ w.r.t. $\Cone'_H$.

\item[{\rm (iv)}] Every Hamiltonian in $\mathscr{U}(H_*)$ inherits the strict positivity of the ground state
from $H_*$. More precisely, for each $H\in \mathscr{U}(H_*)$, there exists a sequence of Hamiltonians $\{H_j\}_{j=1}^N\subset \mathscr{U}(H_*)$ with $H_1=H_*$ and $H_N=H$
 possessing the following properties: 
Let $\mathscr{C}_j$ be the set of all self-dual cones in $\mathfrak{H}_{H_j}[M_*]$ satisfying (iii) :
$\mathscr{C}_j=\{\Cone_{H_j}\, |\, \psi_{H_j}>0\, \mbox{w.r.t. $\Cone_{H_j}$}\}$.
Let $P_{j, j+1}$ be the orthogonal projection from $\mathfrak{H}_{H_{j+1}}[M_*]$ onto $\mathfrak{H}_{H_j}[M_*]$.\footnote{Thus, the Hilbert spaces $\mathfrak{H}_{H_1},\dots, \mathfrak{H}_{H_N}$ satisfy
$
Q\mathfrak{E}_n[M_*]=\mathfrak{H}_{H_1}[M_*]\subseteq \mathfrak{H}_{H_2}[M_*]\subseteq \cdots \subseteq \mathfrak{H}_{H_N}[M_*].
$
}
Then there exist  pairs of self-dual cones
$
(\Cone_2, \Cone_2'), (\Cone_3, \Cone_3'), \dots, (\Cone_{N-1}, \Cone_{N-1}')
$ 
  such that:
  \begin{itemize}
  \item[1. ]
      $(\Cone_j, \Cone_j')\in \mathscr{C}_j\times \mathscr{C}_j, \, j=2,3,  \dots, N-1$.
      \item[2. ] $P_{j, j+1} \Cone_{j+1}' \subseteq \Cone_{j+1}',\ j=1, 2, \dots, N-1$, where $\Cone_N'=\Cone_H$.
      
      \item[3. ] Set  $\Cone_1=\Cone_*$. We have 
$
\Cone_1=P_{1, 2}\Cone_2',\ \Cone_2=P_{2, 3}\Cone_3', \dots, \Cone_{N-1}=P_{N-1, N}\Cone_N'
$.
\end{itemize}
Note that $\Cone_j'$ could be equal to $\Cone_j$.
\end{itemize}

\end{Thm}

\begin{rem}
{\rm 
\begin{itemize}
\item
The family  of Hamiltonians $\mathscr{U}(H_*)$ in the above is called the {\it $H_*$- stability class}.
\item 
The property (iv) implies that 
$\la \psi_*|P_{1, 2}\psi_{H_2}\ra\la \psi_{H_2}|P_{2, 3}\psi_{H_3}\ra\cdots \la \psi_{H_{N-1}}|P_{N-1, N}\psi_{H_N}\ra>0$,
which will  play an important role in Section \ref{ManyElUniTh}.

\item In concrete applications of  the above theorem, we will only use  the properties (i) and (ii)  
in order to make   statements simpler. However, a remarkable point in the above theorem is  that every Hamiltonian in $\mathscr{U}(H_*)$ inherits 
the strict positivity of the ground state from $H_*$; Indeed,  (i) and (ii) can be regarded as  by-products of the properties (iii) and (iv), see Sections \ref{GeneralThy}  and \ref{ManyElUniTh} for details
\item  As we will see in Section \ref{ManyElUniTh}, our construction of $\mathscr{U}(H_*)$ clarifies  conditions when a given Hamiltonian actually belongs to $\mathscr{U}(H)_*$.
\item 
We will also discuss ferromagnetism and long-range orders in the ground state from a viewpoint of  stability class.
$\diamondsuit$
\end{itemize}
}
\end{rem}

To understand the importance of the above theorem, we will give two examples in this section;
Indeed, we will examine 
stability of Lieb's theorem and the Nagaoka-Thouless theorem  in terms of the  stability class.

In Section \ref{ConRemarks}, we will further discuss various stabilities of ferromagnetism.
Typical examples are the following: 
\begin{itemize}
\item Stabilities of ferromagnetism under deformations of parameters including $t_{xy} $ and $U_{xy}$ in (\ref{Hami}).
\item Stabilities of ferromagnetism against  randomness comming from the enviroment.
\end{itemize}

\subsection{Models}\label{Models}

To illustrate applications of the main theorem, we introduce three models.

\subsubsection{The Hubbard model}
The Hamiltonian of the Hubbard model on $\Lambda$ is defined by 
\begin{align}
H_{\mathrm{H}}=\sum_{x,y\in \Lambda}\sum_{\sigma=\uparrow, \downarrow} t_{xy}
c_{x\sigma}^*c_{y\sigma}+\sum_{x, y\in \Lambda}\frac{U_{xy}}{2} (n_x-1)(n_y-1),
\end{align}
where $t_{xy}$ is the hopping matrix, and $U_{xy}$ is the energy of the Coulomb interaction.
We suppose that $\{t_{xy}\}$ and $\{U_{xy}\}$ are real symmetric $|\Lambda|\times |\Lambda|$ matrices.
The Hamiltonian $H_{\mathrm{H}}$ acts on the Hilbert space  $\mathfrak{E}$.

\subsubsection{The Holstein-Hubbard model}
In the present paper, we omit trivial tensor products;
let $\mathfrak{X}_1$ and $\mathfrak{X}_2$ be Hilbert spaces, and let $A$ and $B$
be linear operators on $\mathfrak{X}_1$ and $\mathfrak{X}_2$, respectively.
We identify $A\otimes 1=A$ and $1\otimes B=B$, if no confusion arises.

Let us consider the Holstein-Hubbard model of interacting electrons coupled to dispersionless phonons of 
frequency $\omega>0$.
The Hamiltonian is 
\begin{align}
H_{\mathrm{HH}}=H_{\mathrm{H}}+\sum_{x, y\in \Lambda}
 g_{xy}(n_x-1)(b_y^*+b_y)
+\sum_{x\in \Lambda} \omega b_x^*b_x,
\end{align} 
where $H_{\mathrm{H}}$ is the Hubbard Hamiltonian; $b_x^*$ and $b_x$
are bosonic creation- and annihilation operators at site $x\in \Lambda$,
respectively.  The operators $b_x^*$ and $b_x$ satisfy the canonical commutation
relations:
\begin{align}
[b_x, b_y^*]=\delta_{xy},\ \ [b_x, b_y]=0.
\end{align} 
$g_{xy}$ is the strength of the electron-phonon interaction.
We assume that $\{g_{xy}\}$ is a real symmetric matrix.
$H_{\mathrm{HH}}$ lives in the Hilbert space $\mathfrak{E}\otimes
\Fock_{\mathrm{ph}}$, where $\mathfrak{E}$ is given by (\ref{FermiSp}), and 
$\Fock_{\mathrm{ph}}=\bigoplus_{n=0}^{\infty} \otimes_{\mathrm{s}}^n
\ell^2(\Lambda)$, the bosonic Fock space over $\ell^2(\Lambda)$; here,
$\otimes_{\mathrm{s}}^n \mathfrak{h}$ indicates the $n$-fold symmetric
tensor product of $\mathfrak{h}$.
By the  Kato-Rellich theorem \cite[Theorem X. 12]{ReSi2}, 
$H_{\mathrm{HH}}$ is self-adjoint on $\D(N_{\mathrm{ph}})$ and bounded
from below, where $N_{\mathrm{ph}}=\sum_{x\in \Lambda} b_x^*b_x$ and 
$\D(A)$ indicates  the domain of the linear operator $A$.

\subsubsection{A many-electron system coupled to the quantized radiation field}\label{RadHami}
Consider a many-electron system coupled to the quantized radiation
field.
We suppose that $\Lambda$ is embedded into the region $
V=[-L/2, L/2]^3 \subset \BbbR^3$ with $L>0$.\footnote{
Remark that the  cube is used here  for simplicity;
we can take  general $V$, for example,  $V=[-L_1/2, L_1/2]\times [-L_2/2, L_2/2]\times [-L_3/2, L_3/2]$. 
}
The system is described by the following Hamiltonian:
\begin{align}
H_{\mathrm{rad}}=&\sum_{x, y\in \Lambda}\sum_{\sigma=\uparrow,
 \downarrow}
t_{xy} \exp\Bigg\{
i \int_{C_{xy}} dr\cdot A(r)
\Bigg\}c_{x\sigma}^*c_{y\sigma}
+\sum_{x, y\in \Lambda} \frac{U_{xy}}{2}(n_x-1)(n_y-1)\no
&+\sum_{k\in V^*} \sum_{\lambda=1,2}\omega(k) a(k, \lambda)^*a(k, \lambda).
\end{align} 
The Hamiltonian $H_{\mathrm{rad}}$ acts on the Hilbert space  $\mathfrak{E}\otimes \Fock_{\mathrm{rad}}$, where
$\mathfrak{E}$ is given by (\ref{FermiSp}), and 
$\Fock_{\mathrm{rad}}$
is the bosonic Fock space over $\ell^2(V^*\times \{1, 2\})$
with $V^*=(\frac{2\pi}{L}\BbbZ)^3$.
$a(k, \lambda)^*$ and $a(k, \lambda)$ are bosonic creation- and
annihilation operators, respectively.
As usual, these operators satisfy the following relations:
\begin{align}
[a(k, \lambda), a(k',
 \lambda')^*]=\delta_{\lambda\lambda'}\delta_{kk'},\ \ \ [a(k, \lambda),
 a(k', \lambda')]=0.
\end{align} 
$A(r)\, (r\in V)$ is the quantized vector potential given by 
\begin{align}
A(r)=|V|^{-1/2}\sum_{k\in V^*}
 \sum_{\lambda=1,2}\frac{\chi_{\kappa}(k)}{\sqrt{2\omega(k)}}{\boldsymbol \vepsilon}_{\lambda}(k) \Big(
e^{ik\cdot r}a(k, \lambda)+e^{-ik\cdot r}a(k, \lambda)^*
\Big).
\end{align} 
 $\chi_{\kappa}$ is the indicator function of the ball of
radius $0<\kappa<\infty$ centered at the origin, where $\kappa$ is the ultraviolet cutoff.
The dispersion relation $\omega(k)$ is chosen to be $\omega(k)=|k|$ for
$k\in V^* \backslash \{0\}$, $\omega(0)=m_0$ with $0<m_0<\infty$.
$C_{xy}$ is a piecewise smooth curve from $x$ to $y$.
For concreteness, the polarization vectors are chosen as  
\begin{align}
{\boldsymbol \vepsilon}_{1}(k)
=\frac{(k_2, -k_1, 0)}{\sqrt{k_1^2+k_2^2}}
,\ \ {\boldsymbol  \vepsilon}_2(k)=\frac{k}{|k|} \wedge \vepsilon(k, 1). \label{Pola}
\end{align} 
To avoid ambiguity, we set $
{\boldsymbol \vepsilon}_{\lambda}(k)=0
$ if $k_1=k_2=0$.
 $A(r)$ is essentially self-adjoint. We denote its
closure by the same symbol.
This model was introduced by Giuliani et al. in \cite{GMP}.
Remark that $H_{\mathrm{rad}}$ is essentially self-adjoint and bounded from below.
We denote its closure by the same symbol.

\subsection{Stability of Lieb's theorem}\label{StaLi}
As an application of Theorem \ref{SummaryMain},  we explore stabilities of Lieb\rq{}s theorem.

\subsubsection{Basic definitions}
\begin{define}\label{BasicDefs}
{\rm
Let $\Lambda$ be a finite lattice.
Let $\{M_{xy}\}$ be a real symmetric $|\Lambda|\times |\Lambda|$ matrix.

\begin{itemize}
\item[(i)] We say that $\Lambda$ is {\it connected by $\{M_{xy}\}$}, if, for every $x,y\in \Lambda$, there are $x_1, \dots, x_n\in \Lambda$ such that 
$
M_{xx_1}M_{x_1x_2}\cdots M_{x_ny} \neq 0.
$
\item[(ii)] We say that $\Lambda$ is {\it bipartite  in terms of $\{M_{xy}\}$}, if $\Lambda$ 
can be divided into two disjoint sets $A$ and $B$ such that $M_{xy}=0$,
whenever $x,y\in A$ or $x,y\in B$. $\diamondsuit$
\end{itemize}
}
\end{define}

\subsubsection{The Marshall-Lieb-Mattis  stability class }

The Marshall-Lieb-Mattis Hamiltonian  is defined by  
\begin{align}
H_{\mathrm{MLM}}=\bS_A\cdot \bS_B, \label{MLMHamiltonian}
\end{align} 
where
$
\bS_A=\sum_{x\in A} \bS_x$ and $ \bS_B=\sum_{x\in B}\bS_x.
$
Here,  $A$ and $B$ are subsets of $\Lambda$ in Definition \ref{BasicDefs}.
The Hamiltonian $H_{\mathrm{MLM}}$ acts on $\mathfrak{E}_{n=|\Lambda|}$.
Let $Q$ be  the orthogonal projection    defined by
 \begin{align}
 Q=\prod_{x\in \Lambda} (n_{x\uparrow}-n_{x\downarrow})^2. \label{DefQ}
 \end{align}
 We readily check that $Q$ satisfies {\bf (Q)}.

In Section \ref{MLMLiebSta}, we will prove the following.
\begin{Thm}\label{QAns1}
The Hamiltonian $H_{\mathrm{MLM}}$ satisfies {\bf (U. 1)} and {\bf (U. 2)} with $S_*=\big||A|-|B|\big|/2$.
Hence, by Theorem \ref{SummaryMain}, the $H_{\mathrm{MLM}}$-stability class  $\mathscr{U}(H_{\mathrm{MLM}})$ has the following properties:
For each $H\in \mathscr{U}(H_{\mathrm{MLM}})$, the ground state of $H$ has total spin $S=\big||A|-|B|\big|/2$, and is unique apart from the trivial $(2S+1)$-degeneracy.
\end{Thm}
\begin{rem}
{\rm 
$\mathscr{U}(H_{\mathrm{MLM}})$ is called the {\it  Marshall-Lieb-Mattis  stability class}. $\diamondsuit$
}
\end{rem}

\subsubsection{Lieb's theorem}
We examine the Hubbard model $H_{\mathrm{H}}$.
We assume the following:
\begin{itemize}
\item[{\bf (A. 1)}] $\Lambda$ is connected by $\{t_{xy}\}$;
\item[{\bf (A. 2)}] $\Lambda$ is bipartite in terms of $\{t_{xy}\}$;
\item[{\bf (A. 3)}] $\{U_{xy}\}$ is positive definite.\footnote{
More precisely, $
\sum_{x, y\in \Lambda}z_x^*z_y U_{xy} > 0
$ for all ${\Bs z}=\{z_x\}_{x\in \Lambda} \in \BbbC^{|\Lambda|}$ with ${\Bs z} \neq {\Bs 0}$.
}
\end{itemize}

Note  that the number of electron is conserved, that is, $\Ne$ commutes with $H_{\mathrm{H}}$.
Since we are interested in the half-filled system, we will study the following Hamiltonian:
\begin{align}
H_{\mathrm{H}, |\Lambda|}&=H_{\mathrm{H}} \restriction \mathfrak{E}_{n=|\Lambda|}, \label{HilHalf}
\end{align}

In Section \ref{MLMLiebSta}, we will prove the following.
\begin{Thm}[Lieb's theorem \cite{Lieb}] \label{LiebThm1}
Assume that $|\Lambda|$ is even. Assume  {\bf (A. 1)}, {\bf (A. 2)} and {\bf (A. 3)}.
Then $H_{\mathrm{H}, |\Lambda|}$ belongs to $\mathscr{U}(H_{\mathrm{MLM}})$.
Thus, by Theorem \ref{QAns1}, the ground state of $H_{\mathrm{H}, |\Lambda|}$ has total spin $S=\frac{1}{2}\big|
|A|-|B|
\big|$ and is unique apart from the trivial $(2S+1)$-degeneracy.
\end{Thm}

\subsubsection{Stability of Lieb's theorem I}

Let us consider the Holstein-Hubbard model $H_{\mathrm{HH}}$.
As before, the number of electrons is conserved. We will study the
half-filled case; thus, we focus our attention on the restricted
Hamiltonian:
\begin{align}
H_{\mathrm{HH}, |\Lambda|}=H_{\mathrm{HH}} \restriction
 \mathfrak{E}_{n=|\Lambda|}\otimes \Fock_{\mathrm{ph}}.
\end{align} 

Here, we continue to assume {\bf (A. 1)} and {\bf (A. 2)}. On the other
hand,
the assumption {\bf (A. 3)} will be replaced by a new condition {\bf (A. 5)} below.
As to the electron-phonon interaction, we assume the following:
\begin{itemize} 
\item[{\bf (A. 4)}] $\sum_{x\in \Lambda} g_{xy}$ is a constant
	     independent of $y\in \Lambda$.
\end{itemize} 
An important example satisfying {\bf (A. 4)} is the on-site interaction: $g_{xy}=g\delta_{xy}$.

To state our result, we introduce the effective Coulomb interaction by 
\begin{align}
U_{\mathrm{eff}, xy}=U_{xy}-\frac{2}{\omega} \sum_{z\in \Lambda}g_{xz}g_{yz}.
\end{align} 
Our new assumption in stated as follows.
\begin{itemize} 
\item[{\bf (A. 5)}] 
$
\{
U_{\mathrm{eff}, xy}
\}
$ is positive definite.
\end{itemize} 

\begin{Thm}\label{StaLiThm1}
Assume that $|\Lambda|$ is even. Assume that {\bf (A. 1)}, {\bf (A. 2)},
 {\bf (A. 4)} and {\bf (A. 5)}. Then   $H_{\mathrm{HH}, |\Lambda|}$ belongs to 
 $\mathscr{U}(H_{\mathrm{MLM}})$.
 Hence, by Theorem \ref{QAns1}, the ground state of
 $H_{\mathrm{HH}, |\Lambda|}$ has total spin $S=\frac{1}{2}\big|
|A|-|B|
\big|$ and is unique apart from the trivial $(2S+1)$-degeneracy.
\end{Thm} 

We will provide a proof of Theorem \ref{StaLiThm1} in Section \ref{MLMLiebSta}.

\subsubsection{Stability of Lieb's theorem II}
Next, let us consider a many-electron system coupled to the quantized radiation field.
Note that 
the number of electrons is conserved, as before. We will study the
Hamiltonian at half-filling:
\begin{align}
H_{\mathrm{rad}, |\Lambda|}=H_{\mathrm{rad}}\restriction
 \mathfrak{E}_{n=|\Lambda|}\otimes \Fock_{\mathrm{rad}}.
\end{align}

\begin{Thm}\label{StaLiThm2}
Assume that $|\Lambda|$ is even. Assume that {\bf (A. 1)}, {\bf (A. 2)} and {\bf (A. 3)}. Then
  $H_{\mathrm{rad}, |\Lambda|}$ belongs to  $\mathscr{U}(H_{\mathrm{MLM}})$.
  Thus, by Theorem \ref{QAns1}, the ground state of
 $H_{\mathrm{rad}, |\Lambda|}$ has total spin $S=\frac{1}{2}\big|
|A|-|B|
\big|$ and is unique apart from the trivial $(2S+1)$-degeneracy.
\end{Thm} 

As far as we know,  this theorem is new. We provide a proof of Theorem \ref{StaLiThm2} in the following sections.

\subsection{Stability of the Nagaoka-Thouless  theorem}\label{StaNT}

As an  additional example, we explore the stabilities of the  Nagaoka-Thouless theorem in this subsection.

\subsubsection{Basic definitions}\label{NTBasicDef}
To explain  the Nagaoka-Thouless ferromagnet, we introduce the {\it
hole-connectivity} as follows.

The set of spin configurations with a single hole is denoted by $\mathcal{S}$:
\begin{align}
\mathcal{S}=\Big\{
\bsigma=&\{\sigma_x\}_{x\in \Lambda}\in \{\uparrow, 0,
 \downarrow\}^{\Lambda}\, 
\Big|\, \no
&\mbox{There exists an $x_0\in \Lambda$ such that
 $\sigma_{x_0}=0$ and $\sigma_x\neq 0$ if $x\neq x_0$
}
\Big\}. \label{SetS}
\end{align} 
We say that the $x_0$ in (\ref{SetS}) is the position of the hole.
For each $\bsigma\in \mathcal{S}$, the position of the hole is denoted
by $x_0(\bsigma)$.

For each $\bsigma\in \mathcal{S}$, we denote by $n_{\uparrow}(\bsigma)$ (resp.,
$n_{\downarrow}(\bsigma)$)  the number of up spins
(resp., down spins) in $\bsigma$. For each $M\in \{-(|\Lambda|-1)/2,\,
-(|\Lambda|-3)/2, \dots, (|\Lambda|-1)/2\}$, we set $\mathcal{S}_M
=\{\bsigma\in \mathcal{S}\, |\, n_{\uparrow}(\bsigma)-n_{\downarrow}(\bsigma)=2M\}$.

An element $(x, \bsigma)\in \Lambda\times \mathcal{S}_M$ is called the
{\it hole-spin configuration}, if $x=x_0(\bsigma)$. The set of
all hole-spin configurations is denoted by  $\mathcal{C}_M$.
For each $y\in \Lambda$, we define a map $S_y: \mathcal{C}_M\to \mathcal{C}_M$
 by 
$
S_y(x, \bsigma)=(y, \bsigma'),
$
where $\bsigma'=\{\sigma_z'\}_{z\in \Lambda}\in \mathcal{S}_M$ is given
by
\begin{align}
\sigma_z'=
\begin{cases}
\sigma_y & \mbox{if $z=x$}\\
0 & \mbox{if $z=y$}\\
\sigma_z & \mbox{otherwise}
\end{cases}. 
\end{align} 
If $x=y$, then we set $S_y(x, \bsigma)=(x, \bsigma)$. 

\begin{define}
{\rm
Let $\{M_{xy}\}$ be a real symmetric matrix.
We say that $\Lambda$ has the {\it hole-connectivity associated with} $\{M_{xy}\}$, if  the following
 condition holds:
For every pair $(x, \bsigma), (y, {\boldsymbol \tau})\in \mathcal{C}_M$
 with $(x, \bsigma)\neq (y, {\boldsymbol \tau})$,
there exist sites $x_1, \dots, x_{\ell} \in \Lambda$ such that 
\begin{align}
M_{y x_{\ell}} M_{x_{\ell}x_{\ell-1}}\cdots M_{x_2x_1} M_{x_1x} \neq 0
\end{align} 
and 
\begin{align}
\Big(
S_y\circ S_{x_{\ell}} \circ S_{x_{\ell-1}} \circ \cdots \circ S_{x_1}
\Big) (x, \bsigma)=(y, {\boldsymbol \tau}). \ \ \ \ \diamondsuit
\end{align} 
}
\end{define}

\begin{example}\label{HoleC}
{\rm
It is known that models with the following (i) and (ii) satisfy the hole-connectivity associated with $\{t_{xy}\}$
\cite[Section 4.3]{Tasaki22}:
\begin{itemize}
\item[(i)] $\Lambda$ is a triangular, square cubic, fcc or bcc lattice;
\item[(ii)] $t_{xy}$ is nonvanishing between nearest neighbor sites. $\diamondsuit$
 \end{itemize} 
}
\end{example}

\subsubsection{The Nagaoka-Thouless  theorem}
Let us consider the Hubbard model $H_{\mathrm{H}}$. We assume the
following:
\begin{itemize}
\item[{\bf (B. 1)}] $t_{xy} \ge 0$ for all $x, y\in \Lambda$.
\item[{\bf (B. 2)}] $\Lambda$ has the hole-connectivity associated with $\{t_{xy}\}$.
\end{itemize} 
For simplicity,  let us suppose the following condition:
\begin{itemize}
\item[{\bf (B. 3)}] The on-site Coulomb energy is independent of $x$: $U_{xx}=U$ for all $x\in \Lambda$.
\end{itemize}
We are interested in the $N=|\Lambda|-1$ electron system. Thus, we will
study the restricted Hamiltonian:
\begin{align}
H_{\mathrm{H}, |\Lambda|-1}=H_{\mathrm{H}} \restriction \mathfrak{E}_{n=|\Lambda|-1}
\end{align}

First, we derive the effective Hamiltonian describing the system with
$U=\infty$. To this end, we introduce the Gutzwiller projection by 
\begin{align}
P_{\mathrm{G}}=\prod_{x\in \Lambda}(1-n_{x\uparrow} n_{x\downarrow}). \label{GutzP}
\end{align} 
$P_{\mathrm{G}}$ is the orthogonal projection onto the subspace with no doubly
occupied sites.

\begin{Prop}[\cite{Miyao6,Tasaki2}]\label{EffHami1}
We define the effective Hamiltonian by $H_{\mathrm{H}}^{\infty}
=P_{\mathrm{G}}H_{\mathrm{H}, |\Lambda|-1}^{U=0}P_{\mathrm{G}}
$, where $H_{\mathrm{H}, |\Lambda|-1}^{U=0}$ is the Hamiltonian
 $H_{\mathrm{H}, |\Lambda|-1}$ with $U=0$. For all $z\in \BbbC\backslash
 \BbbR$, we have
\begin{align}
\lim_{U\to \infty} \big(H_{\mathrm{H}, |\Lambda|-1}-z\big)^{-1}
=\big(
H_{\mathrm{H}}^{\infty}-z
\big)^{-1}P_{\mathrm{G}}
\end{align}
in the operator norm topology. 
\end{Prop} 
We set $\mathfrak{E}^{U=\infty}=P_{\mathrm{G}}\mathfrak{E}_{n=|\Lambda|-1}$. 
The restriction of $H_{\mathrm{H}}^{\infty}$ onto
$\mathfrak{E}^{U=\infty}$ is denoted by the same symbol, if no confusion occurs.

In \cite{Tasaki2}, Tasaki extended Nagaoka's theorem as follows.

\begin{Thm}[Generalized Nagaoka-Thouless  theorem]\label{NagaokaTThm1}
Assume {\bf (B. 1)}, {\bf (B. 2)} and {\bf (B. 3)}. 
The ground state of
 $H_{\mathrm{H}}^{\infty}$ has total spin $S=(|\Lambda|-1)/2$ and is
 unique apart from the trivial $(2S+1)$-degeneracy.
\end{Thm}

\subsubsection{The Nagaoka-Thouless  stability class }

In Section \ref{StaNTThm}, we will prove the following theorem,  which is a special case of Theorem \ref{SummaryMain}. Note that we choose $Q$ as $Q=P_{\mathrm{G}}$ in Theorem \ref{QAns3} below.
We readily check that $P_{\mathrm{G}}$ satisfies the condition {\bf (Q)}.

\begin{Thm}\label{QAns3}
Let $H_{\mathrm{H}}^{\infty}$ be the Hamiltonian given in Proposition \ref{EffHami1}. 
The Hamiltonian $
H_{\mathrm{H}}^{\infty}
$ satisfies {\bf (U. 1)} and {\bf (U. 2)} with $S_*=(|\Lambda|-1)/2$. Hence,  by Theorem \ref{SummaryMain},
the  $H_{\mathrm{H}}^{\infty}$-stability class  $\mathscr{U}(H_{\mathrm{H}}^{\infty})$ has the following 
properties: For each $H\in \mathscr{U}(H_{\mathrm{H}}^{\infty})$, the ground state of  $H$ has total spin
$S=(|\Lambda|-1)/2$, and is unique apart from the trivial $(2S+1)$-degeneracy.
\end{Thm}
\begin{rem}
{\rm
$\mathscr{U}(H_{\mathrm{NT}})$ is called the {\it Nagaoka-Thouless  stability class}. $\diamondsuit$
}
\end{rem}

\subsubsection{Stability of the Nagaoka-Thouless  theorem I}
Let us consider the Holstein-Hubbard Hamiltonian $H_{\mathrm{HH}}$.
We will study the $N=|\Lambda|-1$ electron system. Hence, 
we concentrate our attention on the
Hamiltonian 
\begin{align}
H_{\mathrm{HH}, |\Lambda|-1}=H_{\mathrm{HH}}\restriction \mathfrak{E}_{n=|\Lambda|-1}\otimes \Fock_{\mathrm{ph}}.
\end{align}  
As before, we can derive an effective Hamiltonian describing the system
with $U=\infty$.

\begin{Prop}[\cite{Miyao6}]\label{EffHami2}
We define the effective Hamiltonian by $H_{\mathrm{HH}}^{\infty}
=P_{\mathrm{G}}H_{\mathrm{HH}, |\Lambda|-1}^{U=0}P_{\mathrm{G}}
$, where $H_{\mathrm{HH}, |\Lambda|-1}^{U=0}$ is the  Hamiltonian
 $H_{\mathrm{HH}, |\Lambda|-1}$ with $U=0$. For all $z\in \BbbC\backslash
 \BbbR$, we have
\begin{align}
\lim_{U\to \infty} \big(H_{\mathrm{HH}, |\Lambda|-1}-z\big)^{-1}
=\big(
H_{\mathrm{HH}}^{\infty}-z
\big)^{-1}P_{\mathrm{G}}
\end{align}
in the operator norm topology. 
\end{Prop} 

The restriction of $H_{\mathrm{HH}}^{\infty}$ to
$\mathfrak{E}^{U=\infty}\otimes \Fock_{\mathrm{ph}}$  is denoted by the same symbol.
\begin{Thm}[Stability I]\label{StaNT1}
Assume {\bf (B. 1)}, {\bf (B. 2)}and {\bf (B. 3)}. 
Then  $H_{\mathrm{HH}}^{\infty}$ belongs to $\mathscr{U}(H_{\mathrm{H}}^{\infty})$. 
Thus, by Theorem \ref{QAns3}, the ground state of
 $H_{\mathrm{HH}}^{\infty}$ has total spin $S=(|\Lambda|-1)/2$ and is
 unique apart from the trivial $(2S+1)$-degeneracy.
\end{Thm} 

We will prove Theorem \ref{StaNT1} in Section \ref{StaNTThm}.

\subsubsection{Stability of the Nagaoka-Thouless theorem II}

Here, we will study the Hamiltonian $H_{\mathrm{rad}, |\Lambda|-1}
=H_{\mathrm{rad}}\restriction\mathfrak{E}_{n=|\Lambda|-1}\otimes \Fock_{\mathrm{rad}}$.

\begin{Prop}[\cite{Miyao6}]\label{EffHami3}
We define the effective Hamiltonian by $H_{\mathrm{rad}}^{\infty}
=P_{\mathrm{G}}H_{\mathrm{rad}, |\Lambda|-1}^{U=0}P_{\mathrm{G}}
$, where $H_{\mathrm{rad}, |\Lambda|-1}^{U=0}$ is the Hamiltonian
 $H_{\mathrm{rad}, |\Lambda|-1}$ with $U=0$. For all $z\in \BbbC\backslash
 \BbbR$, we have
\begin{align}
\lim_{U\to \infty} \big(H_{\mathrm{rad}, |\Lambda|-1}-z\big)^{-1}
=\big(
H_{\mathrm{rad}}^{\infty}-z
\big)^{-1}P_{\mathrm{G}}
\end{align}
in the operator norm topology. 
\end{Prop} 

As before, we express  the restriction of $H_{\mathrm{rad}}^{\infty}$ to
$\mathfrak{E}^{U=\infty}\otimes \Fock_{\mathrm{rad}}$ by the same symbol.
\begin{Thm}[Stability II]\label{StaNT2}
Assume {\bf (B. 1)}, {\bf (B. 2)} and {\bf (B. 3)}.
Then $H_{\mathrm{rad}}^{\infty}$ belongs to $\mathscr{U}(H_{\mathrm{H}}^{\infty})$. 
  Hence,  by Theorem \ref{QAns3}, the ground state of
 $H_{\mathrm{rad}}^{\infty}$ has total spin $S=(|\Lambda|-1)/2$ and is
 unique apart from the trivial $(2S+1)$-degeneracy.
\end{Thm} 

We will provide a proof of Theorem \ref{StaNT2} in Section \ref{StaNTThm}.

\subsection{Organization}
The organization of the paper is as follows: In Section \ref{GeneralThy}, we first  introduce 
operator theoretic correlation inequalities. 
Using these operator inequalities, we
establish a general  framework of stability of ferromagnetism.

In Section \ref{ManyElUniTh}, we construct a theory of stability of ferromagnetism in many-electron systems.
We will give a proof of Theorem \ref{SummaryMain} stated in Section \ref{SecResults}.
Within this theory, we discuss the existence of ferromagnetism and long-range orders in the ground state.

In Section \ref{MLMLiebSta}, we study Lieb\rq{}s theorem from a viewpoint of  stability class:
We define  the Marshall-Lieb-Mattis  stability class, and show that 
various models belong to this  class.  Stabilities of Lieb\rq{}s theorem are immediate consequences of this fact.

In Section \ref{StaNTThm}, we study the  Nagaoka-Thouless theorem  from a viewpoint of  stability class:
the Nagaoka-Thouless  stability class  is introduced; we prove that some models belong
to this class. As immediate results, we  show Theorems \ref{StaNT1} and \ref{StaNT2}.

In Section \ref{ConRemarks}, we present concluding remarks. 

Appendices \ref{ConstSC}--\ref{PfLi4} are devoted to proving the results in Section \ref{MLMLiebSta}.
We provide a proof of a theorem stated  in Section \ref{ConRemarks} in Appendix \ref{ProofCl1}.
In Appendix \ref{PfStaNTThm}, we prove the theorems stated  in Section \ref{StaNTThm}.
In Appendices \ref{App1} and \ref{App2}, we collect useful propositions
concerning our operator theoretic correlation inequalities. These propositions will be used repeatedly in this study.

\begin{flushleft}
{\bf Acknowledgements.}
\end{flushleft}  
I am  grateful to  J. Fr\"ohlich  and the anoymous referees for useful comments. 
It is a pleasure to thank Keiko Miyao  for drawing the figures.
I  thank the Mathematisches Forschungsinstitut Oberwolfach
for its hospitality.
I also thank A. Arai for financial support.
This work was partially supported by KAKENHI 	15K04888 and  KAKENHI 18K0331508.

\section{General theory of stability of ferromagnetism}\label{GeneralThy}
\setcounter{equation}{0}

\subsection{Preliminaries}
In order to exhibit our results, we introduce some useful terminologies in this subsection.
Many of results  here are taken from  \cite{Miyao1, Miyao2, Miyao4, Miyao5, Miyao7}.

\subsubsection{Self-dual cones}
Let $\h$ be a complex Hilbert space.
By a {\it convex  cone}, we understand a closed convex set  $\Cone\subset \h$
such that $t\Cone \subseteq \Cone$ for all $t\ge 0$ and $\Cone\cap (-\Cone)=\{0\}$.

\begin{define}{\rm
The {\it dual cone   of} $\Cone$ is defined by 
$
\Cone^{\dagger}=\{\eta\in \h\, |\, \la \eta|\xi\ra\ge 0\ \forall \xi\in
\Cone \}.
$
 We say that $\Cone$ is {\it self-dual} if 
$
\Cone=\Cone^{\dagger}. 
$ $\diamondsuit$
}\end{define} 
 
 In what follows, we always assume that $\Cone$ is self-dual.

\begin{define}
{\rm 
\begin{itemize}
\item A vector $\xi$ is said to be  {\it positive w.r.t. $\Cone$} if $\xi\in
 \Cone$.  We write this as $\xi\ge 0$  w.r.t. $\Cone$.

 \item A vector $\eta\in \Cone$ is called {\it strictly positive
w.r.t. $\Cone$},  whenever $\la \xi| \eta\ra>0$ for all $\xi\in
\Cone\backslash \{0\}$. We write this as $\eta>0 $
w.r.t. $\Cone$. $\diamondsuit$

\end{itemize} 
}
\end{define} 

\begin{example}\label{Cone1}{\rm
Let $\mathfrak{X}$ be a finite dimensional Hilbert space. Let $\{x_n\}_{n=1}^N$ be a complete
orthonormal system in $\mathfrak{X}$.
We set $\Cone=\mathrm{Coni}\{x_n\}_{n=1}^N$, where $\mathrm{Coni}(S)$ is  the conical hull of $S$.
Then $\Cone$ is a self-dual cone. $x\ge 0$ w.r.t. $\Cone$, if and only if $\la x_n|x\ra\ge 0$ for all $n\in \BbbN$. Moreover, $x>0$ w.r.t. $\Cone$, if and only if $\la x_n|x\ra>0$ w.r.t. $\Cone$ for all $n\in \BbbN$.
$\diamondsuit$
}
\end{example}

\begin{example}\label{Cone2}
{\rm 
Let $(M, \mathfrak{M}, \mu)$ be a $\sigma$-finite measure space.
We  set 
\begin{align}
L^2(M, d\mu)_+=\{f\in L^2(M, d\mu)\, |\, f(m)\ge 0\ \ \mbox{$\mu$-a.e.}
\}.
\end{align}
$L^2(M, d\mu)_+$ is a self-dual cone in $L^2(M, d\mu)$. (Indeed, the reader can easily check the  
conditions (i)--(iii) of Theorem \ref{SAH}.)
$f\ge 0$ w.r.t. $L^2(M, d\mu)_+$
 if and only if $f(m) \ge 0$ $\mu$-a.e.  Furthermore, $f >0$
 w.r.t. $L^2(M, d\mu)_+$ if and only if $f(m)>0$ $\mu$-a.e. 
 $\diamondsuit$ 
}
\end{example}

\subsubsection{Operator inequalities associated with self-dual cones}

In  subsequent  sections, we  use the following operator inequalities.

We denote by  $\mathscr{B}(\h)$  the set of all bounded linear operators on
$\h$.
\begin{define}{\rm 

Let $A, B\in \mathscr{B}(\h)$. Let $\Cone$ be a self-dual cone in $\h$.

 If $A \Cone\subseteq \Cone,$\footnote{
For each subset $\mathfrak{C}\subseteq \h$, $A\mathfrak{C}$ is
	     defined by $A\mathfrak{C}=\{Ax\, |\, x\in \mathfrak{C}\}$.
} we then 
write  this as  $A \unrhd 0$ w.r.t. $\Cone$.\footnote{This
 symbol was introduced by Miura \cite{Miura}.} In
	     this case, we say that {\it $A$ preserves the
positivity w.r.t. $\Cone$.}
	     $\diamondsuit$
} 
\end{define} 

\begin{rem}\label{Pequiv}
{\rm
$A\unrhd 0$ w.r.t. $\Cone$ $\Longleftrightarrow$ $ \la \xi |A\eta\ra\ge
 0$
for all $\xi, \eta\in \Cone$. $\diamondsuit$
}
\end{rem} 

\begin{example}
{\rm
Let $\Cone$ be a self-dual cone defined in Example \ref{Cone1}.
A matrix $A$ acting on $\mathfrak{X}$ satisfies $A\unrhd 0$ w.r.t. $\Cone$,
if and only if $\la x_m|Ax_n\ra \ge 0$ for all $m,n\in \{1, \dots, N\}$. $\diamondsuit$
}
\end{example}
{\it Proof.} Let $x, y\in \Cone$. We can express $x, y$ as 
$
x=\sum_{n=1}^N\alpha_n x_n
$ and $y=\sum_{n=1}^N\beta_n x_n$ with $\alpha_n, \beta_n\ge 0$.
Since $\la x|Ay\ra=\sum_{m, n=1}^N \la x_m|Ax_n\ra \alpha_n\beta_m$, we conlude the desired assertion by Remark \ref{Pequiv}. $\Box$

\begin{example}
{\rm 
Let us consider the self-dual cone $L^2(M, d\mu)_+$ defined in Example \ref{Cone2}.
Let $F\in L^{\infty}(M, d\mu)$. We identify $F$ with  the multiplication operator by $F$.
Then $F\unrhd 0$ w.r.t. $L^2(M, d\mu)_+$, if and only if $F(m) \ge 0$ $\mu$-a.e. $\diamondsuit$
}
\end{example}

\begin{define}
{\rm 
Let $A\in \mathscr{B}(\h)$.
We write  $A\rhd 0$ w.r.t. $\Cone$, if  $A\xi >0$ w.r.t. $\Cone$ for all $\xi\in
\Cone \backslash \{0\}$. 
 In this case, we say that {\it $A$ improves the
positivity w.r.t. $\Cone$.} $\diamondsuit$
}
\end{define}

\begin{example}\label{HarmonicOS}
{\rm 
Consider a Hilbert space $L^2(\BbbR^n)$. By Example \ref{Cone2}, $L^2(\BbbR^n)_+$ 
is a self-dual cone in $L^2(\BbbR^n)$. Let $E=-\Delta+x^2$, where  $\Delta$
 is the $n$-dimensional Laplacian. It is well-known that $e^{-\beta E} \rhd 0$ w.r.t. $L^2(\BbbR^n)_+$
  for all $\beta > 0$, see, e.g., \cite[Theorems XIII.44 and XIII. 47]{ReSi4}. $\diamondsuit$
}
\end{example}

\subsection{Definitions and results}

Let $\mathfrak{H}$ be a complex Hilbert space. Let $A$ be a linear
 operator on  $\mathfrak{H}$. We say that  $A$ has
 purely discrete spectrum,  if
 essential spectrum of $A$ is empty.
In the present study, the following class of linear operators is important:
\begin{align}
\mathscr{O}(\mathfrak{H})=\{A\  |\,
 \mbox{$A$ is self-adjoint,  and has purely discrete spectrum}\}.
\end{align}

\begin{define}\label{DefCO}
{\rm 
Let $O\in \mathscr{O}(\mathfrak{H}) \cap \mathscr{B}(\mathfrak{H})$.
  We say that a Hamiltonian $H\in \mathscr{O}(\mathfrak{H})$ is in $ \mathscr{C}_0 (O)$, 
if the following conditions are satisfied:
\begin{itemize}
\item[{\bf (H. 1)}]  $H$ is bounded from below;
\item[{\bf (H. 2)}]  $e^{isO} e^{itH}=e^{itH}e^{isO}$ for all $s, t\in \BbbR$.
\end{itemize}
We remark that $H\in \mathscr{C}_0(O)$ could be  unbounded in general. $\diamondsuit$
} 
\end{define}

\begin{define}\label{DefCO2}
{\rm
For a given  self-dual cone $\Cone$  in $\mathfrak{H}$, 
let $\mathscr{C}_0(O, \Cone)$ be the set of all Hamiltonians in $\mathscr{C}_0(O)$
 satisfying the following condition:
\begin{itemize}
\item[{\bf (H. 3)}]  $(H+s)^{-1} \rhd 0$ w.r.t. $\Cone$ for all $s>-E(H)$, where $E(H)=\inf \mathrm{spec}(H)$.
\end{itemize} 
Now we define an important family of Hamiltonians by
$
\mathscr{C}(O)=\bigcup_{\Cone} \mathscr{C}_0(O, \Cone),
$
where the union runs over all self-dual cones in $\mathfrak{H}$.
 $\diamondsuit$
 }
\end{define} 

\begin{rem}\label{PhysMeaning}
{\rm
\begin{itemize}
\item {\bf (H. 1)} implies that $H$ has ground states and $E(H)>-\infty$.
\item {\bf (H. 2)} is equivalent to the condition  that spectral
      measures of $H$ and $O$ commute with  each
      other.
Physically, this means that the eigenvalues of $O$ are  good quantum
      numbers. 

\item By Theorem \ref{PFF},
 {\bf (H. 3)} implies  uniqueness of ground states of $H$. $\diamondsuit$
\end{itemize} 
}
\end{rem} 

\begin{define}\label{FirstSt}
{\rm 
Let $\mathfrak{H}$ be a complex Hilbert space and let $\mathfrak{H}_*$ be a closed subspace of $\mathfrak{H}$.
Let $\Cone$ (resp. $\Cone_*$) be a self-dual cone in $\mathfrak{H}$ (resp. $\mathfrak{H}_*$).
Let $P$ be the  orthogonal projection onto $\mathfrak{H}_*$. We suppose that 
\begin{align}
[O, P]=0.\label{OPC}
\end{align}
Note  that $O\restriction \mathfrak{H_*}\in \mathscr{O}(\mathfrak{H}_*)$ by (\ref{OPC}).

We consider two Hamiltonians $H\in \mathscr{C}(O)$ and $H_*\in \mathscr{C}(O\restriction \mathfrak{H}_*)$.
(Note that $H$ (resp. $H_*$) acts on $\mathfrak{H}$ (resp. $\mathfrak{H}_*$).)
Let $U\in \mathscr{B}(\mathfrak{H})$ be a unitary operator which commutes with $O$.
If $H, H_*$ and $U$ satisfy the
 following (i)--(iv), then 
we write this as $H \leadsto H_* $:
\begin{itemize}
\item[{\rm (i)}] $P\unrhd 0$ w.r.t. $\Cone$;
\item[{\rm (ii)}] $\Cone_*=P\Cone$;
\item[{\rm (iii)}] $U (H+s)^{-1} U^{-1}\rhd 0$ w.r.t. $\Cone$ for all $s >-E(H)$;
\item[{\rm (iv)}] $( H_*+s)^{-1} \rhd 0 $ w.r.t. $\Cone_*$ for all $s>-E(H_*)$. $\diamondsuit$
\end{itemize} 

}
\end{define}

The following proposition will be often useful.

\begin{Prop}\label{UniEQ}
 Let $H\in \mathscr{C}(O)$ and $H_*\in \mathscr{C}(O\restriction \mathfrak{H}_*)$.
 Assume that $H\leadsto H_*$.  Let $V$ be a unitary operator.
  Then one has  
  \begin{align}
  VHV^{-1}\in \mathscr{C}(VOV^{-1}),\ VH_*V^{-1}\in \mathscr{C}(VOV^{-1}\restriction V\mathfrak{H}_*).\label{CC}
  \end{align}
   In addition,  $VHV^{-1}  \leadsto VH_*V^{-1}$.
  
\end{Prop}
{\it Proof.} 
Note  that $V\Cone$ and $V\Cone_*$ are self-dual cones in $V\h$ and $V\h_*$, respectively.
To show (\ref{CC}) is easy. To prove that  $VHV^{-1}  \leadsto VH_*V^{-1}$, 
we will confirm all conditions in Definition \ref{FirstSt} with the following correspondence:
\begin{align}
&(H, \ H_*;\  \mathfrak{H},\  \mathfrak{H}_*;\  P;\  \mathfrak{P}, \ \mathfrak{P}_*; O; U)\no
\leftrightarrow&(VHV^{-1}, \ VH_*V^{-1};\  V\mathfrak{H},\  V\mathfrak{H}_*;\  VPV^{-1};\  V\mathfrak{P}, \ V\mathfrak{P}_*; VOV^{-1}; VUV^{-1}).
\end{align}
By (i) of Definition \ref{FirstSt}, we have $VPV^{-1} \unrhd 0$ w.r.t. $V\Cone$.  In addition, we have $V\Cone_*= VPV^{-1} V\Cone$, so that (ii) of Defintion of \ref{FirstSt} is satisfied. 
We readily check that $
VU ( H+s)^{-1}U^{-1}V^{-1} \rhd 0
$ w.r.t. $V\Cone$  for all $s>-E(H)$ and $V ( H_*+s)^{-1} V^{-1} \rhd 0$ w.r.t. $V\Cone_*$ for all $s>-E(H_*)$. Thus, (iii) and (iv) of Definition \ref{FirstSt} are fulfilled.
 $\Box$
\medskip

Let $H\in \mathscr{C}(O)$ be a Hamiltonian.
A nonzero vector in $\ker(H-E(H))$ is called a {\it ground state of}
$H$.
As mentioned in Remark \ref{PhysMeaning}, $\dim \ker(H-E(H))=1$.
Now, let $\psi\in \ker(H-E(H))$ be  the   normalized ground state of $H$.
Since $O$ is conserved by {\bf (H. 2)}, there exists a $\mu(H)\in \mathrm{spec}(O)$
such that 
\begin{align}
O\psi=\mu(H)\psi.
\end{align} 
Our purpose in this section is to study properties of $\mu(H)$.

\begin{Thm}\label{AbstTh}
If  $H  {\leadsto }H_* $,  then $\mu(H)=\mu(H_*)$.
\end{Thm} 

\begin{rem}
{\rm 
We explain the definition  of $\mu(H_*)$ just in case.
By (iv) of Definition \ref{FirstSt}, ground states of $H_*$ is unique. 
Let $\psi_*$ be the unique ground state of $H_*$.
  Because $H_*$ commutes with $O\restriction \mathfrak{H}_*$, $\psi_*$ must be
 an eigenvector of $O$.  The corresponding eigenvalue is  denoted by  $\mu(H_*)$. $\diamondsuit$

}
\end{rem}

We will provide a proof of Theorem \ref{AbstTh} in Section \ref{PfUni}.

Next, we will generalize Definition \ref{FirstSt} for later use.
Let $P$ be an orthogonal projection on $\mathfrak{H}$ such that 
\begin{align}
[O, P]=0. \label{OPO}
\end{align}
We set $O\restriction P:= O\restriction \R(P)$.
We introduce  the following set of Hamiltonians:
\begin{align}
\mathscr{P}_0(O)=\bigcup_{P\ \mathrm{satisfying } \ (\ref{OPO})} \mathscr{C}(O\restriction P).
\end{align}
We extend the binary relation \lq\lq{}$\leadsto$\rq\rq{} to $\mathscr{P}_0(O)$ as follows.

\begin{define}\label{FirstSt2}
{\rm 
Let $H_1, H_2\in \mathscr{P}_0(O)$. Then there exist orthogonal projections $P_1$ and $P_2$
such that $
H_1\in \mathscr{C}(O\restriction P_1)
$
and 
$
H_2\in \mathscr{C}(O\restriction P_2)
$.
Let $\Cone_1$ and $\Cone_2$ be self-dual cones in $\R(P_1)$ and $\R(P_2)$, respectively.
Let $U\in \mathscr{B}(\R(P_1))$ be a unitary operator which commutes with $O\restriction P_1$.
If  the
 following (i)--(v) are satisfied, then 
we write this as $H_1 \leadsto H_2 $:
\begin{itemize}
\item[{\rm (i)}] $P_1\ge P_2;$
\item[{\rm (ii)}] $P_2\unrhd 0$ w.r.t. $\Cone_1$;
\item[{\rm (iii)}] $\Cone_2=P_2\Cone_1$;
\item[{\rm (iv)}] $
U (H_1+s)^{-1}U^{-1} \rhd 0$ w.r.t. $\Cone_1$ for all $s>-E(H_1)$;
\item[{\rm (v)}] $(H_2+s)^{-1}\rhd 0 $ w.r.t. $\Cone_2$ for all $s>-E(H_2)$.
 $\diamondsuit$
\end{itemize} 

}
\end{define}

\begin{rem}\label{EquivDD}
{\rm
Definition \ref{FirstSt2} can be regarded as 
a special case of Defintion \ref{FirstSt} with
\begin{align}
(H, H_*; \mathfrak{H}, \mathfrak{H}_*; P; \Cone, \Cone_*; O; U)
=(H_1, H_2; \R(P_1), \R(P_2); P_2; \Cone_1, \Cone_2; O\restriction P_1; U). \ \ \ \diamondsuit
\end{align}

}
\end{rem}

By Theorem \ref{AbstTh} and Remark \ref{EquivDD}, we have the following:

\begin{Thm}
Let $H_1, H_2\in \mathscr{P}_0(O)$. If $H_1\leadsto H_2$, then
$\mu(H_1)=\mu(H_2)$.
\end{Thm}

\begin{coro}\label{OBS}
Let $H_1, \dots, H_n\in \mathscr{P}_0(O)$. If 
\begin{align}
H_1\leadsto H_2, H_2\leadsto H_3, \dots, H_{n-1}\leadsto H_n,\label{ChainO}
\end{align}
 then $\mu(H_1)=\mu(H_n)$.
\end{coro}

In what follows, we write the condition (\ref{ChainO}) simply as 
\begin{align}
H_1\leadsto H_2\leadsto \dots \leadsto H_{n-1}\leadsto H_n.
\end{align}

By the above observation, we arrive at the following definition:
\begin{define}\label{DefLeads}
{\rm 
Let $H, H_*\in \mathscr{P}_0(O)$. If there exists a sequence
of Hamiltonians  $H_1, \dots, H_n\in \mathscr{P}_0(O)$ such that 
$
H \leadsto H_1\leadsto H_2 \leadsto\dots \leadsto  H_{n}\leadsto H_*,
$
then we write this as $H\Leads H_*$.
Note that if $H\Leads H_*$, then $\mu(H)=\mu(H_*)$
by Corollary \ref{OBS}. 
Even if there is no sequence satisfying the above condition but $H\leadsto H_*$, we still express this as 
  $H\Leads H_*$. 
$\diamondsuit$
}
\end{define}

The following proposition is important.
\begin{Prop}\label{naminami}
The binary relation \lq\lq{} $\Leads$\rq\rq{} is a preorder on $\mathscr{P}_0(O)$; namely, we have the following:
\begin{itemize}
\item[{\rm (i)}] $H\Leads H$; 
\item[{\rm (ii)}] $H_1\Leads H_2,\ \  H_2 \Leads  H_3 \Longrightarrow H_1\Leads
 H_3 $.
\end{itemize} 
\end{Prop}
{\it Proof.}  (i)
Because $H\in \mathscr{P}_0(O)$, there exists an orthogonal projection $P$ such that $H\in \mathscr{C}(O\restriction P)$. In particular, there is a self-dual cone $\Cone$ in $\R(P)$ such that $(H+s)^{-1}\rhd 0$ w.r.t. $\Cone$ for all $s>-E(H)$.
It is easy to  check all conditions in Definition \ref{FirstSt} with
\begin{align}
(H, H_*; \mathfrak{H}, \mathfrak{H}_*; P; \Cone, \Cone_*; O; U)
=(H, H; \mathfrak{H}, \mathfrak{H}; 1; \Cone, \Cone; O; 1).
\end{align}
(Or we can directly check all conditions in   Definition \ref{FirstSt2}.) Thus $H\leadsto H$, which implies (i).
 (ii) immediately follows from Definition \ref{DefLeads}. $\Box$

\begin{define}\label{HEquiv}
{\rm 
Let $H_1, H_2\in \mathscr{P}_0(O)$.
If $H_1\Leads H_2$ and $H_2\Leads H_1$, then we write this as  $H_1\equiv H_2$.
The binary relation \lq\lq{} $\equiv$ \rq\rq{} is an equivalence relation on $\mathscr{P}_0(O)$.

Let $\mathscr{P}(O)$ be the set of equivalence classes:  $\mathscr{P}(O)=\mathscr{P}_0(O) / \equiv$.
The equivalence class containing $H$ is denoted  by $[H]$.
The binary relation \lq\lq{}$\Leads$\rq\rq{} on $\mathscr{P}(O)$ is naturally  defined by 
$
[H_1] \Leads [H_2]\ \ \ \mbox{if}\ \ H_1\Leads H_2.
$
This is a {\it partial order} on $\mathscr{P}(O)$; namely, we have, by Proposition \ref{naminami},
\begin{itemize}
\item[(i)] $[H] \Leads [H]$;
\item[(ii)] $[H_1] \Leads [H_2],\ [H_2] \Leads [H_1]\Longrightarrow [H_1] =[H_2]$; 
\item[ (iii)] $[H_1]\Leads [H_2],\   [H_2] \Leads  [H_3] \Longrightarrow [H_1]\Leads
 [H_3]$. $\diamondsuit$
\end{itemize}
}
\end{define}

We abbreviate $[H]$ simply  as $H$, if no confusion occurs.

Let $[H]\in \mathscr{P}(O)$.  Remark that if $H_1\equiv H_2$, then $\mu(H_1)=\mu(H_2)$
by Theorem \ref{AbstTh}. Thus, it is natural to define $\mu([H])$ by $\mu([H]):=\mu(H)$.
We also abbreviate $\mu([H])$ as $\mu(H)$.

By Corollary  \ref{OBS}, we immediately obtain the following:

\begin{Thm}\label{Coro1}
Let $H_1, H_2\in \mathscr{P}(O)$.
 If  $H_1 \Leads H_2$,  
 then $\mu(H_1)=\mu(H_2)$.

\end{Thm}

\begin{define}\label{StrictO}
{\rm
We define a binary relation  \lq\lq{}$\bl $\rq\rq{}
on $\mathscr{P}(O)$
by 
\begin{align}
H_1\bl H_2\ \mbox{ if $H_1\Leads H_2$ and $H_1\neq H_2.$}
\end{align}
The binary relation ``$\bl$''  is a {\it strict order}. Namely, 
\begin{itemize}
\item[(i)] $H\bl H$ does not hold for any $H\in \mathscr{P}(O)$; 
\item[(ii)] If $H_1\bl H_2$, then $H_2 \bl H_1$ does not hold; 
\item[(iii)] If $H_1\bl H_2$ and $ H_2 \bl H_3 $,  then $H_1\bl
 H_3 $ holds.
$\diamondsuit$
\end{itemize} 
}
\end{define}

\begin{Thm}\label{Coro2}
Let $H, H_*\in \mathscr{P}(O)$. Assume that  $H\bl H_*$.
Then we have 
   $\mu(H)=\mu(H_*)$.
\end{Thm}
{\it Proof.}   Use Theorem  \ref{Coro1}.
 $\Box$

\begin{define}\label{DefUniO}
{\rm
Let $H_*\in \mathscr{P}(O)$. The {\it $H_*$-stability class} $\mathscr{U}_O(H_*)$ is defined by 
\begin{align}
\mathscr{U}_O(H_*)=\{
H\in \mathscr{P}(O)\, |\, H\bl H_*
\} \cup \{H_*\}.
\end{align}
Remark that $H_*\notin  \{
H\in \mathscr{P}(O)\, |\, H\bl H_*
\}$ by (i) of Definition \ref{StrictO}. 
We can also express $\mathscr{U}_O(H_*)$ as 
$
\mathscr{U}_O(H_*)=\{H\in \mathscr{P}(O)\, |\, H \Leads H_*\}
$.
$\diamondsuit$
}
\end{define}

By Theorem  \ref{Coro2}, we conclude the following:
\begin{Thm}\label{Coro3}
For every Hamiltonian $H\in \mathscr{P}(O)$ in the $H_* $-stability class, we have $\mu(H)=\mu(H_*)$.
\end{Thm} 

This theorem is a prototype of Theorem \ref{SummaryMain} in Section \ref{SecResults}; Indeed, 
we will apply the idea here to construct  stability classes in many electron systems.

\subsection{Proof of Theorem \ref{AbstTh}} \label{PfUni}

We begin with the following proposition:

\begin{Prop}\label{VecP2}
Let $A\in \mathscr{B}(\h)$ with $A\neq 0$. Assume that  $u>0$ w.r.t. $\Cone$.
If $A\unrhd 0$ w.r.t. $\Cone$, then $Au\neq 0$.
\end{Prop} 
{\it Proof.}
We will divide our proof into two parts.

{\bf Step 1.}
In this step, we prove the following claim:
 Let $A\in \mathscr{B}(\h)$. If $Au= 0$ for all $u\in \Cone$, then $A=0$.

 By Corollary \ref{DecRI}, each $u\in \h$ can be written as 
$
u=v_1-v_2+i(w_1-w_2)
$, where $v_1, v_2, w_1, w_2\in \Cone$ such that $\la v_1|v_2\ra=0$ and
$\la w_1|w_2\ra=0$. Thus, the assumption implies that $Au=0$
 for {\it all} $u\in \h$. 

{\bf Step 2.} In this step, we will  complete the proof of Proposition
\ref{VecP2}.

Assume that $Au=0$. Then,  $\la v|Au\ra=0$ for all $v\in \Cone$,
implying that  $\la A^*v|u\ra=0$.
Since $u>0$  and $A^* v\ge 0$ w.r.t. $\Cone$, we conclude that 
$A^*v$ must be zero. Because $v$ is arbitrary, 
$A^*=0$  by STEP 1. 
This contradicts with the assumption $A\neq 0$.
$\Box$
\medskip

\begin{flushleft}
{\it Completion of proof of Theorem \ref{AbstTh}}
\end{flushleft}

By (iii) of Definition \ref{FirstSt}, and Theorem \ref{PFF}, the ground state of
$UHU^*$ is unique and strictly positive w.r.t. $\Cone$.
Similarly, 
by (iv) of Definition \ref{FirstSt}, the ground state of $H_*$ is unique and strictly positive
w.r.t. $\Cone_*$.

Let $\psi$ be the ground state of $UHU^*$. Since $\psi>0$ w.r.t. $\Cone$ and
$P\unrhd 0$ w.r.t. $\Cone$ ((i) of Definition \ref{FirstSt}), we have $P\psi\neq 0$ by Proposition \ref{VecP2}.
Let $\psi_*$ be the ground state of $H_*$. Since $\psi_*>0$
w.r.t. $\Cone_*=P\Cone$ ((ii) of Definition \ref{FirstSt}) and $P\psi\ge 0$ w.r.t. $P\Cone$, we have 
$\la P \psi| \psi_*\ra>0$. Hence,
$
\mu(UHU^*) \la P \psi| \psi_*\ra =\la PO\psi|\psi_*\ra=\la P \psi| O
 \psi_*\ra
=\mu(H_*) \la P \psi| \psi_*\ra,
$
which implies that $\mu(UHU^*)=\mu(H_*)$.

Because $U^*\psi$ is the  unique ground state of $H$, we have 
$
O(U^*\psi)=\mu(H) U^* \psi
$. Since $U$ commutes with $O$, we have
$
\mu(UHU^*) U^*\psi=U^*O\psi=O(U^*\psi)=\mu(H) U^*\psi,
$
which implies that $\mu(H)=\mu(UHU^*)$. Combining this and the result in the last paragraph, we conclude
 that $\mu(H)=\mu(H_*)$. 
   $\Box$

\section{Stability of ferromagnetism in  many-electron systems}\label{ManyElUniTh}
\setcounter{equation}{0}
In this section, we construct  a theory of stability of ferromagnetism  in many-electron systems.
In addition, by taking crystal structures of $\Lambda$ into account, we study  ferromagnetism
 and long-range orders in the ground state from the perspective of the new theory
\subsection{Preliminaries}
First, we will introduce some basic terminologies and symbols.

\subsubsection{The $N$-electron subspaces}
Let $\mathfrak{X}$ be a complex Hilbert space.
  We consider a coupled system $\mathfrak{E}\otimes \mathfrak{X}$ in this section.
  Recall that $\mathfrak{E}$ can be regarded as a subspace of $\mathfrak{E}\otimes \mathfrak{X}$ by the 
  identification $\mathfrak{E}\cong \mathfrak{E}\otimes x_0$, where $x_0$ is some normalized vector in $\mathfrak{X}$, see Section \ref{BFMS} for details.
  Note that, if $\mathfrak{X}=\BbbC$, then we understand that 
  $
  \mathfrak{E}\otimes \BbbC=\mathfrak{E}. 
  $
   Let $\mathfrak{E}_n$ be the $n$-electron subspace defined by (\ref{Nsub}).
We also call $Q\mathfrak{E}_n$ and $\mathfrak{E}_n \otimes \mathfrak{X}$ the $n$-electron subspace, where $Q$ is an  orthogonal projection on $\mathfrak{E}_n$ satisfying {\bf (Q)}.

In the present paper, we will use the following notation:
\begin{Not}{\rm
Let $\mathfrak{Z}_n$ be  the $n$-electron subspace of $\mathfrak{Z}$:
 $\mathfrak{Z}_n=Q\mathfrak{E}_n$ or $\mathfrak{E}_n$ or $\mathfrak{E}_n\otimes \mathfrak{X}$.
 Let $X$ be a linear operator on 
$\mathfrak{Z}$ which commutes with $\Ne$.
We set 
$
X_{n}=X\restriction \mathfrak{Z}_n.
$
Remark that $X=\bigoplus_{n=0}^{2|\Lambda|} X_n$. $\diamondsuit$
}
\end{Not}

\subsubsection{The $M$-subspaces}
Here, we recall the definitions of the $M$-subspace and introduce some notations.
Let us consider the $n$-electron subspace $\mathfrak{E}_n$.
Note  that $
\mathrm{spec}(S^{(3)}_{\mathrm{tot}, n})=\{-n/2, -n/2+1, \dots, n/2\}
$. Thus, we have the following decomposition:
\begin{align}
\mathfrak{E}_n=\bigoplus_{M\in \mathrm{spec}(S^{(3)}_{\mathrm{tot}, n})} \mathfrak{E}_n[M],\ \
 \ \mathfrak{E}_n[M]=\ker(S^{(3)}_{\mathrm{tot}, n}-M). \label{HilDec}
\end{align} 
The subspace $\mathfrak{E}_n[M]$ is called the $M$-subspace.
The $M$-subspace of  $Q\mathfrak{E}_n$  (resp. $\mathfrak{E}_{n}\otimes \mathfrak{X}$) is  defined  by 
$
Q \mathfrak{E}_{n}[M] $ (resp. $
\mathfrak{E}_{n}[M] \otimes \mathfrak{X}
$).

\begin{Not}
{\rm 
Let $\mathfrak{Z}_n=Q\mathfrak{E}_n$ or $\mathfrak{E}_n$ or $\mathfrak{E}_n\otimes \mathfrak{X}$.
Let $\mathfrak{Z}_n[M]$ be  the $M$-subspace of $\mathfrak{Z}_n$.
Let $X_n$ be a linear operator on $\mathfrak{Z}_n$. Suppose that $X_n$
 commutes with $S_{\mathrm{tot}, n}^{(3)}$.
We set
$
X_n[M]=X \restriction \mathfrak{Z}_n[M].
$
Note  that $X_n=
\bigoplus_{M\in \mathrm{spec}(S_{\mathrm{tot}, n}^{(3)})} X_n[M]
$. $\diamondsuit$
}
\end{Not}

\subsection{ Stability classes in  many-electron systems}

We introduce an important  class of Hamiltonians.
To this end, let $P$ be an orthogonal projection on $\mathfrak{E}_n\otimes \mathfrak{X}$
 such that 
 \begin{align}
 [S_{\mathrm{tot}, n}^{(j)}, P]=0,\ \ \ j=1,2,3. \label{CPrj}
 \end{align}
 We set $
  S_{\mathrm{tot}, n}^{(j)} \restriction P:=S_{\mathrm{tot}, n}^{(j)} \restriction \R(P).
 $
 Define
 \begin{align}
 \mathscr{C}_{ \mathfrak{X}}^{\#}=\bigcup_{P\  {\rm satisfying}\  (\ref{CPrj})} \bigcap_{j=1}^3\mathscr{C}\Big(S_{\mathrm{tot}, n}^{(j)} \restriction P\Big),\label{CXD}
 \end{align}
 where $\mathscr{C}(O)$ is given by Definition \ref{DefCO2}.
 Here, we emphasize the $\mathfrak{X}$-dependence by the subscript $\mathfrak{X}$ in the LHS of 
 (\ref{CXD}).
 Clearly, $\mathscr{C}_{\mathfrak{X}}^{\#} \subseteq \mathscr{C}(S_{\mathrm{tot}, n}^2)$.

In what follows, we suppose that every Hamiltonian commutes with $\Ne$.
Of course, $H_n$ indicates  the restriction of $H$ onto the $n$-electron subspace. 
\begin{define}\label{ElUniD}
{\rm 
Let $H_{*, n}\in \mathscr{C}_{\mathfrak{X}=\BbbC}^{\#}$ be a Hamiltonian.
Assume that the ground state of $H_{*, n}$ has total spin $S_*$\footnote{Note that the ground state of $H_{*, n}$ is unique apart from the trivial $(2S_*+1)$-degeneracy.}.
  The $H_{*, n}${\it -stability class} $\mathscr{U}(H_{*, n})$ is defined by 
\begin{align}
&\mathscr{U}(H_{*, n}) \no
=&\{H_{*, n}\}\cup \bigcup _{\mathfrak{X}}\Big\{
H_n\in \mathscr{C}_{\mathfrak{X}}^{\#}\, \Big|\, \exists M\in \mathrm{spec}\Big(S_{\mathrm{tot}, n}^{(3)}\Big) 
 \mathrm{s.t.}\ \mbox{$|M| \le S_* $ and} \ H_n[M]
\bl H_{*, n}[M]
\Big\}.
\end{align}
where the union runs over all separable Hilbert spaces.
Note  that
$\mathscr{U}(H_{*, n})$ is a kind of generalization of $\mathscr{U}_O(H_{*. n})$ with
   $O=S_{\mathrm{tot}, n}^2[M]$ given in Definition \ref{DefUniO}.
$\diamondsuit$
}
\end{define}

The following proposition is convenient in the subsequent sections.
\begin{Prop}\label{Daig}
Let $\mathfrak{X}$ be a separable Hilbert space.
Let $H, H_1, \dots, H_K\in \mathscr{C}_{ \mathfrak{X}}^{\#}$ be Hamiltonians.
If  there exists an  $M\in \mathrm{spec}(S_{\mathrm{tot}, n}^{(3)})$ such that  $H[M]\leadsto H_1[M] \leadsto \cdots \leadsto H_K[M]\leadsto H_{*, n}[M]$, then $H$ belongs to the $H_{*, n}$-stability class.
\end{Prop}
{\it Proof.} The proposition immediately follows from  the definition of the strict order ``$\bl$'' (Definition \ref{StrictO}). $\Box$

\begin{example}\label{DiaEx}
{\rm 
Let us consider the following diagram:
\[
\xymatrix@C=6pt{
&& & & H_{*, n} \ar@{<~}[lld] \ar@{<~}[ld] \ar@{}[d]|(.6){\dots} \ar@{<~}[rd] \ar@{<~}[rrd]
\\
&&\cdots H_{1,1}\ar@{<~}[lld] \ar@{<~}[ld] \ar@{}[d]|(.6){\dots}  & H_{1,2} & \dots\dots & H_{1, m-1} & H_{1,m } \cdots\\
\cdots H_{2, 1}  & H_{2, 2} & \dots\dots &&
}
\]
Here, the symbol 
$
\xymatrix@C=24pt{
H_{i+1, j} \ar@{~>}[r]&H_{i,k}
}
$ indicates that there exists an $M\in \mathrm{spec}(S_{\mathrm{tot}, n}^{(3)})$ such that
$|M| \le S_*$ and 
  $
H_{i+1, j}[M] \leadsto H_{i, k}[M]
$. By  Proposition \ref{Daig}, every Hamiltonian $H_{i, j}\ (i, j=1,2,\dots)$ belongs to   the $H_{*, n}$- stability class. $\diamondsuit$
}
\end{example}

The following theorem  is an  abstract form of the stability of magnetism in  many-electron systems.

\begin{Thm}\label{ElUniTh}
Let $H_n\in \mathscr{U}(H_{*, n})$.
Assume that the ground state of $H_{*, n}$ has total spin $S_*$.
(Remark that the ground state of $H_{*, n}$ is unique apart from the trivial  $(2S_*+1)$-degeneracy.)
Then the ground state of $H_n$ has total spin $S_*$ and 
is unique apart from the trivial  $(2S_*+1)$-degeneracy, 
  as well.　
\end{Thm}
{\it Proof.} 
We divide the proof into several steps. 

{\bf Step 1.}
Throughout this proof, we assume that $H$ acts on a Hilbert space $\mathfrak{H}=\mathfrak{E}\otimes \mathfrak{K}$ for notational simplicity.
Note  that because $H_n\in \bigcap_{j=1}^3 \mathscr{C}\Big(S_{\mathrm{tot}, n}^{(j)}\Big)$, we have $H_n[M] \in \mathscr{C}\Big(
S_{\mathrm{tot}, n}^{2}[M]
\Big)$ for all $M$.
Suppose that  there is an $M\in \mathrm{spec}\Big(S^{(3)}_{\mathrm{tot}, n}\Big)$ such that
$|M| \le S_*$ and 
  $H_n[M]\bl H_{*, n}[M]$.
By Theorem \ref{Coro2}, the ground state of $H_n[M]$ is unique and has total spin $S_*$.
\medskip

{\bf Step 2.} 
Let us introduce linear operators $S_{\mathrm{tot}, n}^{(\pm)}$ by 
$S_{\mathrm{tot}, n}^{(\pm)}=S_{\mathrm{tot}, n}^{(1)}+iS_{\mathrm{tot}, n}^{(2)}$.
As usual, $\mathfrak{H}[M]$ indicates  the $M$-subspace of $\mathfrak{H}$.
Let $\vphi_M$ be the unique ground state of $H_n[M]:\ H_n[M]\vphi_M=E_M\vphi_M$.
Set $\vphi_{M+1} =S_{\mathrm{tot}, n}^{(+)} \vphi_M$.
Since $H_n[M+1]S_{\mathrm{tot}, n}^{(+)}=S_{\mathrm{tot}, n}^{(+)}H_n[M]$, we have 
$H_n[M+1]\vphi_{M+1} =E_M \vphi_{M+1}$.
Because $\vphi_{M+1}$ belongs to the $(M+1)$-subspace, we have $E_{M+1} \le E_M$.
Next, let $\psi_{M+1} $ be a ground state of $H_n[M+1]:\ H_n[M+1]\psi_{M+1}=E_{M+1} \psi_{M+1}$.
(At this stage, we have  not yet  proved the uniqueness.)
We set $\psi_M=S_{\mathrm{tot}, n}^{(-)} \psi_{M+1} \in \mathfrak{H}[M]$.
As before, we have $H_n[M]\psi_M=E_{M+1} \psi_M$, which implies $E_M\le E_{M+1}$.
Thus, $E_M=E_{M+1}$.
\medskip

{\bf Step 3.}
We will prove the uniqueness of the ground state of $H_n[M+1]$.
Assume that
there exist two ground states of $H_n[M+1]$, say
  $\phi_1$ and $\phi_2$,  such that 
\begin{align}
\la \phi_1|\phi_2\ra=0. \label{Perp}
\end{align}
Set $\Phi_1=S_{\mathrm{tot}, n}^{(-)} \phi_1$ and   $\Phi_2=S_{\mathrm{tot}, n}^{(-)} \phi_2$.
Then $\Phi_1$ and $\Phi_2$ are ground states of $H_n[M]$.
Because the ground state of $H_n[M]$ is unique, there is a constant $c\neq 0$
such that $\Phi_1=c\Phi_2$. Since $S_{\mathrm{tot}, n}^{(-)}$ is a bijective map
from $\mathfrak{H}[M+1]$ onto $\mathfrak{H}[M]$, we obtain $\phi_1=c\phi_2$, which contradicts with (\ref{Perp}). 
Thus, the ground state of $H_n[M+1]$ must be unique.
\medskip

{\bf Step 4.}
Repeating the arguments in {\bf Step 2} and {\bf Step 3}, we have $E_M=E_{M+\ell}$ and 
the ground state of $H_n[M+\ell]$ is unique for all $\ell\ge 0$ with $\ell+M\le S_*$.
Similarly, we can prove that $E_M=E_{M-\ell}$ and the ground state of $H_n[M-\ell]$ is unique
 for each $\ell\ge 0$ with $M-\ell \ge -S_*$. 
\medskip

 {\bf Step 5.}
 By {\bf Step 4}, we know that the ground state of $H_n[M\rq{}]$ is unique for all $M\rq{}\in  \mathrm{spec}\Big(S^{(3)}_{\mathrm{tot}, n}\Big)$ with $|M'| \le S_*$. Let  $\psi_{M\rq{}}$ be the unique ground state of $H_n[M']$.
 In this step, we show that each  $\psi_{M\rq{}}$ has total spin $S=S_*
 $.
 
 To this end, we observe that $\psi_{M+\ell}=\mathrm{const}. \Big(S_{\mathrm{tot}, n}^{(+)}\Big)^{\ell} \psi_M$ by the uniqueness.
 Since $S_{\mathrm{tot}, n}^2$ commutes with $S_{\mathrm{tot}, n}^{(+)}$,
 $\psi_{M+\ell}$ has total spin $S=
 S_*
 $ as well, provided that $M+\ell \le S_*$. Similarly, we can prove that $\psi_{M-\ell}$ has same total spin,
 provided that $-S_*\le M-\ell$. $\Box$

\begin{example}
{\rm 
Let us consider the diagram given in Example \ref{DiaEx}.
Assume that the ground state of $H_{*, n}$ has total spin $S_*$.
By Theorem \ref{ElUniTh},  the ground state of $H_{i, j}$ has total spin $S_*$ and 
is unique apart from the trivial  $(2S_*+1)$-degeneracy for each $i, j=1,2,\dots.$
$\diamondsuit$
}
\end{example}

\subsection{Ferromagnetism in the ground state}\label{FerrGr}

Let $\mathscr{P}$ be a Bravais lattice with the set of primitive vectors
$\{\ab_1, \dots, \ab_d\}$. Let $\Lambda_{\mathrm{B}}$ be a subset of $\mathscr{P}$
given by $
\Lambda_{\mathrm{B}}=
\{
n_1\ab_1+\cdots + n_d\ab_d\, |\, n_j\in \BbbZ\, \ -L+1 \le n_j\le L
\}$. We impose the periodic boundary condition:
\begin{align}
n_1\ab_1+\cdots+(L+1)\ab_j+\cdots+n_d\ab_d
\equiv n_1\ab_1+\cdots+(-L+1)\ab_j+\cdots+n_d\ab_d
\end{align}
  for each  $j=1, \dots, d$.
A crystal structure $\Lambda$ is determined by $\Lambda_{\mathrm{B}}$
and a basis $\{{\Bs e}_j\, |\, j=1, \dots, n-1\}$:
\begin{align}
\Lambda=\{{\boldsymbol r}+{\Bs e}_j\, |\, {\boldsymbol r}\in
 \Lambda_{\mathrm{B}},\ j=1, \dots, n-1\}\cup \Lambda_{\mathrm{B}},
\end{align} 
where we understand that $\Lambda=\Lambda_{\mathrm{B}}$ if $n=1$.
We assume that the basis $\{{\Bs e}_j\, |\, j=1, \dots, n-1\}$ is  consistent with the periodic boundary condition.

In the remainder of this section, we will take the crystal structure into consideration.
To emphasize $\Lambda$ dependence, we write $\mathfrak{E}_n$ as $\mathfrak{E}_{n, \Lambda}$.
Suppose that the Hilbert space $\mathfrak{X}$ also depends on $\Lambda$ (or $L$).
Therefore, we consider a family of Hilbert spaces $\{\mathfrak{E}_{n, \Lambda}\otimes \mathfrak{X}_{\Lambda}\}$. 
Corresponding to this, every Hamiltonians in the remainder of this section depends on $\Lambda$ or $L$.
We will specify the $\Lambda$-dependence as $H_n=H_{n, \Lambda}$.
In addition, 
we simply write $\mathscr{C}^{\#}_{\mathfrak{X}_{\Lambda}}$ as  $\mathscr{C}^{\#}_{\Lambda}$ to clarify the $\Lambda$-dependence.

\begin{example}[The Lieb lattice]\label{LiebL}
{\rm 
Consider a 2-dimensional Bravais lattice 
 $
\Lambda_{\mathrm{B}}=
\{
n_1\ab_1+ n_2\ab_2\, |\, n_1, n_2\in \BbbZ \cap [-L+1, L]
\}$ with $
{\Bs a}_1=(2, 0) $ and $ {\Bs a}_2=(0, 2)
$.
The {\it Lieb lattice} is defined by 
$
\Lambda=\{
{\Bs r}+{\Bs e}_j\, |\, {\Bs r} \in \Lambda, j=1,2
\} \cup \Lambda_{\mathrm{B}},
$
where ${\Bs e_1}=(1, 0),\ {\Bs e}_2=(0, 1)$,
 see Figure \ref{Lieblattice}. 
\begin{figure}[ht]
\centering\includegraphics[height=4cm, width=8cm, bb=0 0 800 400]{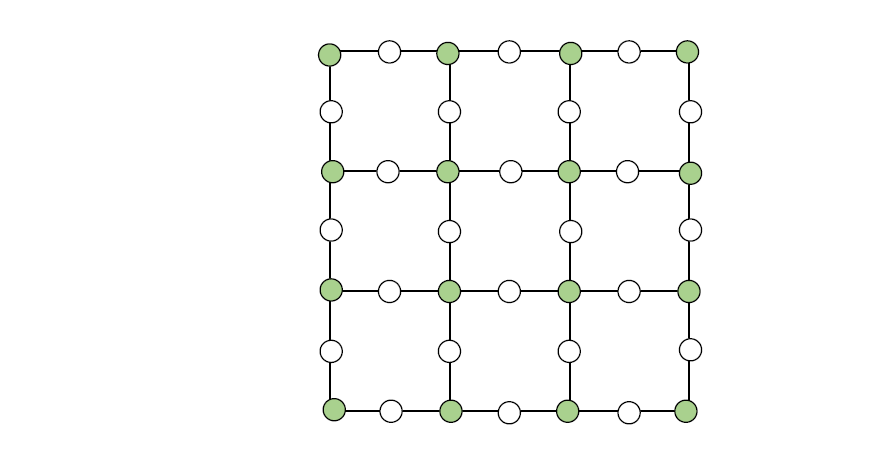}
\caption{The 2D Lieb lattice}
\label{Lieblattice}
\end{figure}
 $\diamondsuit$
}
\end{example}

\begin{define}
{\rm 
We say that a state  exhibits {\it ferromagnetism}, if it has total spin $S$ such that $S=c|\Lambda|+o(|\Lambda|)$ with $c>0$. $\diamondsuit$

}
\end{define}

\begin{Thm}\label{FerroUni}
Let $H_{*, n, \Lambda}\in \mathscr{C}_{ \mathfrak{X}=\BbbC, \Lambda}^{\#}$ be given for all $L\in \BbbN$.
Suppose that
$H_{n, \Lambda}\in \mathscr{U}(H_{*, n,  \Lambda})$ for all $L$.
If  the ground state of $H_{*, n, \Lambda}$ exhibits ferromagnetism,
 then the ground state of  $H_{n, \Lambda}$ exhibits ferromagnetism as well.
\end{Thm}
{\it Proof.}
Let $S$ and $S_{*}$ be total spins of ground states of $H_{n, \Lambda}$ and $H_{*, n, \Lambda}$, respectively. 
By Theorem  \ref{ElUniTh}, we have $S=S_{*}$ for all $L\in \BbbN$.
Since $S_{*}=c|\Lambda|+o(|\Lambda|)$ with $c>0$, we conclude that $S=c|\Lambda|+o(|\Lambda|)$. $\Box$

\subsection{Existence of long-range orders}\label{LRODef}

We suppose that $H_{n, \Lambda} \in \mathscr{C}_{ \Lambda}^{\#}$ acts on $\mathfrak{E}_{n, \Lambda}\otimes \mathfrak{X}_{\Lambda}$
 for each $L\in \BbbN$.
We say  that $H_{n, \Lambda}$ is {\it translation invariant}, if it holds that 
\begin{align}
\tau_{a} H_{n, \Lambda}\tau_{a}^{-1}=H_{n, \Lambda},\ \ \ a\in \Lambda_{\mathrm{B}}.
\end{align}
Here,  $\tau_a\, (a\in \Lambda)$ on $\mathfrak{E}_{n, \Lambda}\otimes \mathfrak{X}_{\Lambda}$ is the translation such that 
\begin{align}
\tau_{a} c_{x\sigma}\tau_{a}^{-1}
=c_{x+a\,  \sigma},\ \ \ \tau_a \Omega=\Omega.
\end{align}

\begin{example}\label{TransInv}
{\rm 
We say that a matrix $\{M_{xy}\}_{x, y\in \Lambda}$ is {\it translation invariant}, if $M_{xy}=M_{x-y, o}$ for each $x, y\in \Lambda$.
If $\{g_{xy}\},\ \{J_{xy}\},\ \{t_{xy}\}$ and $\{U_{xy}\}$ are translation invariant, then
$H_{\mathrm{Heis}}, H_{\mathrm{H}},  H_{\mathrm{HH}}$ and $H_{\mathrm{rad}}$ are translation invariant.
 $\diamondsuit$
}
\end{example}

\begin{example}
{\rm
Let us consider the Marshall-Lieb-Mattis Hamiltonian $H_{\mathrm{MLM}}$.
Assume that $A+x=A$ and $B+x=B$ for all $x\in \Lambda_{\mathrm{B}}$.
Because $\tau_a {\Bs S}_A\tau_a^{-1}={\Bs S}_{A+a}$ and  $\tau_a {\Bs S}_B\tau_a^{-1}={\Bs S}_{B+a}$,
$H_{\mathrm{MLM}}$ is translation invariant. $\diamondsuit$
}
\end{example}
In what follows, we assume that $H_{n, \Lambda}$ is translation invariant. 

Let $\psi$ be the unique  ground state of $H_{n, \Lambda}$ in the $M=0$-subspace.
For each $x\in \Lambda_{\mathrm{B}}$, we define
\begin{align}
\mathbb{S}_x^{(\pm)}=S_{x}^{(\pm)}+\sum_{j=1}^{n-1} S_{x+{\Bs e}_j}^{(\pm)},
\end{align}
where $S_x^{(\pm)}=S_x^{(1)}+iS_x^{(2)}$.
Here,  we understand that $\mathbb{S}_x^{(\pm)}=S_{x}^{(\pm)}$ if $n=1$.
We introduce the two-point correlation  function  by 
\begin{align}
G_{\Lambda}(x)=\Big\la \psi\Big|\mathbb{S}_x^{(+)} \mathbb{S}_o^{(-)} \psi\Big\ra
=\Big\la \mathbb{S}_x^{(+)} \mathbb{S}_o^{(-)}\Big\ra.
\end{align}  
By the translation invariance, we have
\begin{align}
G_{\Lambda}(x-y)=\Big\la \mathbb{S}_x^{(+)} \mathbb{S}_y^{(-)}\Big\ra, \label{Two}
\end{align}
provided that $x, y\in \Lambda_{\mathrm{B}}$. 
Let $\Lambda^*$ be the dual lattice of $\Lambda$.
Let $\hat{G}_{\Lambda}(p)$ be the Fourier transformation of
$G_{\Lambda}(x)$:
\begin{align}
\hat{G}_{\Lambda}(p)=(2\pi)^{-d/2} \sum_{x\in \Lambda_{\mathrm{B}}} e^{-ip\cdot x}
 G_{\Lambda}(x),\ \ p\in \Lambda^*.
\end{align} 
We have, by (\ref{Two}), 
\begin{align}
\hat{G}_{\Lambda}(p)=\Big\la \tilde{\mathbb{S}}_{ -p}^{(+)} \tilde{ \mathbb{S}}_{ p}^{(-)}\Big\ra,
\end{align} 
where
$\displaystyle 
\tilde{\mathbb{S}}_{ p}^{(\pm )}=|\Lambda|^{-1/2}\sum_{x\in \Lambda_{\mathrm{B}}} e^{-ip\cdot x}
 \mathbb{S}_x^{(\pm)}.$
Since $(\tilde{\mathbb{S}}_{p}^{(+)}
)^*=\tilde{\mathbb{S}}_{-p}^{(-)}$, we have $\hat{G}_{\Lambda}(p)
\ge 0$ for all $p\in \Lambda^*$.

\begin{define}
{\rm
We say that 
$H_{n, \Lambda}$ has  {\it long-range order with ordering wave
 vector $p$} (simply, LRO[$p$]) in the ground state,
  if  it holds that 
\begin{align}
\liminf_{L\to \infty} \frac{\hat{G}_{\Lambda}(p)}{|\Lambda|}>0. 
\end{align} 
As for a physical interpretation of LRO, see, e.g.,  \cite{Tasaki4}. $\diamondsuit$
}
\end{define} 

\begin{Prop}\label{EigV}
  If 
 the ground state of $H_{n, \Lambda}$ exhibits ferromagnetism, then $H_{n, \Lambda}$ has
 {\rm LRO[$p=0$]} in the ground state.
\end{Prop} 
{\it Proof.}
Since
$\displaystyle 
\hat{G}_{\Lambda}(0)=|\Lambda|^{-1} \Big\la S^{(+)}_{\mathrm{tot}, n}
S^{(-)}_{\mathrm{tot}, n}
\Big\ra=
\frac{|\Lambda|^{-1}}{2} \Big\la S_{\mathrm{tot}, n}^2[M=0]
\Big\ra,
$
we obtain the desired result. 
$\Box$
\medskip\\

By Theorem \ref{FerroUni}
 and Proposition \ref{EigV}, we obtain the following.

\begin{Thm}\label{LROUni}
Assume 
that $H_{n, \Lambda}\in \mathscr{U}(H_{*, n, \Lambda})$ for all $L$.
If  the ground state of $H_{*, n, \Lambda}$ 
exhibits ferromagnetism,
then $H_{n, \Lambda}$
 has  {\rm LRO[$0$]} in the ground state.
\end{Thm}

\subsection{Proof of Theorem \ref{SummaryMain}}

Here, we prove Theorem \ref{SummaryMain} in Section \ref{SecResults}.

(i)
This immediately follows from Theorem \ref{ElUniTh}.

(ii)
We consider an extended Hilbert space $Q\mathfrak{E}_n\otimes \BbbC^2$.
We define  a Hamiltonian $H_1$ acting on $Q\mathfrak{E}_n\otimes \BbbC^2$ by 
$
H_1=H_*\otimes 1-1\otimes \sigma_1,
$
where $\sigma_1$ is the standard Pauli matrix given by $
\sigma_1=\begin{pmatrix}
0 & 1\\
1 & 0
\end{pmatrix}
$. Remark the following fact: 
$\BbbR^2_+=\Big\{
\begin{pmatrix}
x\\
y 
\end{pmatrix}
\in \BbbC^2\, \Big|\, x, y\ge 0\Big\}$ is a self-dual cone in $\BbbC^2$.
 Now we define a self-dual cone in $Q\mathfrak{E}_n[M_*]\otimes \BbbC^2$ by 
 $
 \Cone_1=\Big\{ \Psi_1\otimes 
 \begin{pmatrix}
1\\
0 
\end{pmatrix}
+\Psi_2\otimes 
\begin{pmatrix}
0\\
1 
\end{pmatrix}
\Big|\, 
\Psi_1, \Psi_2
 \in \Cone_*
 \Big\}.
 $
 \begin{lemm}
 For all $s>-E(H_1[M_*])$, we have $\big(H_1[M_*]+s\big)^{-1} \rhd 0$ w.r.t. $\Cone_1$.
 \end{lemm}
 {\it Proof.} Let $\eta=\frac{1}{\sqrt{2}}
 \begin{pmatrix}
1\\
1 
\end{pmatrix}
 $. We readily confirm that $\eta$ is the unique ground state of $-\sigma_1$ and $\eta>0$ w.r.t. $\BbbR_+^2$. Let $\psi_*$ be the ground state of $H_*[M_*]$. By {\bf (U. 2)}, it holds that $\psi_*>0$
 w.r.t. $\Cone_*$. Trivially, $\psi_*\otimes \eta$ is the unique ground state of $H_1[M_*]$ and, by 
 applying Corollary \ref{TPV}, $\psi_*\otimes \eta>0$ w.r.t. $\Cone_1$.
Hence, by using Theorem \ref{PFF}, we obtain the desired result. $\Box$
\medskip

We introduce an orthogonal projection $P$ by $P\Psi\otimes r= \Psi\otimes (0, r_2)^T$ for each $\Psi\in Q\mathfrak{E}_n[M_*]$ and $r=(r_1, r_2)^T \in \BbbC^2$, where $a^T$ indicates the transpose of $a$.
We can identify $\R(P) $ with $Q\mathfrak{E}_n[M_*]$
by the isometry $\tau: \R(P)\ni \Psi\otimes (0, r_2)^T\mapsto r_2\Psi \in \mathfrak{E}_n[M_*]$.
By definition, we have $P\unrhd 0$ w.r.t. $\Cone_1$ and $P \Cone_1=\Cone_*$ by the aforementioned identification.
Hence, we can readily  confirm that $H_1[M_*] \leadsto H_*[M_*]$.

Next, let us consider a further extended Hilbert space 
$(Q\mathfrak{E}_n\otimes \BbbC^2)\otimes \BbbC^2$.
Define a Hamiltonian $H_2$ by 
$
H_2=H_1\otimes 1-1\otimes \sigma_1
$, and define a self-dual cone $\Cone_2$  by $
\Cone_2=\Big\{ \Phi_1\otimes 
 \begin{pmatrix}
1\\
0 
\end{pmatrix}
+\Phi_2\otimes 
\begin{pmatrix}
0\\
1 
\end{pmatrix}
\Big|\, 
\Phi_1, \Phi_2
 \in \Cone_1
 \Big\}.
$   Using  arguments similar to those in the  last paragraph. we can  confirm that $
H_2[M_*] \leadsto H_1[M_*]
$.
Repeating this procedure, we can construct a sequence of Hamiltonians $\{H_{\ell}\}$ such that $H_{\ell}[M_*] \bl H_*[M_*]$.
Therefore, $\mathscr{U}(H_*)$ contains at least countably infinite number of Hamiltonians.

 (iii) and (iv) immediately follow from the definition of $\mathscr{U}(H_{*, n})$.
   $\Box$

\section{The Marshall-Lieb-Mattis stability class and stability of Lieb's  theorem}\label{MLMLiebSta}
\setcounter{equation}{0}

\subsection{The Heisenberg model}

Consider the spin-$1/2$ antiferromagnetic Heisenberg model on a finite lattice $\Lambda$.
The Hamiltonian is
\begin{align}
H_{\mathrm{Heis}}=\sum_{x,y\in \Lambda} J_{xy}{\boldsymbol S}_x\cdot {\boldsymbol S}_y,
 \end{align} 
 where $J_{xy} \ge 0$ and ${\boldsymbol S}_x=(S_x^{(1)}, S_x^{(2)}, S_x^{(3)})$.

We assume  the following:
\begin{itemize}
\item[{\bf (C. 1)}] $\Lambda$ is connected by $\{J_{xy}\}$;
\item[{\bf (C. 2)}] $\Lambda$ is bipartite in terms of $\{J_{xy}\}$.
 \end{itemize} 
 
$H_{\mathrm{Heis}}$ acts 
 on  the Hilbert space 
 $
 \mathfrak{H}_Q=
 Q \mathfrak{E}_{n=|\Lambda|},
 $
 where the orthogonal projection $Q$ is  defined by (\ref{DefQ}). 

\begin{rem}
{\rm 
The Marshall-Lieb-Mattis Hamiltonian $H_{\mathrm{MLM}}$ given by (\ref{MLMHamiltonian})  is a special case of the Heisenberg Hamiltonian with
$J_{xy}=1$ if $x\in A,\ y\in B$ or $x\in B, y\in A$, $J_{xy}=0$, otherwise.
Of course, this interaction satisfies {\bf (C. 1)} and {\bf (C. 2)}. $\diamondsuit$
}
\end{rem}

\subsection{The Marshall-Lieb-Mattis   stability class}

We will study  the half-filled system, so that 
the  Hilbert space of electron states  is $\mathfrak{E}_{n=|\Lambda|}$ or $\mathfrak{H}_Q$.
In addition, we assume that $|\Lambda|$ is even in the remainder of this section.
 Note that $H_{\mathrm{MLM}} \in \mathscr{C}_{ \mathfrak{X}=\BbbC}^{\#}$.

\begin{define}\label{MLMUniDef}
{\rm 
The $H_{\mathrm{MLM}}$- stability class $\mathscr{U}(H_{\mathrm{MLM}})$
 is called the {\it Marshall-Lieb-Mattis   stability class}.
 $\diamondsuit$
}
\end{define}

The following  is a general form of stability theorems:
\begin{Thm}[Theorem \ref{QAns1}]\label{GeneSt}
Suppose that  $H$ is in the Marshall-Lieb-Mattis   stability class.
  The ground state of
 $H$
 has total spin $S=\frac{1}{2}\big||A|-|B|\big|$ and is unique apart from the trivial $(2S+1)$-degeneracy.
\end{Thm}

We will provide a proof of Theorem \ref{GeneSt} in Appendix \ref{PfLi1}.

\begin{example}
{\rm 
Let us consider the following diagram:
\[
\xymatrix@C=6pt{
&& & & H_{\mathrm{MLM}} \ar@{<~}[lld] \ar@{<~}[ld] \ar@{}[d]|(.6){\dots} \ar@{<~}[rd] \ar@{<~}[rrd]
\\
&&\cdots H_{1,1}\ar@{<~}[lld] \ar@{<~}[ld] \ar@{}[d]|(.6){\dots}  & H_{1,2} & \dots\dots & H_{1, n-1} & H_{1,n } \cdots\\
\cdots H_{2,1}  & H_{2,2} & \dots\dots &&
}
\]
By Theorem \ref{GeneSt}, the ground state of each Hamiltonian $H_{i, j}\, (i, j=1, 2, \dots)$ has the total spin $S=\frac{1}{2}\big||A|-|B|\big|$
and is unique apart from  the trivial $(2S+1)$-degeneracy. $\diamondsuit$
}
\end{example}

\subsection{Ferromagnetism and long-range orders in the ground state}
In this subsection, we assume that $\Lambda$ has a crystal structure and every Hamiltonian is translation invariant (as for  basic definitions, see  Sections \ref{FerrGr} and \ref{LRODef}).
We will specify the $\Lambda$-dependence of the Hamiltonian $H$ as $H_{\Lambda}$.
Recall that the size of $\Lambda$ is determined by $L$, see Section \ref{FerrGr}.

By Theorems \ref{FerroUni}, \ref{LROUni} and \ref{GeneSt}, we have the following.
\begin{Thm}
Suppose that $H_{\Lambda}$ belongs to  the Marshall-Lieb-Mattis   stability class for each $L\in \BbbN$.
If  there exists a constant $c>0$ independent of $\Lambda$ such that $\big|
|A|-|B|
\big|=c|\Lambda|$ , then the ground state of $H_{\Lambda}$ exhibits ferromagnetism.
In addition, $H_{\Lambda}$ has {\rm LRO[0]} in the ground state.
\end{Thm}

\begin{example}
{\rm 
Let us consider the Lieb lattice given in Example \ref{LiebL}.
We choose $A=\{
{\Bs r}+{\Bs e}_j\, |\, {\Bs r} \in \Lambda_{\mathrm{B}}, j=1,2
\} $ and $ B= \Lambda_{\mathrm{B}}$.
Then we can easily check that  $\big|
|A|-|B|
\big|
=\frac{1}{3}|\Lambda|
$.
Thus, if  $H_{\Lambda}$ belongs to the Marshall-Lieb-Mattis  stability class, then the ground state of $H$ exhibits ferromagnetism.
In addition, $H_{\Lambda}$ has {\rm LRO[0]} in the ground state. $\diamondsuit$
}
\end{example}

\subsection{The Marshall-Lieb-Mattis theorem}

We will explain the well-known Marshall-Lieb-Mattis theorem from the viewpoint of  the  stability class.

In this subsection, we assume {\bf (C. 1)} and {\bf (C. 2)}. 

\begin{Thm}\label{HeisMLM}
$H_{\mathrm{Heis}}$ is equivalent to $H_{\mathrm{MLM}}$ in the sense that 
$H_{\mathrm{Heis}}[M] \equiv H_{\mathrm{MLM}}[M]$ for all $M\in \mathrm{spec}\Big(
S_{\mathrm{tot}, |\Lambda|}^{(3)}
\Big)$. (As for the definition of the binary relation \lq\lq{}\,$\equiv$\rq\rq{}, see Definition \ref{HEquiv}.)
In particular, $H_{\mathrm{Heis}}$ belongs to the Marshall-Lieb-Mattis  stability class.
\end{Thm}
We will prove Theorem \ref{HeisMLM} in Appendix \ref{PfLi1}.

Combining Theorems \ref{GeneSt} and \ref{HeisMLM}, we obtain the following:
\begin{coro}[The Marshall-Lieb-Mattis theorem\cite{Marshall, LiebMattis}]

The ground state of $H_{\mathrm{Heis}} $
  has total spin $S=\frac{1}{2}\big||A|-|B|\big|$ and is unique apart from the trivial $(2S+1)$-degeneracy.
\end{coro}

  \subsection{Lieb\rq{}s theorem}

Here, we will interpret Lieb\rq{}s theorem in the context of  stability class defined   in Section \ref{ManyElUniTh}.

In this subsection, we assume the following:
\begin{itemize}
\item {\bf (A. 1)}--{\bf (A. 3)}.
\item {\bf (C. 1)}, {\bf (C. 2)}.
\end{itemize}

\begin{Thm}\label{GeneLieb}
 $H_{\mathrm{H}, |\Lambda|}[M] \leadsto H_{\mathrm{Heis}}[M]$ for all $M\in \mathrm{spec}\Big(S_{\mathrm{tot}, |\Lambda|}^{(3)}\Big)$.
\end{Thm} 
We will provide a sketch of  proof of Theorem \ref{GeneLieb} in Appendix \ref{PfLi2}.

\begin{coro}[Theorem \ref{LiebThm1}] \label{HubbardMLM}
$H_{\mathrm{H}, |\Lambda|}$ belongs to the Marshall-Lieb-Mattis   stability class.
\end{coro}
{\it Proof.} By Theorems \ref{HeisMLM} and \ref{GeneLieb}, we have  the following chain:
\begin{align}
H_{\mathrm{H}, |\Lambda|}[M] \leadsto H_{\mathrm{Heis}}[M] \leadsto
H_{\mathrm{MLM}}[M] 
\end{align}
for all $M\in \mathrm{spec}\Big(S_{\mathrm{tot}, |\Lambda|}^{(3)}\Big)$. Thus, we conclude the assertion in the corollary. $\Box$

\begin{coro}[Theorem \ref{LiebThm1}]
The ground state of $H_{\mathrm{H}, |\Lambda|}$
  has total spin $S=\frac{1}{2}\big||A|-|B|\big|$ and is unique apart from the trivial $(2S+1)$-degeneracy.

\end{coro}
{\it Proof.} This immediately follows from Theorem \ref{GeneSt}  and Corollary \ref{HubbardMLM}. $\Box$

\subsection{Stability of Lieb's theorem  I}\label{StLI}
Let us consider the Holstein-Hubbard model at half-filling: $H_{\mathrm{HH}, |\Lambda|}$.
In this subsection, we assume the following:
\begin{itemize}
\item {\bf (A. 1)}--{\bf (A. 5)}.
\item {\bf (C. 1)}, {\bf (C. 2)}.
\end{itemize}

Let $\Omega_{\mathrm{ph}}$ be the Fock vacuum in $\Fock_{\mathrm{ph}}$.
Let us define a closed subspace of $\mathfrak{E}_{n=|\Lambda|}[M] \otimes \Fock_{\mathrm{ph}}$ by 
$
\mathfrak{E}_{n=|\Lambda|}[M]\otimes \Omega_{\mathrm{ph}}
=\{\vphi\otimes \Omega_{\mathrm{ph}}\, |\, \vphi\in \mathfrak{E}_{n=|\Lambda|}[M] \}
$.
Now, we define an isometry $\tau$ from $
\mathfrak{E}_{n=|\Lambda|}[M] \otimes \Omega_{\mathrm{ph}}
$  onto $
\mathfrak{E}_{n=|\Lambda|}[M]
$ by $\tau\vphi\otimes \Omega_{\mathrm{ph}}=\vphi$.
Hence, we can naturally identify $
\mathfrak{E}_{n=|\Lambda|}[M]
$
 with $
 \mathfrak{E}_{n=|\Lambda|}[M]\otimes \Omega_{\mathrm{ph}}
 $.
 By  this fact, $
 \mathfrak{E}_{n=|\Lambda|}[M]
 $
 can be regarded as a closed subspace of $
 \mathfrak{E}_{n=|\Lambda|}[M] \otimes \Fock_{\mathrm{ph}}
 $.

\begin{Thm}\label{GeneHH}
 $H_{\mathrm{HH}, |\Lambda|}[M] \leadsto H_{\mathrm{H}, |\Lambda|}[M]
 $ for all $M\in \mathrm{spec}\Big(S_{\mathrm{tot}, |\Lambda|}^{(3)}\Big)$.
\end{Thm} 

\begin{rem}
{\rm
The Coulomb interactions in $H_{\mathrm{H}}$ and $H_{\mathrm{HH}}$ can be chosen, independently. $\diamondsuit$
}
\end{rem}
We will provide a sketch of  proof of Theorem \ref{GeneHH} in Appendix  \ref{StabHH}.

\begin{coro}[Theorem \ref{StaLiThm1}]\label{GeneHH2}
$H_{\mathrm{HH}, |\Lambda|}$ belongs to the Marshall-Lieb-Mattis   stability class.
\end{coro}
{\it Proof.}
By Theorems \ref{HeisMLM}, \ref{GeneLieb} and \ref{GeneHH}, we have the following chain:
\begin{align}
H_{\mathrm{HH}, |\Lambda|}[M] \leadsto H_{\mathrm{H}, |\Lambda|}[M] 
\leadsto H_{\mathrm{Heis}}[M]  \leadsto H_{\mathrm{MLM}}[M]
\end{align}
for all $M$. Thus, $H_{\mathrm{HH}, |\Lambda|}$ belongs to the Marshall-Lieb-Mattis   stability class. 
$\Box$

\begin{coro}[Theorem \ref{StaLiThm1}]
The ground state of $H_{\mathrm{HH}, |\Lambda|}$
  has total spin $S=\frac{1}{2}\big||A|-|B|\big|$ and is unique apart from the trivial $(2S+1)$-degeneracy.
\end{coro}
{\it Proof.} By Theorem \ref{GeneSt} and Corollary \ref{GeneHH2}, we obtain the desired assertion. $\Box$

\subsection{Stability of Lieb's theorem II}\label{StaLiIIExh}

Let us consider the many-electron system coupled to the quantized radiation field.
In this subsection, we assume the following:
\begin{itemize}
\item {\bf (A. 1)}--{\bf (A. 3)}.
\item {\bf (C. 1)}, {\bf (C. 2)}.
\end{itemize}

Let $\Omega_{\mathrm{rad}}$ be the Fock vacuum in $\Fock_{\mathrm{rad}}$. 
As before, we can  regard $\mathfrak{E}_{n=|\Lambda|}[M]$ as a closed subspace of $
\mathfrak{E}_{n=|\Lambda|}[M]\otimes \Fock_{\mathrm{rad}}
$ by the identification $
\mathfrak{E}_{n=|\Lambda|}[M]=\mathfrak{E}_{n=|\Lambda|}[M]\otimes \Omega_{\mathrm{rad}}
$.

\begin{Thm}\label{SpinHrad}
 $H_{\mathrm{rad}, |\Lambda|}[M] \leadsto H_{\mathrm{H}, |\Lambda|}[M] $ for all $M\in \mathrm{spec}\Big(S_{\mathrm{tot}, |\Lambda|}^{(3)}\Big)$.
\end{Thm} 

We will provide a proof of Theorem \ref{SpinHrad} in Appendix \ref{PfLi4}.

\begin{coro}[Theorem \ref{StaLiThm2}]\label{SpinHrad2}
$H_{\mathrm{rad}, |\Lambda|}$ belongs to the Marshall-Lieb-Mattis   stability class.
\end{coro}
{\it Proof.}
By Theorems \ref{HeisMLM}, \ref{GeneLieb} and \ref{SpinHrad}, we have the following chain:
\begin{align}
H_{\mathrm{rad}, |\Lambda|}[M] \leadsto H_{\mathrm{H}, |\Lambda|}[M] 
\leadsto H_{\mathrm{Heis}}[M]  \leadsto H_{\mathrm{MLM}} [M]
\end{align}
for all $M$. Thus, $H_{\mathrm{rad}, |\Lambda|}$ belongs to the Marshall-Lieb-Mattis   stability class. 
$\Box$
\medskip\\

By Theorem \ref{GeneSt} and Corollary \ref{SpinHrad2}, we obtain the following:
\begin{coro}[Theorem \ref{StaLiThm2}]
The ground state of $H_{\mathrm{rad}, |\Lambda|}$
  has total spin $S=\frac{1}{2}\big||A|-|B|\big|$ and is unique apart from the trivial $(2S+1)$-degeneracy.

\end{coro}

\subsection{Summary of Section \ref{MLMLiebSta}}

Our results in this section are summarized in the following diagram:
\[
\xymatrix{
& &H_{\mathrm{HH}}\ar@{~>}[ld]\\
H_{\mathrm{MLM}}\equiv H_{\mathrm{Heis}} \ar@{<~}[r]& H_{\mathrm{H}} &\\
 & &H_{\mathrm{rad}} \ar@{~>}[ul]
}
\]

\begin{example}\label{LiebFerr}
{\rm 
Let us consider the Lieb lattice in Example \ref{LiebL}. 
Suppose that $\{g_{xy}\},\ \{J_{xy}\},\ \{t_{xy}\}$ and $\{U_{xy}\}$ satisfy the conditions in Example \ref{TransInv}.
By the above diagram,
the ground states of $H_{\mathrm{Heis}},\ H_{\mathrm{H}, |\Lambda|},\ H_{\mathrm{HH}, |\Lambda|}$ and $H_{\mathrm{rad}, |\Lambda|}$ exhibit  ferrogmanetism  and LRO[0].  $\diamondsuit$
}
\end{example}

\section{The Nagaoka-Thouless  stability class and stability of  the Nagaoka-Thouless theorem}\label{StaNTThm}

\setcounter{equation}{0}

In this section, we explain the Nagaoka-Thouless theorem from the viewpoint  of   stability class  discussed in Section \ref{ManyElUniTh}. 
\subsection{The Nagaoka-Thouless stability   class}
Let us consider the    many-electron system with one electron less than half-filling and 
infinitely large Coulomb strength. The Hilbert space of electrons is 
$
\mathfrak{H}_{\mathrm{NT}}=P_{\mathrm{G}} \mathfrak{E}_{n=|\Lambda|-1},
$
where $P_{\mathrm{G}}$ is the Gutzwiller projection defined by (\ref{GutzP}).
Recall that $P_{\mathrm{G}}$ satisfies the condition {\bf (Q)}.
In this section, we assume 
\begin{itemize}
\item {\bf (B. 1)}, {\bf (B. 2)}, {\bf (B. 3)}.
\end{itemize}

For notational simplicity, we set
\begin{align}
\mathcal{S}^2=S_{\mathrm{tot}}^2\restriction \mathfrak{H}_{\mathrm{NT}}, \ \  \ 
\mathcal{S}^{(j)}=S_{\mathrm{tot}}^{(j)}\restriction \mathfrak{H}_{\mathrm{NT}},\ \ j=1,2,3.
\end{align}
Note  that $\mathrm{spec}(\mathcal{S}^{(3)})=\{
-(|\Lambda|-1)/2, -(|\Lambda|-3)/2, \dots, (|\Lambda|-1)/2
\}$.
As before, the $M$-subspace of $\mathfrak{H}_{\mathrm{NT}}$ is defined by 
$
\mathfrak{H}_{\mathrm{NT}}[M]=\ker(\mathcal{S}^{(3)}-M) $
for 
$ M\in \mathrm{spec}(\mathcal{S}^{(3)}).
$

Let $X$ be a linear operator on $\mathfrak{H}_{\mathrm{NT}}$ which commutes  with $\mathcal{S}^{(3)}$.
For each $M\in \mathrm{spec}(\mathcal{S}^{(3)})$, we set 
$
X[M]=X\restriction  \mathfrak{H}_{\mathrm{NT}}[M]
$ as before.

Let $H_{\mathrm{H}}^{\infty}$ be the effective Hamiltonian defined in Proposition \ref{EffHami1}.

\begin{define}\label{NTUniDef}
{\rm
The $H_{\mathrm{H}}^{\infty}$-stability class $\mathscr{U}(H_{\mathrm{H}}^{\infty})$
 is called the {\it Nagaoka-Thouless   stability class}.
 $\diamondsuit$
}
\end{define}

Before we proceed, recall the Nagaoka-Thouless theorem (Theorem \ref{NagaokaTThm1}).
The following theorem is a general form of stability  of the  Nagaoka-Thouless theorem:

\begin{Thm}[Theorem \ref{QAns3}]\label{HHNT}
Suppose that $H$ is in the Nagaoka-Thouless stability  class. The ground state of $H$
has total spin $S=(|\Lambda|-1)/2$ and is unique apart from the trivial $(2S+1)$-degeneracy.
\end{Thm}

We will prove Theorem \ref{HHNT} in Appendix \ref{PfStaNTThm}.

\begin{example}
{\rm 
Let us consider the following diagram:
\[
\xymatrix@C=6pt{
&& & & H_{\mathrm{H}}^{\infty} \ar@{<~}[lld] \ar@{<~}[ld] \ar@{}[d]|(.6){\dots} \ar@{<~}[rd] \ar@{<~}[rrd]
\\
&&\cdots H_{1,1}\ar@{<~}[lld] \ar@{<~}[ld] \ar@{}[d]|(.6){\dots}  & H_{1,2} & \dots\dots & H_{1, n-1} & H_{1,n } \cdots\\
\cdots H_{2,1}  & H_{2,2} & \dots\dots &&
}
\]
By Theorem \ref{HHNT}, the ground state of each Hamiltonian $H_{i, j}\, (i, j=1, 2, \dots)$ has the total spin $S=(|\Lambda|-1)/2$
and is unique apart from  the trivial $(2S+1)$-degeneracy. $\diamondsuit$
}
\end{example}

\subsection{Ferromagnetism and long-range orders in the ground state}
In this subsection, we assume that $\Lambda$ has a crystal structure and every Hamiltonian is translation invariant.
We will specify the $\Lambda$-dependence of the Hamiltonian $H$ as $H_{\Lambda}$.

By Theorems \ref{FerroUni}, \ref{LROUni} and \ref{HHNT}, we have the following.
\begin{Thm}
Suppose that $H_{\Lambda}$ is in the Nagaoka-Thouless   stability class for all $L\in \BbbN$.
Then the ground state of $H_{\Lambda}$ exhibits ferromagnetism.
In addition, $H_{\Lambda}$ has {\rm LRO[0]} in the ground state.
\end{Thm}

\begin{example}\label{ConcNT}
{\rm 
Suppose that $\Lambda$ and $t_{xy}$ satisfy the conditions in Example \ref{HoleC}.
If $H_{\Lambda}$ belongs to the Nagaoka-Thouless  stability class, then the ground state of $H_{\Lambda}$ exhibits ferromagnetism and  {\rm LRO[0]}. $\diamondsuit$
} 
\end{example}

\subsection{Stability of  the Nagaoka-Thouless theorem I}\label{SectionStaNT1}

Let $\Omega_{\mathrm{ph}}$ be the Fock vacuum in $\Fock_{\mathrm{ph}}$. Let us define a closed subspace of $\mathfrak{H}_{\mathrm{NT}}[M]\otimes \Fock_{\mathrm{ph}}$ by 
$
\mathfrak{H}_{\mathrm{NT}}[M] \otimes \Omega_{\mathrm{ph}}
=\{\vphi\otimes \Omega_{\mathrm{ph}}\, |\, \vphi\in \mathfrak{H}_{\mathrm{NT}}[M]\}.
$
In a similar way as in Section \ref{StLI}, we can identify $\mathfrak{H}_{\mathrm{NT}}[M]$ with $
\mathfrak{H}_{\mathrm{NT}}[M] \otimes \Omega_{\mathrm{ph}}
$. Thus, $\mathfrak{H}_{\mathrm{NT}}[M]$ can be viewed as a closed subspace of $
\mathfrak{H}_{\mathrm{NT}}[M] \otimes\Fock_{\mathrm{ph}}
$.

Let $H_{\mathrm{HH}}^{\infty}$ be the effective Hamiltonian defined in Proposition \ref{EffHami2}.

\begin{Thm} [Theorem \ref{StaNT1}]\label{HHNT1}
$
H_{\mathrm{HH}}^{\infty}[M]\leadsto H_{\mathrm{H}}^{\infty}[M] 
$ for all $M$. In particular, 
$H_{\mathrm{HH}}^{\infty}$ belongs to  the Nagaoka-Thouless  stability class.
\end{Thm} 

We will prove Theorem \ref{HHNT1} in Appendix \ref{PfStaNTThm}.

By Theorems \ref{HHNT} and  \ref{HHNT1}, we obtain the following:

\begin{coro}[Theorem \ref{StaNT1}]\label{StNaga1}
The ground state of $H_{\mathrm{HH}}^{\infty}$
 has  total spin $S=\frac{1}{2}(|\Lambda|-1)$ and is unique apart from the trivial $(2S+1)$-degeneracy.
\end{coro}

\subsection{Stability of the  Nagaoka-Thouless theorem II}

In this subsection, we regard $\mathfrak{H}_{\mathrm{NT}}[M]$ as a closed subspace of $
\mathfrak{H}_{\mathrm{NT}}[M] \otimes \Fock_{\mathrm{rad}}
$, as we did in Section \ref{SectionStaNT1}.

Let $H_{\mathrm{rad}}^{\infty}[M]$ be the effective Hamiltonian defined in Proposition \ref{EffHami3}.

\begin{Thm}[Theorem \ref{StaNT2}]\label{HHNT2}
$
H_{\mathrm{rad}}^{\infty}[M] \leadsto H_{\mathrm{H}}^{\infty}[M] 
$
 for all $M$. In particular, 
$H_{\mathrm{rad}}^{\infty}$ belongs to  the Nagaoka-Thouless  stability class.
\end{Thm} 

We will provide a proof of Theorem \ref{HHNT2} in Appendix \ref{PfStaNTThm}.

By Theorems \ref{HHNT} and  \ref{HHNT2}, we obtain the following:

\begin{coro}[Theorem \ref{StaNT2}]\label{StNaga2}
The ground state of $H_{\mathrm{rad}}^{\infty}$
 has  total spin $S=\frac{1}{2}(|\Lambda|-1)$ and is unique apart from the trivial $(2S+1)$-degeneracy.
\end{coro}

\subsection{Summary of Section \ref{StaNTThm}}

Our results in this section are summarized in the following diagram:
\[
\xymatrix{
 & H_{\mathrm{H}}^{\infty} &\\
H_{\mathrm{HH}}^{\infty}\ar@{~>}[ru] &  &H_{\mathrm{rad}}^{\infty} \ar@{~>}[ul]
}
\]
\begin{example}

{\rm 
Assume that $\Lambda$ and $\{t_{xy}\}$ satisfy the conditions in Example \ref{HoleC}.
By Example \ref{ConcNT} or the above diagram, the ground states of 
$
H_{\mathrm{H}}^{\infty},\ H_{\mathrm{HH}}^{\infty}
$ and $H_{\mathrm{rad}}^{\infty}$ exhibit ferromagnetism and   LRO$[0]$. $\diamondsuit$
}
\end{example}

\section{Concluding remarks}\label{ConRemarks}
\subsection{Why are the on-site Coulomb interactions important?}
\setcounter{equation}{0}

The Hubbard model with the on-site interaction has occupied  an important place 
in the study of magnetism. In this subsection, we explain a reason for this  from the viewpoint of   stability class.

Let $H_{\mathrm{H}}^{(U)}$ be the Hubbard Hamiltonian $H_{\mathrm{H}}$ with the on-site Coulomb interaction:
\begin{align}
U\sum_{x\in \Lambda} (n_x-1)^2
\end{align}
with $U>0$.
In a similar way, we can define $H_{\mathrm{HH}}^{(U)}$ and $H_{\mathrm{rad}}^{(U)}$.
In this subsection, we assume
\begin{itemize}
\item  {\bf (A. 1)}, {\bf  (A. 2)}.
\end{itemize}

\begin{Thm}\label{Cl1}
We have the following:
\begin{itemize}
\item[{\rm (i)}] Whenever $U>0$ and $\{U_{xy}\}$ is positive definite, we have
$H_{\mathrm{H}, |\Lambda|}^{(U)}[M] \equiv H_{\mathrm{H}, |\Lambda|}[M]$ for all $M\in \mathrm{spec}
\Big(
S_{\mathrm{tot}, |\Lambda|}^{(3)}
\Big)$.
\item[{\rm (ii)}] Whenever $U>0$ and $\{U_{\mathrm{eff}, xy}\}$ is positive definite, we have
$H_{\mathrm{HH}, |\Lambda|}^{(U)}[M] \equiv H_{\mathrm{HH}, |\Lambda|}[M]$ for all $M\in \mathrm{spec}
\Big(
S_{\mathrm{tot}, |\Lambda|}^{(3)}
\Big)$.
\item[{\rm (iii)}] Whenever $U>0$ and $\{U_{xy}\}$ is positive definite, we have
$H_{\mathrm{rad}, |\Lambda|}^{(U)}[M] \equiv H_{\mathrm{rad}, |\Lambda|}[M]$ for all $M\in \mathrm{spec}
\Big(
S_{\mathrm{tot}, |\Lambda|}^{(3)}
\Big)$.
\end{itemize}
\end{Thm}
We will prove Theorem \ref{Cl1} in Appendix \ref{ProofCl1}.

\begin{coro}\label{Cl2}
Under the same assumptions in Theorem \ref{Cl1}, we have the following:
\begin{itemize}
\item[{\rm (i)}] If the ground state of $H_{\mathrm{H}, |\Lambda|}^{(U)}$  has total spin $S$, so does $H_{\mathrm{H}, |\Lambda|}$.
\item[{\rm (ii)}] If the ground state of $H_{\mathrm{HH}, |\Lambda|}^{(U)}$  has total spin $S$, so does $H_{\mathrm{HH}, |\Lambda|}$.
\item[{\rm (iii)}] If the ground state of $H_{\mathrm{rad}, |\Lambda|}^{(U)}$  has total spin $S$, so does $H_{\mathrm{rad}, |\Lambda|}$.
\end{itemize}
\end{coro}
{\it Proof.} Apply Theorems \ref{AbstTh} and \ref{Cl1}. $\Box$

\begin{rem}
{\rm
\begin{itemize}
\item Theorem \ref{Cl1} and Corollary \ref{Cl2}
indicate that  total spin of the ground state is stable under the deformations of the Coulomb interaction.

\item Lieb\rq{}s theorem tells us that $S=\frac{1}{2}\big||A|-|B|\big|$, see Theorem \ref{LiebThm1}.

\item The Hamiltonian with the on-site Coulomb interaction is a good representative of the equivalence  class, because its structure is simpler.

\item Similar argument tells us that the total spin of the ground state is stable under the deformation of the hopping matrix.

\item  Let us consider  random Coulomb interactions and random hopping matrices.
Whenever the randomness is weak such that {\bf (A. 1)}--{\bf (A. 3)} are satisfied almost  surely, we can also prove that the total spin of the ground state 
is stable against the randomness.

\item Similar observations  hold true for $H_{\mathrm{H}}^{\infty},\ H_{\mathrm{HH}}^{\infty}$ and $H_{\mathrm{rad}}^{\infty}$.
  $\diamondsuit$

\end{itemize}

}
\end{rem}

\subsection{The Su-Schrieffer-Heeger model}
There are many other models which belong to the Marshall-Lieb-Mattis  stability class.
In this subsection, we provide an example.

The Su-Schrieffer-Heeger model is a one-dimensional model for polyacetylene \cite{SSH}.
The Hamiltonian is 
\begin{align}
H_{\mathrm{SSH}}=&\sum_{j=1}^{L-1} \sum_{\sigma=\uparrow, \downarrow}\{t-\delta(Q_{j+1}-Q_j)\}
(c_{j+1\, \sigma}^*c_{j\sigma}+c_{j\sigma}^* c_{j+1\, \sigma}) +U\sum_{j=1}^L (n_j-1)^2\no
&+\sum_{j=1}^L \frac{P_j^2}{2M}+\sum_{j=1}^{L-1} \frac{K}{2}(Q_{j+1}-Q_j)^2,
\end{align}
where $Q_j$ and $P_j$ are the local phonon coordinate and momentum at site $j$.
After removal of center-of-mass motion of the phonons, we obtain the following Hamiltonian  \cite{Miyao2}:
\begin{align}
\hat{H}_{\mathrm{SSH}}=&\sum_{j=1}^{L-1} \sum_{\sigma=\uparrow, \downarrow}f_j({\Bs q})
(c_{j+1\, \sigma}^*c_{j\sigma}+c_{j\sigma}^* c_{j+1\, \sigma}) +U\sum_{j=1}^L (n_j-1)^2\no
&+\sum_{j=1}^{L-1} \Big( \frac{p_j^2}{2M}+ \frac{M\omega_j^2}{2}q_j^2\Big),
\end{align}
where $p_n=-i\partial/\partial q_n$ and 
\begin{align}
f_j({\Bs q})&=t-2\delta \sqrt{\frac{2}{L}} \sum_{n=1}^{L-1} q_n \sin(i \theta_n) \sin \frac{\theta_n}{2},\ \ \ \theta_n=\frac{n\pi}{L},\\
\omega_n&=\sqrt{\frac{4K}{M}} \sin \frac{\theta_n}{2}.
\end{align}
Using the result in \cite{Miyao2}, we obtain the following:
\begin{Thm}
 Let us consider the half-filled model $\hat{H}_{\mathrm{SSH}, |\Lambda|}$.
 If $U>0$, then $\hat{H}_{\mathrm{SSH}, |\Lambda|}$ belongs to the the Marshall-Lieb-Mattis
  stability class.
\end{Thm}

Because the nearest neighbour hopping matrix in the one-dimensional chain satisfies $|A|=|B|$, we have the following:

\begin{coro}
The ground state of $\hat{H}_{\mathrm{SSH}, |\Lambda|}$
is unique and 
  has total spin $S=0$.
\end{coro}

\subsection{A  stability class of the attractive Hubbard model }

\subsubsection{The attractive Hubbard model}
In \cite{Lieb}, Lieb studied the attractive Hubbard model, which forms an important  stability class.
We consider a model with the on-site Coulomb interaction as a representative:
\begin{align}
H_{\mathrm{AH}}^{(-U)}=\sum_{x, y\in \Lambda} \sum_{\sigma=\uparrow, \downarrow} t_{xy} c_{x\sigma}^*c_{y\sigma}-U\sum_{x\in \Lambda}(n_x-1)^2,\ \ \ U>0.
\end{align}
We denote by $H_{\mathrm{AH}}$
the (general) attractive Hubbard model, i.e., the Hubbard model (\ref{Hami})
with the Coulomb interaction replaced by $\{-U_{xy}\}$.

From the viewpoint of   stability class, Lieb\rq{}s result can be expressed as follows:

\begin{Thm}\label{AttH}
Assume {\bf (A. 1)} and {\bf (A. 3)}. If  $U>0$, then $H_{\mathrm{AH}, n}$ is equivalent to $H_{\mathrm{AH}, n}^{(-U)}$ for every $n$ even.
Moreover, the ground state of $H_{\mathrm{AH}, n}$ is unique and has total spin $S=0$.
\end{Thm}

\subsubsection{The Holstein model}

The Holstein model \cite{Holstein} is defined by 
\begin{align}
H_{\mathrm{Hol}}=\sum_{x, y\in \Lambda} \sum_{\sigma=\uparrow, \downarrow} t_{xy} c_{x\sigma}^*c_{y\sigma}
+g \sum_{x\in \Lambda} n_x(b_x+b_x^*)
+\omega\sum_{x\in \Lambda}b_x^*b_x. 
\end{align}
Using the result in \cite{FL}, we obtain the following:
\begin{Thm}
Assume {\bf (A. 1)}. Assume $U>0$.
$H_{\mathrm{Hol}, n}$ belongs  to $\mathscr{U}(H_{\mathrm{AH}, n}^{(-U)})$ for each $n$ even.
\end{Thm}
By Theorem \ref{AttH}, we have the following:
\begin{coro}\label{HolS0}
Assume {\bf (A. 1)}. 
The ground state of $H_{\mathrm{Hol}, n}$ is unique and has total spin $S=0$.
\end{coro}
Remark that Corollary \ref{HolS0} is a main theorem in \cite{FL}.

\subsection{Other models}
Here, we mention some other models.
In \cite{TSU, UTS}, Ueda, Tsunetsugu and Sigirist studied the periodic Anderson model and Kondo lattice model.
They applied Lieb\rq{}s spin reflection positivity to these models and clarified the total spin in the ground states. Roughly speaking, the above-mentioned models belong to the (generalized) Marshall-Lieb-Mattis
 stability class. 
By using the methods in the present paper, 
the results in \cite{TSU, UTS}  can be extended even if the electron-phonon interaction is taken into account.
We will publish elsewhere the extensions. 

There are several extensions of the  Nagaoka-Thouless theorem; we remark that  results in \cite{KT,KSV,Kohno} can be explained in terms of the (generalized)  Nagaoka-Thouless  stability class.

In \cite{Shen,Shen2,Tian, Tsune}, reader can find various models which belong to the  stability classes studied in the present paper.

Finally, we conjecture that Mielke-Tasaki\rq{}s flat-band ferromagnetism \cite{Mielke, Mielke2,MT,Tasaki22,Tasaki3} is  related to  
 a new  stability class.

\subsection{Coexistence of long-range orders}

In \cite{Shen3}, Shen, Qiu and Tian proved that ferromagnetic and antiferromagnetic long-range orders coexist in the ground state of the Hubbard model.
From a viewpoint of our theory, this result can be extended as follows:
In the ground state of every Hamiltonian in  a certain subclass of the Marshall-Lieb-Mattis  stability class,
 ferromagnetic and antiferromagnetic long-range orders coexist in the ground state.
We will publish elsewhere this extension.

\subsection{How to find a  representative of  stability class}
Let us consider a  stability class $\mathscr{U}(H_*)$. 
The reader may wonder  how  to find the  representative $H_*$ of  stability class in actual  applications.
In many cases,  $H_*$   can be obtained by a certain scaling limit:
For example, the Heisenberg model is a large $U$-limit of the Hubbard model, and 
the Hubbard model  is a large $\omega$-limit of the Holstein-Hubbard model \cite{Miyao7}.
Remark that this approach was initiated  by Lieb \cite{Lieb}.
We can also  find a  representative of the   stability class of  the Kondo and Anderson models by a suitable scaling arguments.
 We will publish elsewhere relations   between   stability classes and scaling limits.

It is noteworthy that there are some interesting similarities between the idea of scaling limit in \cite{Miyao7}
and that of the gapped quantum liquid phases \cite{ZCZW}.
To study this aspect is a future problem.

\appendix

\section{Construction of self-dual cones  }\label{ConstSC}
\setcounter{equation}{0}
In this section, we define some self-dual cones which are important in the proofs of theorems in Section \ref{MLMLiebSta}.

\subsection{A canonical cone in $\mathscr{L}^2(\h )$}\label{L1Define}

Let $\h $ be a complex Hilbert space. The set of all
Hilbert--Schmidt class operators on $\h $ is denoted  by
$\mathscr{L}^2(\h )$, i.e.,  
$
\mathscr{L}^2(\h )=\{
\xi\in \mathscr{B}(\h )\, |\, \Tr[\xi^* \xi]<\infty
\}$.
Henceforth, we regard $\mathscr{L}^2(\h )$ as a Hilbert space equipped
with  the inner product $\la \xi| \eta \ra_{\mathscr{L}^2}=\Tr[\xi^*
\eta],\,   \xi, \eta\in \mathscr{L}^2(\h )$. 

For each $A\in \mathscr{B}(\h )$, the {\it left multiplication
operator} is defined by
\begin{align}
\mathcal{L}(A)\xi=A\xi,\ \ \xi\in \mathscr{L}^2(\h ).
\end{align} 
Similarly, the {\it right multiplication operator} is defined by 
\begin{align}
\mathcal{R}(A)\xi=\xi A, \ \ \xi\in \mathscr{L}^2(\h ).
\end{align} 
Note that $\mathcal{L}(A)$ and
 $\mathcal{R}(A)$
belong to $\mathscr{B}(\mathscr{L}^2(\h))$.  
It is not  difficult to check that 
\begin{align}
\mathcal{L}(A)\mathcal{L}(B)=\mathcal{L}(AB),\ \
 \mathcal{R}(A)\mathcal{R}(B)=\mathcal{R}(BA),\ \ A, B\in \mathscr{B}(\h ).
\end{align}

 Let $\vartheta$ be an
antiunitary  operator on $\h$.\footnote{
We say that a bijective map  $\vartheta$  on $\h$ is  {\it antiunitary}
 if $\la \vartheta x|\vartheta y\ra=\overline{\la x|y\ra}$ for all $x,
 y\in\h$.
} 
Let $\Phi_{\vartheta}$ be an isometric
isomorphism  from $\mathscr{L}^2(\h)$ onto $\h\otimes \h$ defined by
$
\Phi_{\vartheta}(|x\ra\la y|)=x\otimes \vartheta y\ \ \ \forall x,y\in \h. \label{VecEquiv}
$
Then,
\begin{align}
\mathcal{L}(A) =\Phi_{\vartheta}^{-1} A\otimes 1\Phi_{\vartheta} ,\ \
 \ \mathcal{R}(\vartheta
 A^*\vartheta)=\Phi_{\vartheta}^{-1} 1 \otimes A\Phi_{\vartheta} \label{Bare}
\end{align} 
for each $A\in \mathscr{B}(\h)$. We  write these facts simply  as 
\begin{align}
\h\otimes \h=\mathscr{L}^2(\h),\ \ A\otimes 1 =\mathcal{L}(A),\ \ 1 \otimes
 A=\mathcal{R}(\vartheta A^*\vartheta), \label{Ident}
\end{align} 
if no confusion arises.
If $A$ is self-adjoint,    then
 $\mathcal{L}(A)$ and $\mathcal{R}(A)$ are self-adjoint.

Recall that a  bounded linear operator $\xi$ on $\h$ is said to be {\it positive} if $\la x|
\xi x\ra_{\h} \ge 0$ for all $x\in \h$. We write this as $\xi\ge 0$.

\begin{define}\label{L2Define}{\rm 
A canonical   cone in $\mathscr{L}^2(\h )$ is given by
\begin{align}
\mathscr{L}^2(\h )_+= \Big\{\xi\in \mathscr{L}^2(\h )\, \Big|\,\mbox{$\xi$ is
 self-adjoint and $\xi\ge 0$
  as
 an operator on $\h $} \Big\}.\ \ \ \diamondsuit
\end{align} 
}
\end{define} 

\begin{Thm}\label{SDL2}
$\mathscr{L}^2(\h)_+$ is a self-dual cone in $\mathscr{L}^2(\h)$.
\end{Thm}  
{\it Proof.}
 Apply \cite[Proposition 2.5]{Miyao5} and  Theorem  \ref{SAH}. $\Box$

\begin{Prop}\label{GeneralPP}
Let $A\in \mathscr{B}(\h )$. We have 
 $\mathcal{L}(A^*)\mathcal{R}(A)\unrhd 0$ w.r.t. $\mathscr{L}^2(\h )_+$.
\end{Prop} 
{\it Proof.} For each $\xi\in \mathscr{L}^2(\h )_+$,  we have 
$
\mathcal{L}(A^*)\mathcal{R}(A)\xi=A^*\xi A \ge 0.
$ $\Box$

\begin{rem}
{\rm 
\begin{itemize}
\item
In \cite{Gloss}, Gross studied a theory of noncommutative integration in the fermionic Fock space; his framework is related to   this subsection.
\item Proposition \ref{GeneralPP} is closely connected  with reflection positivity, see, e.g., \cite{FSS, FILS}.
$\diamondsuit$
 \end{itemize} 
}
\end{rem}
\subsection{The hole-particle transformations}
Before we proceed, we introduce an important unitary operator $W $ as follows:
The {\it hole-particle transformation} is a unitary operator $W$ on $\mathfrak{E}$ 
satisfying the following (i) and (ii):
\begin{itemize}
\item[(i)] For all $x\in \Lambda$,
$
W c_{x\uparrow} W^* =c_{x\uparrow} $ and $ Wc_{x \downarrow}W^*=\gamma(x) c_{x\downarrow}^*,
$
 where $\gamma(x)=1$ if $x\in A$, $\gamma(x)=-1$ if $x\in B$. 
 \item[(ii)]
$\displaystyle 
W \Omega=\prod_{x\in \Lambda}^{\#} c_{x\downarrow}^*\Omega,
$
where $\Omega$ is the fermionic Fock vacuum in $\mathfrak{E}$,  and $\prod_{x\in \Lambda}^{\#}$
 indicates the product taken over all sites in $\Lambda$ with an arbitrarily fixed order.
\end{itemize}

The hole-particle transformations on 
$\mathfrak{E} \otimes 
\Fock_{\mathrm{ph}}$ and $\mathfrak{E}\otimes \Fock_{\mathrm{rad}}$ 
are defined by $W\otimes 1$.
\begin{Not}\label{TilNot}{\rm
Let $X$ be a linear operator on $\mathfrak{Z}$.
Suppose that $X$ commutes with $\Ne$ and  $S_{\mathrm{tot}}^{(3)}$.
We will use the following notations:
\begin{itemize}
\item
$\tilde{X}=WXW^{-1}$;
\item 
$\tilde{X}_{n}=WX_{n}W^{-1}$;

\item 
$
\tilde{X}_{n}[M]=W X_{n}[M]W^{-1}=\tilde{X}\restriction W \mathfrak{Z}_{n}[M]$. $\diamondsuit$
\end{itemize}
}
\end{Not}

\subsection{Some useful expressions of $W\mathfrak{E}_{n=|\Lambda|}[M]$ and $W\mathfrak{H}_{Q}$}

Let $\mathfrak{E}(\mathfrak{X})$ be the fermionic Fock space over $\mathfrak{X}$.
Remark the factorization property:
$
\mathfrak{E}(\mathfrak{X}_1\oplus \mathfrak{X}_2)=\mathfrak{E}(\mathfrak{X}_1)\otimes \mathfrak{E}(
\mathfrak{X}_2)
$.
By this fact, we have 
\begin{align}
\mathfrak{E}=\mathfrak{E}(\ell^2(\Lambda) \oplus \ell^2(\Lambda))
=\mathfrak{E}(\ell^2(\Lambda))\otimes \mathfrak{E}(\ell^2(\Lambda)). \label{FermiFact}
\end{align}
In this expression, the $N$-electron subspace becomes
\begin{align}
\mathfrak{E}_N=\bigoplus_{N_1+N_2=N} \mathcal{E}^{N_1} \otimes \mathcal{E}^{N_2},
\end{align}
where $\mathcal{E}^{K}=\bigwedge^{K} \ell^2(\Lambda)$. Moreover, 
the $M$-subspace can be  written as  
\begin{align}
\mathfrak{E}_N[M]=\mathcal{E}^{\frac{N}{2}+M} \otimes \mathcal{E}^{\frac{N}{2}-M}.\label{ExMSub}
\end{align}

Because 
$
\tilde{N}_{\mathrm{el}}=W\Ne W^*=2S^{(3)}_{\mathrm{tot}}+|\Lambda|
$ and $
  \tilde{S}_{\mathrm{tot}}^{(3)}=
WS_{\mathrm{tot}}^{(3)} W^*=\frac{\Ne}{2}-\frac{|\Lambda|}{2}, 
$
 we obtain, by (\ref{ExMSub}), that 
\begin{align}
W \mathfrak{E}_{n=|\Lambda|} [M]=\mathfrak{E}_{n=2M+|\Lambda|}[M=0]=\mathcal{E}^{\Md} \otimes \mathcal{E}^{\Md}, \label{HalfTens}
\end{align}
where   $\Md=M+|\Lambda|/2$.
This expression will play an important role in the present paper.
  Using this,   we have 
\begin{align}
W\mathfrak{H}_{Q}[M]=\tilde{Q} \mathcal{E}^{\Md} \otimes \mathcal{E}^{\Md}. \label{ElIdn}
\end{align}

The following formula  will be useful:
 \begin{align}
\tilde{Q}=WQW^* =\prod_{x\in \Lambda} (n_x-1)^2.\label{GwP}
\end{align}

We derive a convenient expression of $W\mathfrak{H}_{Q}[M]$ for later use.
To this end, we set $\mathcal{S}=\{0, 1\}^{\Lambda}$.
For each ${\Bs m } =({\Bs m}_{\uparrow}, {\Bs m}_{\downarrow}) \in \mathcal{S} \times \mathcal{S}$
with ${\Bs m}_{\sigma}=\{m_{x\sigma}\}_{x\in \Lambda}\, (\sigma=\uparrow, \downarrow)$, we define
\begin{align}
|{\Bs m}\ra =
\prod^{\#}_{x\in \Lambda} \Big(
c_{x\uparrow}^*
\Big)^{m_{x\uparrow}}
 \Big(
c_{x\downarrow}^*
\Big)^{m_{x\downarrow}}
\Omega. \label{Vecm}
\end{align}
Clearly, $\{ |{\Bs m}\ra\, |\, {\Bs m} \in \mathcal{S} \times \mathcal{S}\}$
 is a complete orthonormal system (CONS) of $\mathfrak{E}$.  Using  $|{\Bs m}\ra$,  we can represent $W\mathfrak{H}_{Q}[M]$ as 
 \begin{align}
W \mathfrak{H}_Q[M]=
\mathrm{Lin} \Big\{
| {\Bs m} \ra\ \Big|\ {\Bs m}_{\uparrow}={\Bs m}_{\downarrow}, \ \  |{\Bs m}_{\uparrow}|=
|{\Bs m} _{\downarrow}|=\Md \Big\}, \label{HeisEx}
\end{align}
where $|{\Bs m_{\sigma}}|=\sum_{x\in \Lambda}  m_{x\sigma}$.

\subsection{A self-dual cone in $W\mathfrak{E}_{n=|\Lambda|}[M]$}\label{EIdenP}
 Corresponding to the identification (\ref{FermiFact}), we have
\begin{align}
c_{x\uparrow}=\mathsf{c}_x\otimes 1,\ \ \ c_{x\uparrow}=(-1)^{\mathsf{N}}\otimes \mathsf{c}_x,\label{cTensorRP}
\end{align}
where
$\mathsf{c}_x$ is the annihilation operator on $\mathfrak{E}(\ell^2(\Lambda))$, and 
  $\mathsf{N}=\sum_{x\in \Lambda}
\mathsf{n}_x
$ with $\mathsf{n}_x=\mathsf{c}_x^* \mathsf{c}_x$.

Let us  define a natural self-dual cone in $W\mathfrak{E}_{n=|\Lambda|}[M]$.
By the identifications  (\ref{Ident}) and (\ref{HalfTens}), we have
\begin{align}
W\mathfrak{E}_{n=|\Lambda|} [M]=\mathscr{L}^2(\mathcal{E}^{\Md}) \label{ElIdn2}.
\end{align}
Here, the antiunitary $\vartheta$ for this expression is defined as 
\begin{align}
\vartheta \mathsf{c}_x\vartheta=\mathsf{c}_x,\ \ \ \ \vartheta \Omega=\Omega, \label{AniTheta}
\end{align}
where $\Omega$ is the Fock vacuum in $\mathfrak{E}(\ell^2(\Lambda))$. (As for $\vartheta$, see Section \ref{L1Define}.)
Now, we define a natural self-dual cone $\mathfrak{P}_{\mathrm{H}}[M]$ in $\mathscr{L}^2(\mathcal{E}^{\Md})$
 by 
 \begin{align}
 \mathfrak{P}_{\mathrm{H}} [M]=\mathscr{L}^2(\mathcal{E}^{\Md})_+. \label{SCH}
 \end{align}

\subsection{A self-dual cone in $W\mathfrak{E}_{n=|\Lambda|} [M]\otimes \Fock_{\mathrm{ph}}$}\label{SchRP}
We set 
$
p_x=i \sqrt{\frac{\omega}{2}}(b_x^*-b_x) $ and $ q_x=\frac{1}{\sqrt{2\omega}}(b_x^*+b_x).\label{bPQ}
$
Both operators are essentially self-adjoint. We denote their closures by the same symbols.
Remark the following identification:
$
\Fock_{\mathrm{ph}}=L^2(\mathcal{Q}, d{\Bs q})=L^2(\mathcal{Q}),
$
where $\mathcal{Q}=\BbbR^{|\Lambda|},\ d{\Bs q} =\prod_{x\in \Lambda} dq_x$ is the $|\Lambda|$-dimensional
Lebesgue measure on $\mathcal{Q}$, and $L^2(\mathcal{Q})$ is the Hilbert space of the square 
integrable functions on $\mathcal{Q}$.
Under this identification, $q_x$ and $p_x$ can be viewed as multiplication and partial differential operators, 
respectively. 
This expression of $p_x$ and $q_x$ in $L^2(\mathcal{Q})$ is called the {\it Schr\"odinger representation}. The Hilbert space $W\mathfrak{E}_{n=|\Lambda|}[M] \otimes \Fock_{\mathrm{ph}}$
 can be identified with $
 W\mathfrak{E}_{n=|\Lambda|}[M] \otimes L^2(\mathcal{Q})
 $ in this representation.

A natural self-dual cone in $W\mathfrak{E}_{n=|\Lambda|}[M] \otimes \Fock_{\mathrm{ph}}$
 is defined by 
 \begin{align}
 \mathfrak{P}_{\mathrm{HH}}[M]=\int_{\mathcal{Q}}^{\oplus} \mathfrak{P}_{\mathrm{H}}[M] d{\Bs q}, \label{DefPHH}
 \end{align}
 where the RHS of (\ref{DefPHH}) is the direct integral of $\mathfrak{P}_{\mathrm{H}}[M]$, see Appendix \ref{App2} for details.

\subsubsection{A self-dual cone in $W\mathfrak{H}_{Q}[M]$}
Using the expression (\ref{HeisEx}), we define a self-dual cone in $W \mathfrak{H}_Q[M]$ by 
\begin{align}
\mathfrak{P}_{\mathrm{Heis}}[M]
=
\mathrm{Coni} \Big\{
|{\Bs m}\ra\ \Big|\ {\Bs m}_{\uparrow}={\Bs m}_{\downarrow},\ 
|{\Bs m}_{\uparrow}|=|{\Bs m}_{\downarrow}| =\Md
\Big\}.\label{DefPHeis}
\end{align}

The following property will be useful:
\begin{Prop}\label{HeisHP}
Under the identification (\ref{ElIdn2}), we have 
$
\mathfrak{P}_{\mathrm{Heis}}[M]=\tilde{Q} \mathfrak{P}_{\mathrm{H}}[M], 
$
where $\tilde{Q}$ is given by (\ref{GwP}).
\end{Prop}
{\it Proof.} Let $\vphi\in W\mathfrak{E}_{n=|\Lambda|}[M]$.  $\vphi$ can be expressed as 
$
\vphi=\sum_{{\Bs m}_{\uparrow},
{\Bs m}_{\downarrow}\in \mathcal{S}(\Md)} \vphi_{
{\Bs m}_{\uparrow},
{\Bs m}_{\downarrow}}|{\Bs m}\ra,
$
where $\mathcal{S}(\Md)=\{
{\Bs m\in \mathcal{S}\, |\, |{\Bs m}|=\Md}
\}$.
We say that $\{
\vphi_{
{\Bs m}_{\uparrow},
{\Bs m}_{\downarrow}}
\}$ is positive semidefinite, if 
$
\sum_{{\Bs m}_{\uparrow},
{\Bs m}_{\downarrow}\in \mathcal{S}(\Md)} \vphi_{
{\Bs m}_{\uparrow},
{\Bs m}_{\downarrow}} z^*_{{\Bs m}_{\uparrow}} z_{{\Bs m}_{\downarrow}} \ge 0
$
for all ${\Bs z}=\{
z_{{\Bs m}}
\}_{{\Bs m} \in \mathcal{S}(\Md)} \in \BbbC^{|\mathcal{S}(\Md)|}$. Remark that 
$
\vphi_{
{\Bs m}_{\uparrow},
{\Bs m}_{\downarrow} }\ge 0
$
if ${\Bs m}_{\uparrow}={\Bs m}_{\downarrow}$ in this case.

We note  that $\vphi\in \Cone_{\mathrm{H}}[M]$ if and only if 
$
\{\vphi_{
{\Bs m}_{\uparrow},
{\Bs m}_{\downarrow}}
\}
$ is positive semidefinite.
To see this, we just remark that, by the definition of $\Phi_{\vartheta}$ in Section \ref{L1Define}, 
$\vphi$ can be written as 
$
\vphi=\sum_{{\Bs m}_{\uparrow},
{\Bs m}_{\downarrow}\in \mathcal{S}(\Md)} \vphi_{
{\Bs m}_{\uparrow},
{\Bs m}_{\downarrow}}|{\Bs m}_{\uparrow}\ra \la {\Bs m}_{\downarrow}|.
$
Here, we used the fact that $|\vartheta {\Bs m}_{\downarrow}\ra=|{\Bs m}_{\downarrow}\ra$
by (\ref{AniTheta}).

For each $x\in \Lambda$, we observe that 
$
(n_x-1)^2|{\Bs m} \ra\neq 0
$, if and only if $m_{x\uparrow}=m_{x\downarrow}$.
Hence, $\tilde{Q}|{\Bs m}\ra\neq 0$,  if and only if ${\Bs m}_{\uparrow}={\Bs m}_{\downarrow}$.
Therefore, for each $\vphi\in \mathfrak{P}_{\mathrm{H}}[M]$, we obtain
\begin{align}
\tilde{Q} \vphi=\sum_{{\Bs m}_{\uparrow}=
{\Bs m}_{\downarrow}\in \mathcal{S}(\Md)} \vphi_{
{\Bs m}_{\uparrow},
{\Bs m}_{\downarrow}}|{\Bs m}\ra.\label{QAct}
\end{align}
Because $
 \vphi_{
{\Bs m}_{\uparrow},
{\Bs m}_{\downarrow}}
\ge 0
$ provided that  $
{\Bs m}_{\uparrow}={\Bs m}_{\downarrow}
$, the RHS of (\ref{QAct}) belongs to $
\mathfrak{P}_{\mathrm{Heis}}[M]
$. $\Box$

\section{Proof of Theorems \ref{GeneSt} and  \ref{HeisMLM}}\label{PfLi1}
\setcounter{equation}{0}
In this appendix, we will prove Theorems \ref{GeneSt} and  \ref{HeisMLM}.
Because several notations are defined in Appendix \ref{ConstSC},
the reader is suggested to study  Appendix \ref{ConstSC} first.

\subsection{Basic properties of $H_{\mathrm{Heis}}$ }
For each $x, y\in \Lambda$ with $x\neq y$ and $\sigma\in \{\uparrow, \downarrow\}$, set
$
A_{xy\sigma} =c_{x\sigma}c_{y\sigma}^*. \label{Axy}
$
One can express $H_{\mathrm{Heis}}$ as 
\begin{align}
H_{\mathrm{Heis}}
&=\sum_{x, y\in \Lambda}\frac{J_{xy}}{2}
\bigg\{-(A_{xy \uparrow}A_{xy
 \downarrow}^*+A_{xy\uparrow}^* A_{xy \downarrow})
+\frac{1}{2}(n_{x\uparrow}-n_{x\downarrow})(n_{y\uparrow}-n_{y\downarrow})\bigg\}.
\end{align}
By the hole-particle transformation $W$, we have
$
\tilde{H}_{\mathrm{Heis}}=WH_{\mathrm{Heis}} W^{-1}=-T+V,
$
where
\begin{align}
T=\sum_{x, y\in \Lambda}\frac{J_{xy}}{2}
(A_{xy \uparrow}^*A_{xy
 \downarrow}^*+A_{xy\uparrow} A_{xy \downarrow}),\ \ \ 
 V= \sum_{x, y\in \Lambda} \frac{J_{xy}}{4} (n_x-1)(n_y-1).
\end{align}

 For each $x, y\in \Lambda$, we set
$
T_{xy}=\frac{J_{xy}}{2} A_{xy\uparrow} A_{xy \downarrow}.
$
Trivially, we have $T=\sum_{x,y\in \Lambda}(T_{xy}+T_{xy}^*)$.

\begin{lemm}\label{TxyP}
Let $x, y\in \Lambda$ with $x\neq y$.
For all ${\Bs m} =({\Bs m}_{\uparrow}, {\Bs m}_{\downarrow}) \in \mathcal{S} \times \mathcal{S}$ with ${\Bs m}_{\uparrow}={\Bs m}_{\downarrow}$, 
we have the following:
\begin{itemize}
\item[{\rm (i)}] $\displaystyle
T_{xy}|{\Bs m}\ra=
\frac{J_{xy}}{2}
\big|{\Bs m_{\uparrow}-{\Bs\delta}_{x}+{\Bs \delta}_{y}},
{\Bs m_{\downarrow}-{\Bs\delta}_{x}+{\Bs \delta}_{y}}
\big\ra 
$, where ${\Bs \delta}_a=\{ \delta_{ax}\}_{x\in \Lambda}$.
\item[{\rm (ii)}] $\displaystyle
T_{xy}^* |{\Bs m}\ra=
\frac{J_{xy}}{2}
\big|{\Bs m_{\uparrow}+{\Bs\delta}_{x}-{\Bs \delta}_{y}},
{\Bs m_{\downarrow}+{\Bs\delta}_{x}-{\Bs \delta}_{y}}
\big\ra 
$.
\end{itemize}
In the above, we understand that 
\begin{align}
\big|{\Bs m_{\uparrow}-{\Bs\delta}_{x}+{\Bs \delta}_{y}},
{\Bs m_{\downarrow}-{\Bs\delta}_{x}+{\Bs \delta}_{y}}
\big\ra =0
\end{align}
if $m_{\uparrow x}=0$ or $m_{\uparrow y}=1$ or $m_{\downarrow x}=0$ or $m_{\downarrow y}=1$, and 
\begin{align}
\big|{\Bs m_{\uparrow}+{\Bs\delta}_{x}-{\Bs \delta}_{y}},
{\Bs m_{\downarrow}+{\Bs\delta}_{x}-{\Bs \delta}_{y}}
\big\ra =0
\end{align}
if $m_{\uparrow x}=1$ or $m_{\uparrow y}=0$ or $m_{\downarrow x}=1$ or $m_{\downarrow y}=0$.
\end{lemm}
{\it Proof.} This lemma immediately follows from (\ref{Vecm}) and (\ref{Axy}). $\Box$
\medskip\\

Let $\Cone_{\mathrm{Heis}}[M]$ be the self-dual cone in $W\mathfrak{H}_Q[M]$ defined by (\ref{DefPHeis}).
\begin{lemm}\label{TPos}
We have  $T_{xy} \unrhd 0$ w.r.t. $\Cone_{\mathrm{Heis}}[M]$. Therefore, we have
$
T\unrhd 0 $ and $ e^{\beta T} \unrhd 0
$
w.r.t. $\Cone_{\mathrm{Heis}}[M]$ for all $\beta \ge 0$.
\end{lemm}
{\it Proof.} By Lemma \ref{TxyP},  we immediately get $T_{xy} \unrhd 0$ w.r.t. $\mathfrak{P}_{\mathrm{Heis}}[M]$, which implies that $T\unrhd 0$ w.r.t. $\Cone_{\mathrm{Heis}}[M]$.
Thus, applying Proposition \ref{PSum}, we conclude that $e^{\beta T} \unrhd 0$ w.r.t. $\mathfrak{P}_{\mathrm{Heis}}[M]$ for all $\beta \ge 0$. $\Box $

\begin{lemm}\label{VPos}
$e^{-\beta V} \unrhd 0$ w.r.t. $\Cone_{\mathrm{Heis}}[M]$ for all $\beta \ge 0$.
\end{lemm}
{\it Proof.} Since 
$
n_x|{\Bs m}\ra=(m_{x\uparrow}+m_{x\downarrow}) |{\Bs m}\ra
$, every $|{\Bs m}\ra$ is an eigenvector of $V$: $V|{\Bs m}\ra=V({\Bs m}) |{\Bs m}\ra$,
where $V({\Bs m})$ is the corresponding eigenvalue.
Hence, $
e^{-\beta V} |{\Bs m}\ra=e^{-\beta V({\Bs m})} |{\Bs m}\ra
$, which implies $e^{-\beta V} \unrhd 0$ w.r.t. $\Cone_{\mathrm{Heis}}[M]$. $\Box$

\begin{Prop}
$
e^{-\beta \tilde{H}_{\mathrm{Heis}}[M]} \unrhd 0
$ w.r.t. $\Cone_{\mathrm{Heis}}[M]$ for all $\beta \ge 0$.
\end{Prop}
{\it Proof.} By Lemmas \ref{TPos}, \ref{VPos} and Theorem  \ref{TK}, we arrive at the assertion in the proposition. $\Box$

\begin{Thm} \label{PIHeis}
Assume {\bf  (C. 1)} and {\bf (C. 2)}. We have
$
\big( \tilde{H}_{\mathrm{Heis}}[M]+s\big)^{-1} \rhd 0
$ w.r.t. $\Cone_{\mathrm{Heis}}[M]$ for all $s >-E\big(
\tilde{H}_{\mathrm{Heis}}[M]
\big)$.
In particular, $
\big( \tilde{H}_{\mathrm{MLM}}[M]+s\big)^{-1} \rhd 0
$ w.r.t. $\Cone_{\mathrm{Heis}}[M]$ for all $s>-E\big(
\tilde{H}_{\mathrm{MLM}}[M]
\big)$.
\end{Thm}
{\it Proof.}
Before we will enter the proof, we remark the following. Let $\psi\in \Cone_{\mathrm{Heis}}[M]$.
Set $S_M=\{
{\Bs m}=({\Bs m}_{\uparrow}, {\Bs m}_{\downarrow}) \in \mathcal{S}\times \mathcal{S}\, |\, 
{\Bs m}_{\uparrow}={\Bs m}_{\downarrow},\ |{\Bs m}_{\uparrow}|=|{\Bs m}_{\downarrow}|=\Md
\}$.
We can express $\psi$ as 
$
\psi=\sum_{{\Bs m} \in S_M} \psi({\Bs m}) |{\Bs m}\ra \label{Expsi}
$
with $\psi({\Bs m}) \ge 0$ for all ${\Bs m}\in S_M$.

We will apply Theorem \ref{PEq} with $A=-T$ and $B=V$.
To this end, we will check all assumptions in the theorem.

The assumption (b) is satisfied by Lemma  \ref{VPos}.
To check (a) and (c), we set 
$
V_n=(1-e^{-n})V, \ n\in \BbbN
$. Trivially, $-T+V_n$ converges to $\tilde{H}_{\mathrm{Heis}}[M]$, and $ \tilde{H}_{\mathrm{Heis}}[M]-V_n$ converges to $-T$
in the uniform topology as $n\to \infty$. Thus, the assumption (a) is fulfilled.
To see (c), take $\psi, \psi'\in \Cone_{\mathrm{Heis}}[M]$. Suppose that $\la \psi|\psi'\ra=0$.
We can express these as 
\begin{align}
\psi=\sum_{{\Bs m} \in S_M} \psi({\Bs m}) |{\Bs m}\ra,\ \ \ \psi\rq{}=\sum_{{\Bs m} \in S_M} \psi\rq{}({\Bs m}) |{\Bs m}\ra \label{TwoVecR}
\end{align}
with $\sum_{{\Bs m} \in S_M} \psi({\Bs m}) \psi'({\Bs m})=0$. Because $\psi({\Bs m})$ and $\psi'({\Bs m})$
 are nonnegative, we conclude $\psi(\Bs m) \psi'(\Bs m)=0$ for all ${\Bs m} \in S_M$. Thus, 
 \begin{align}
 \la \psi|e^{-\beta V_n} \psi'\ra
 =\sum_{{\Bs m} \in S_M} \psi(\Bs m) \psi'(\Bs m) e^{-\beta(1-e^{-n})V({\Bs m})}=0,
 \end{align}
 where $V({\Bs m})$ is defined in the proof of Lemma \ref{VPos}.
Therefore, (c) is satisfied.

Choose $\psi, \psi\rq{} \in \Cone_{\mathrm{Heis}}[M] \backslash \{0\}$, arbitrarily.
Again we  express these vectors  as (\ref{TwoVecR}).
Since $\psi$ and $\psi\rq{}$ are nonzero, there exist ${\Bs m}, {\Bs m}\rq{} \in S_M$ such that 
$\psi({\Bs m})>0$ and $\psi\rq{}({\Bs m\rq{}})>0$.
Because $\psi \ge \psi({\Bs m}) |{\Bs m}\ra$ and $\psi\rq{} \ge \psi\rq{}({\Bs m}\rq{}) |{\Bs m}\rq{}\ra$
 w.r.t. $\Cone_{\mathrm{Heis}}[M]$, we see that 
 \begin{align}
 \la \psi|e^{\beta T} |\psi\rq{}\ra \ge \psi({\Bs m}) \psi\rq{}({\Bs m}\rq{}) \la {\Bs m}|e^{\beta T} |{\Bs m}\rq{}\ra \label{LowT1}
 \end{align}
 by Lemma \ref{TPos}.

By Lemma \ref{TPos} again, we have
\begin{align}
e^{\beta T} =\sum_{n  =0}^{\infty} \frac{\beta^{n }}{n!} T^{n}
 \unrhd 
 \frac{\beta^{\ell}}{\ell!} T^{\ell}\label{LowT2}
 \end{align}
w.r.t. $\Cone_{\mathrm{Heis}}[M]$ for all $\beta \ge 0$ and $\ell\in \BbbN_0$.
On the other hand, using the assumption {\bf (C. 1)} and Lemma \ref{TPos}, there are  sequencces $
\{x_1, x_2, \dots, x_{\ell}; y_1, y_2, \dots, y_{\ell}
\}
$  and $\{\#_1, \dots, \#_{\ell}\}\in \{{\pm}\}^{\ell}$such that 
\begin{align}
\Big\la {\Bs m} \Big| T_{x_1y_1}^{\#_1} T_{x_2y_2}^{\#_2}\cdots T_{x_{\ell} y_{\ell}}^{\#_{\ell}}\Big|{\Bs m}\rq{}\Big\ra>0,  \label{LowT3}
\end{align}
where $a^{\#}=a$ if $\#=-$,  $a^{\#}=a^*$ if $\#=+$.
Since $T \unrhd T_{xy}$ for all $x, y\in \Lambda$, we have
\begin{align}
T^{\ell}\unrhd  T_{x_1y_1}^{\#_1} T_{x_2y_2}^{\#_2}\cdots T_{x_{\ell}y_{ \ell}}^{\#_{\ell}} \label{LowT4}
\end{align}
w.r.t. $\Cone_{\mathrm{Heis}}[M]$ for all $\beta \ge 0$.
Combining (\ref{LowT2}),  (\ref{LowT3}) and (\ref{LowT4}), we obtain
\begin{align}
\la {\Bs m}|e^{\beta T} |{\Bs m}\rq{}\ra 
 \ge 
 \frac{\beta^{\ell}}{\ell!}
 \Big\la {\Bs m} \Big| T_{x_1y_1}^{\#_1} T_{x_2y_2}^{\#_2}\cdots T_{x_{\ell} y_{\ell}}^{\#_{\ell}}\Big|{\Bs m}\rq{}\Big\ra>0. 
\end{align}
By this and (\ref{LowT1}), we  conclude that  $
e^{\beta T} \rhd 0
$ w.r.t. $\Cone_{\mathrm{Heis}}[M]$ for all $\beta >0$.
Therefore,
we conclude that, by Theorem \ref{PEq},
  $
e^{-\beta \tilde{H}_{\mathrm{Heis}}[M]} \rhd 0
$ w.r.t. $\Cone_{\mathrm{Heis}}[M]$ for all $\beta >0$.
Because $H_{\mathrm{MLM}} $ is a special case of the Heisenberg model, we also get 
 $
e^{-\beta \tilde{H}_{\mathrm{MLM}}[M]} \rhd 0
$ w.r.t. $\Cone_{\mathrm{Heis}}[M]$ for all $\beta >0$. 
By applying Proposition \ref{PIEquiv}, we obtain the desired result in Theorem \ref{PIHeis}.
  $\Box$ 

\subsection{Proof of Theorem \ref{GeneSt}}

We begin with the following proposition \cite{Marshall, LiebMattis}:
\begin{Prop}\label{MLMSpin}
The ground state of $H_{\mathrm{MLM}}$
  has total spin $S=\frac{1}{2}\big||A|-|B|\big|$ and is unique 
    apart from the trivial $(2S+1)$-degeneracy.
\end{Prop}
{\it Proof.}
By Theorems \ref{PFF} and \ref{PIHeis},  the ground state of $H_{\mathrm{MLM}}[M]$ is unique for all $M\in \mathrm{spec}\Big(S_{\mathrm{tot}, |\Lambda|}^{(3)}\Big)$.
Let us work in the $M=0$ subspace.
Since $2H_{\mathrm{MLM}}=({\Bs S}_A+ {\Bs S}_B)^2-{\Bs S}_A^2-{\Bs S}_B^2$,
we  see that the ground state of $H_{\mathrm{MLM}}[M=0]$ has total spin $
S=\frac{1}{2}\big||A|-|B|\big|$.  We denote by $\vphi_0$ the ground state of $H_{\mathrm{MLM}}[0]$.
Let 
$
S_{\mathrm{tot}}^{(\pm)}=S_{\mathrm{tot}}^{(1)}\pm i S_{\mathrm{tot}}^{(2)}.
$
We set
$\vphi_M=(S_{\mathrm{tot}}^{(+)})^M \vphi_0$ for $M>0$ and  $\vphi_M=(S_{\mathrm{tot}}^{(-)})^{|M|} \vphi_0$ for $M<0$. 
Because $H_{\mathrm{MLM}}$ commutes with $S_{\mathrm{tot}}^{(\pm)}$,
we know that $\vphi_M$ is the unique ground state of $H_{\mathrm{MLM}}[M]$
which has total spin   $
S=\frac{1}{2}\big||A|-|B|\big|$. 
Thus, the ground state of $H_{\mathrm{MLM}}$ is unique apart from the trivial $(2S+1)$-degeneracy.
$\Box$
\medskip

\begin{flushleft}
{\it 
Completion of proof of Theorem \ref{GeneSt}
}
\end{flushleft}
 By applying Theorem \ref{ElUniTh} and Proposition \ref{MLMSpin}, one  obtains Theorem \ref{GeneSt}. $\Box$
 
\subsection{Proof of Theorem \ref{HeisMLM}}
By Theorem \ref{PIHeis}, we can  check all conditions in  Definition \ref{FirstSt} with 
\begin{align*}
&(H, \ H_*;\  \mathfrak{H}, \ \mathfrak{H}_*;\  P;\  \mathfrak{P}, \ \mathfrak{P}_*; O; U)\no
=&
\Big(\tilde{H}_{\mathrm{Heis}}[M],\,  \tilde{H}_{\mathrm{MLM}}[M];\, W\mathfrak{H}_{Q}[M],\  
W\mathfrak{H}_{Q}[M]; 1; \, \mathfrak{P}_{\mathrm{Heis}}[M],\,   \mathfrak{P}_{\mathrm{Heis}}[M];
\tilde{S}_{\mathrm{tot}, |\Lambda|}^2[M]; 1\Big).
\end{align*}
Hence, $
\tilde{H}_{\mathrm{Heis}}[M] \leadsto \tilde{H}_{\mathrm{MLM}}[M]
$ for all $M\in \mathrm{spec}\Big(S_{\mathrm{tot}, |\Lambda|}^{(3)}\Big)$.

By interchanging the roles of $H_{\mathrm{Heis}}$ and $H_{\mathrm{MLM}}$ in the above, we have 
$\tilde{H}_{\mathrm{MLM}}[M] \leadsto
\tilde{H}_{\mathrm{Heis}}[M] 
$ for all $M\in \mathrm{spec}\Big(S_{\mathrm{tot}, |\Lambda|}^{(3)}\Big)$. 
Thus, $\tilde{H}_{\mathrm{MLM}}[M] \equiv
\tilde{H}_{\mathrm{Heis}}[M] $.
By applying Proposition \ref{UniEQ} with $V=W$, we conclude that $H_{\mathrm{MLM}}[M] \equiv
H_{\mathrm{Heis}}[M] $.
Especially, $H_{\mathrm{Heis}}$  belongs to the Marshall-Lieb-Mattis  stability class.
$\Box$

\section{Proof of Theorem \ref{GeneLieb}}\label{PfLi2}
\setcounter{equation}{0}

Because several notations are defined in Appendix \ref{ConstSC}, the reader is suggested to study  Appendix \ref{ConstSC} first.
\subsection{Basic materials}

\subsubsection{Properties of $\tilde{Q}$}

Let $\Cone_{\mathrm{H}}[M]$ be the self-dual cone in $W\mathfrak{E}_{n=|\Lambda|}[M]$ defined by (\ref{SCH}).
Recall that $\tilde{Q}$ is defined by $\tilde{Q}=WQW^{-1}$, see Notation \ref{TilNot} for more details.
\begin{Prop}\label{ProjPP}
 We have $\tilde{Q} \unrhd 0$ w.r.t. $\Cone_{\mathrm{H}}[M]$ for all $M\in \mathrm{spec}\Big(S_{\mathrm{tot}, |\Lambda|}^{(3)}\Big)$.
\end{Prop}
{\it Proof.}
Recall the formula (\ref{GwP}).
Set $q_x=(1-n_x)^2$. Since 
\begin{align}
q_x=1-n_x+2n_{x\uparrow} n_{x\downarrow}
=1-\mathcal{L}({\mathsf{n}}_x)-\mathcal{R}(\mathsf{n}_x)+2\mathcal{L}(\mathsf{n}_x) \mathcal{R}(\mathsf{n}_x),
\end{align}
we have, by Theorem \ref{AbstRP}, 
\begin{align}
e^{tq_x} \unrhd 0\ \ \ \mbox{ w.r.t. $\Cone_{\mathrm{H}}[M]$ for all $t\ge 0$.} \label{epP}
\end{align}
Because $q_x$ is an orthogonal projection, we have 
\begin{align}
e^{tq_x} =1+q_x(e^t-1)=q_x^{\perp} +q_x e^t. \label{epP2}
\end{align}
By (\ref{epP}) and (\ref{epP2}), we obtain 
$
e^{-t}q_x^{\perp} +q_x \unrhd 0
$ w.r.t. $\Cone_{\mathrm{H}}[M]$. Taking $t\to \infty$, we arrive at 
$q_x\unrhd 0$ w.r.t. $\Cone_{\mathrm{H}}[M]$, which implies that 
$\displaystyle 
\tilde{Q} =\prod_{x\in \Lambda} q_x \unrhd 0
$
w.r.t. $\Cone_{\mathrm{H}}[M]$. $\Box$

\subsubsection{Properties of $\tilde{H}_{\mathrm{H}, |\Lambda|}[M]$}

First, we remark that, by the hole-particle transformation, 
the Coulomb interaction term becomes
\begin{align}
W \sum_{x,y\in \Lambda} \frac{U_{xy}}{2} (n_x-1)(n_y-1) W^{-1}
=
 \sum_{x,y\in \Lambda} \frac{U_{xy}}{2} (n_{x\uparrow}-n_{x\downarrow})(n_{y\uparrow}-n_{y\downarrow}).
 \end{align}
 Using the identification (\ref{ElIdn2}), we can express $\tilde{H}_{\mathrm{H}, |\Lambda|}[M]$ as 
$
\tilde{H}_{\mathrm{H}, |\Lambda|}[M] =\mathbb{T}-\mathbb{U},
$
where
\begin{align}
\mathbb{T}=\mathcal{L}(T)+\mathcal{R}(T),\ \ \ 
T=\sum_{x, y\in \Lambda} \bigg(
t_{xy} \mathsf{c}_{x}^* \mathsf{c}_y+\frac{U_{xy}}{2} \mathsf{n}_x\mathsf{n}_y
\bigg)
\end{align}
and
$
\mathbb{U}=\sum_{x, y\in \Lambda} U_{xy} \mathcal{L}(\mathsf{n}_x) \mathcal{R}(\mathsf{n}_y).
$
Here, recall that $\mathsf{c}_x$ and $\mathsf{n}_x$ are defined in Section \ref{EIdenP}.
Also, recall that $\mathcal{L}(\cdot)$ and $\mathcal{R}(\cdot)$ are defined in Section \ref{L1Define}.

\begin{lemm}\label{Coulomb}
We have  $\mathbb{U} \unrhd 0$ w.r.t. $\Cone_{\mathrm{H}}[M]$.
\end{lemm}
{\it Proof.}
Since  $\mathbf{U}=\{U_{xy}\}_{x, y}$ is a positive definite matrix by {\bf (A. 3)}, all eigenvalues $\{\lambda_x\}_{x\in \Lambda}$ of $\mathbf{U}$ are strictly positive,  and  there exists an orthogonal matrix $\mathbf{P}$ 
such that $\mathbf{U}=\mathbf{P} \mathbf{D} \mathbf{P}^{\mathrm{T}}$, where $
\mathbf{D}=\mathrm{diag}(\lambda_x)$. We set $\mathbf{n}=\{\mathsf{n}_x\}_{x\in \Lambda}$
and set $\hat{\mathbf{n}}=\mathbf{P}^{\mathrm{T}} \mathbf{n}$.
Denoting $\hat{\mathbf{n}}=\{\hat{\mathsf{n}}_x\}_{x\in \Lambda}$, we have 
\begin{align}
\mathbb{U} =\big\la \mathcal{L}(\mathbf{n})|\mathbf{U} \mathcal{R}(\mathbf{n}) \big\ra
=\big\la \mathcal{L}(\hat{\mathbf{n}})|\mathbf{D} \mathcal{R}(\hat{\mathbf{n}}) \big\ra
=\sum_{x\in \Lambda} \lambda_x \mathcal{L}(\hat{\mathsf{n}}_x)\mathcal{R}(\hat{\mathsf{n}}_x)\unrhd 0.
\label{CouEx}
\end{align}
This completes the proof.  $\Box$

\begin{Prop} \label{HubbardPP}
We have 
$\big( \tilde{H}_{\mathrm{H}, |\Lambda|}[M]+s\big)^{-1} \unrhd 0$
 w.r.t. $\Cone_{\mathrm{H}}[M]$ for all $s>-E\big(
 \tilde{H}_{\mathrm{H}, |\Lambda|}[M]
 \big)$.
\end{Prop}
{\it Proof.} We remark that $\mathbb{U} \unrhd 0$ w.r.t. $\Cone_{\mathrm{H}}[M]$ by  Lemma \ref{Coulomb},  and 
\begin{align}
e^{-\beta \mathbb{T}} =\mathcal{L} (e^{-\beta T}) \mathcal{R} (e^{-\beta T}) \unrhd 0 \ \ \mbox{w.r.t. $\Cone_{\mathrm{H}}[M]$}.
\end{align}
Therefore, by Theorem \ref{StandPP}, we have 
$
e^{-\beta 
\tilde{H}_{\mathrm{H}, |\Lambda|}[M]
} \unrhd 0
$ w.r.t.  $\Cone_{\mathrm{H}}[M]$ for all $\beta >0$.
By applying Proposition \ref{PPEquiv}, we conclude 
  the assertion in Proposition \ref{HubbardPP}. $\Box$
\medskip\\

In \cite[Theorem 4.6]{Miyao2}, we  proved the following stronger result: 
\begin{Thm} \label{HPP}
 We have $\big(\tilde{H}_{\mathrm{H}, |\Lambda|}[M]+s\big)^{-1} \rhd 0$ w.r.t. $\Cone_{\mathrm{H}}[M]$ for all $s >-E\big(\tilde{H}_{\mathrm{H}, |\Lambda|}[M]\big)$.
\end{Thm}

\subsection{Completion of   proof of Theorem \ref{GeneLieb}}
We will  check all conditions in Definition \ref{FirstSt} with 
\begin{align*}
&(H, \ H_*;\  \mathfrak{H},\  \mathfrak{H}_*;\  P;\  \mathfrak{P}, \ \mathfrak{P}_*; O; U)\no
=&
\Big(\tilde{H}_{\mathrm{H}, |\Lambda|}[M], \, \tilde{H}_{\mathrm{Heis}}[M];\,  W\mathfrak{E}_{n=|\Lambda|}[M],\  
W\mathfrak{H}_Q[M];\,  \tilde{Q};\,
\mathfrak{P}_{\mathrm{H}}[M],\,  \mathfrak{P}_{\mathrm{Heis}}[M];\ 
\tilde{S}_{\mathrm{tot}, |\Lambda|}^2[M]; 1\Big).
\end{align*}

By Proposition \ref{ProjPP}, (i) of Definition \ref{FirstSt} is satisfied.
(ii) of Definition \ref{FirstSt} follows from Proposition \ref{HeisHP}.
(iii) and (iv) of Definition \ref{FirstSt} follow from Theorem \ref{HPP} and 
Theorem \ref{PIHeis}, respectively.
Taking Proposition \ref{UniEQ} (with $V=W$) into consideration, we obtain Theorem \ref{GeneLieb}. $\Box$

\section{Proof of Theorem \ref{GeneHH}} \label{StabHH}\label{PfLi3}
\setcounter{equation}{0}

\subsection{Basic materials}

\subsubsection{Properties  of $P_{\mathrm{ph}}$}

Let $P_{\mathrm{ph}}=1\otimes |\Omega_{\mathrm{ph}} \ra\la \Omega_{\mathrm{ph}}|$.
Let $\Cone_{\mathrm{HH}}[M]$ be the self-dual cone in $W\mathfrak{E}_{n=|\Lambda|}[M]\otimes \Fock_{\mathrm{ph}}$ defined by (\ref{DefPHH}).
\begin{Prop}\label{HHProjP}
We have $P_{\mathrm{ph}} \unrhd 0$ w.r.t. $\Cone_{\mathrm{HH}}[M]$
for all $M\in \mathrm{spec}\Big(S_{\mathrm{tot}, |\Lambda|}^{(3)}\Big)$.
\end{Prop}
{\it Proof.} 
Remark that, in the Schr\"{o}dinger representation, 
the Fock vacuum $\Omega_{\mathrm{ph}}$ can be expressed as $
\Omega_{\mathrm{ph}}({\Bs q})=\pi^{-|\Lambda|/4}\exp(-{\Bs q}^2/2)
$, which is strictly positive for all ${\Bs q}$.
Hence, for each $\vphi\in L^2(\mathcal{Q})_+$, we have
$\la \Omega_{\mathrm{ph}}|\vphi\ra \ge 0$, which implies that 
$|\Omega_{\mathrm{ph}}\ra \la \Omega_{\mathrm{ph}}| \unrhd 0$ w.r.t. $L^2(\mathcal{Q})_+$.
Thus, by Corollary \ref{BasicDP2}, we conclude the assertion in  Proposition \ref{HHProjP}. $\Box$

\begin{Prop}\label{HHProjP2}
Under the identification by $\tau$ in Section \ref{StLI}, we have 
$P_{\mathrm{ph}} \mathfrak{P}_{\mathrm{HH}}[M]=\mathfrak{P}_{\mathrm{H}}[M]$
 for all $M\in \mathrm{spec}\Big(S_{\mathrm{tot}, |\Lambda|}^{(3)}\Big)$.
\end{Prop}
{\it Proof.} It is easily checked that $P_{\mathrm{ph}} \mathfrak{P}_{\mathrm{HH}}[M]
=\mathfrak{P}_{\mathrm{H}}[M]\otimes \Omega_{\mathrm{ph}}
$.
Hence, by the identification in Section \ref{StLI}, we obtain the desired result. $\Box$

\subsubsection{Properties of $\tilde{H}_{\mathrm{HH}, |\Lambda|} [M]$}\label{PropHHH}

We work in the Schr\"{o}dinger representation introduced in Section \ref{SchRP}.
Let 
\begin{align}
L=-i \sqrt{2} \omega^{-3/2}\sum_{x, y\in \Lambda} g_{xy} (n_{x\uparrow}-n_{x\downarrow}) p_y.
\end{align}
$L$ is essentially  anti-self-adjoint.\footnote{Namely, $iL$ is essentially self-adjoint.} 
We denote its closure by the same symbol.
A unitary operator  $e^L$ is called the {\it Lang-Firsov transformation} \cite{LF}.
We remark the following properties:
\begin{align}
e^L c_{x\uparrow} e^{-L} &=\exp\Bigg\{+
i\sqrt{2} \omega^{-3/2} \sum_{y\in \Lambda} g_{xy} p_y
\Bigg\} c_{x\uparrow},  \label{LFP1}\\
e^L c_{x\downarrow} e^{-L} &=\exp\Bigg\{-
i\sqrt{2} \omega^{-3/2} \sum_{y\in \Lambda} g_{xy} p_y
\Bigg\} c_{x\downarrow},\\
e^L b_x e^{-L} &= b_x-\omega^{-1} \sum_{y\in \Lambda}g_{xy} (n_{y\uparrow}-n_{y\downarrow}).\label{LFP2}
\end{align}
The following facts will be  used:
\begin{align}
e^{-i \frac{\pi}{2}N_{\mathrm{ph}}} q_x e^{i \frac{\pi}{2}N_{\mathrm{ph}}}
=\omega^{-1}p_x,\ \ \  e^{-i \frac{\pi}{2}N_{\mathrm{ph}}} p_x e^{i \frac{\pi}{2}N_{\mathrm{ph}}}
=\omega q_x, \label{Fourier}
\end{align}
where  $N_{\mathrm{ph}}=\sum_{x\in \Lambda} b_x^*b_x=\frac{1}{2}\sum_{x\in \Lambda} (p_x^2+ \omega^2q_x^2-1)$.
\begin{lemm}\label{HHEx}
Set $\mathscr{U}= e^{-i\frac{\pi}{2} N_{\mathrm{ph}}} e^L$. We have
\begin{align}
\mathscr{U}\tilde{H}_{\mathrm{HH}, |\Lambda|} \mathscr{U}^*
=T_{+g, \uparrow}+T_{-g, \downarrow}+\tilde{U}_{\mathrm{eff}} +E_{\mathrm{ph}} +\mathrm{Const.}, \label{FourH}
\end{align}
where
\begin{align}
T_{\pm g, \sigma} &= \sum_{x, y\in \Lambda} t_{xy} \exp\big\{ \pm i \Phi_{xy}\big\} c_{x\sigma}^*c_{y\sigma},\\
\Phi_{xy} &= \sqrt{2} \omega^{-1/2} \sum_{z\in \Lambda} (g_{xz} -g_{yz}) q_z,\\
E_{\mathrm{ph}} &= \frac{1}{2} \sum_{x\in \Lambda} (p_x^2+\omega^2q_x^2),\\
\tilde{U}_{\mathrm{eff}}&= \frac{1}{2} \sum_{x, y\in \Lambda} U_{\mathrm{eff}, xy} (n_{x\uparrow}-n_{x\downarrow})(n_{y\uparrow}-n_{y\downarrow}).
\end{align}
\end{lemm}
{\it Proof.} By the hole-particle transformation, we have
\begin{align}
\tilde{H}_{\mathrm{HH}, |\Lambda|} =\tilde{H}_{\mathrm{H}, |\Lambda|}+\sum_{x, y\in \Lambda} g_{xy} (n_{x\uparrow}-n_{x\downarrow}) \sqrt{2\omega} q_y +E_{\mathrm{ph}}+\frac{ |\Lambda| \omega}{2}.
\end{align}
By  applying  (\ref{LFP1})--(\ref{Fourier}), we obtain the desired expression in Lemma \ref{HHEx}. $\Box$
\medskip\\

In what follows, we ignore the constant term in (\ref{FourH}).
By Lemma \ref{HHEx}  and taking the identification (\ref{ElIdn2}) into consideration, we arrive at the following:
\begin{coro}\label{coroidenHH}
Let $\HH=\mathscr{U} \tilde{H}_{\mathrm{HH}, |\Lambda|}[M] \mathscr{U}^*$. We  have
$
\HH=\mathbb{T}_{\mathrm{HH}}-\mathbb{U}_{\mathrm{eff}} +E_{\mathrm{ph}},
$
where 
\begin{align}
\mathbb{T}_{\mathrm{HH}}&= 
\int^{\oplus}_{\mathcal{Q}} \mathcal{L}\big(\mathbf{T}_{+g}({\Bs q}) \big)d{\Bs q}
+\int^{\oplus}_{\mathcal{Q}} \mathcal{R}\big(\mathbf{T}_{+g}({\Bs q}) \big)d{\Bs q}, \label{THHQ}\\
\mathbf{T}_{+g}({\Bs q})&=\sum_{x, y\in \Lambda} t_{xy} \mathsf{c}^*_x \mathsf{ c}_y\exp\big\{
+i\Phi_{xy} ({\Bs q})
\big\}+\frac{1}{2} \sum_{x, y\in \Lambda} U_{\mathrm{eff}, xy} \mathsf{n}_x \mathsf{n}_y,\\
\Phi_{xy} ({\Bs q})&=\sqrt{2} \omega^{-1/2} \sum_{z\in \Lambda} (g_{xz} -g_{yz}) q_z,\ \ \mbox{ ${\Bs q}=\{q_x\}_{x\in \Lambda} \in \mathcal{Q}$}
\end{align}
and 
\begin{align}
\mathbb{U}_{\mathrm{eff}} =\sum_{x, y\in \Lambda} U_{\mathrm{eff}, xy} \mathcal{L}(\mathsf{n}_x) \mathcal{R}(\mathsf{n}_y). \label{UeEx}
\end{align}
\end{coro}
{\it Proof.} By the identifications (\ref{cTensorRP}) and (\ref{ElIdn2}), we have 
\begin{align}
c_{x\uparrow}^* c_{y\uparrow}
&=\mathsf{c}_x^*\mathsf{c}_{y} \otimes 1=\mathcal{L}(
\mathsf{c}_x^*\mathsf{c}_{y}),\label{LC}\\
c_{x\downarrow}^* c_{y\downarrow}
&=1\otimes \mathsf{c}_x^*\mathsf{c}_{y} =\mathcal{R}\big( \vartheta (
\mathsf{c}_x^*\mathsf{c}_{y})^*\vartheta \big)
=\mathcal{R}(
\mathsf{c}_y^*\mathsf{c}_{x}).\label{RC}
\end{align}
Here, we used the fact $\vartheta \mathsf{c}_x\vartheta=\mathsf{c}_x$.

By (\ref{LC}), we have 
\begin{align}
T_{+g, \uparrow}&= \sum_{x, y\in \Lambda} t_{xy} \int^{\oplus}_{\mathcal{Q}} \exp\big\{
+i \Phi_{xy}({\Bs q})
\big\}
\mathcal{L}(\mathsf{c}_{x}^*\mathsf{c}_y) d{\Bs q}\no
&=\int^{\oplus}_{\mathcal{Q}}
\mathcal{L}\Bigg(
  \sum_{x, y\in \Lambda} t_{xy} \exp\big\{
+i \Phi_{xy}({\Bs q})
\big\}
\mathsf{c}_{x}^*\mathsf{c}_y \Bigg) \, d{\Bs q},
\end{align}
where we used the linearity of $\mathcal{L}(\cdot):\ \mathcal{L}(aX+bY)=a \mathcal{L}(X)+b\mathcal{L}(Y)$.

On the other hand, we have, by (\ref{RC}),
\begin{align}
T_{-g, \downarrow}&=\sum_{x, y\in \Lambda} t_{xy} \int^{\oplus}_{\mathcal{Q}} \exp\big\{
-i \Phi_{xy}({\Bs q})
\big\}
\mathcal{R}(\mathsf{c}_{y}^*\mathsf{c}_x) d{\Bs q}\no
&=\int^{\oplus}_{\mathcal{Q}}
\mathcal{R}\Bigg(
  \sum_{x, y\in \Lambda} t_{xy} \exp\big\{
-i \Phi_{xy}({\Bs q})
\big\}
\mathsf{c}_{y}^*\mathsf{c}_x \Bigg) \, d{\Bs q}\\
&=\int^{\oplus}_{\mathcal{Q}}
\mathcal{R}\Bigg(
  \sum_{x, y\in \Lambda} t_{yx} \exp\big\{
+i \Phi_{yx}({\Bs q})
\big\}
\mathsf{c}_{y}^*\mathsf{c}_x \Bigg) \, d{\Bs q}\no
&=\int^{\oplus}_{\mathcal{Q}}
\mathcal{R}\Bigg(
  \sum_{x, y\in \Lambda} t_{xy} \exp\big\{
+i \Phi_{xy}({\Bs q})
\big\}
\mathsf{c}_{x}^*\mathsf{c}_y \Bigg) \, d{\Bs q}.\label{AAA}
\end{align}
Here, we used the following properties: $t_{xy}=t_{yx}$ and $\Phi_{yx}({\Bs q})=-\Phi_{xy}({\Bs q})$.
(Note that the last equality in (\ref{AAA}) comes from the relabeling.)
Using these observations, we can prove (\ref{THHQ}).
$\Box$

\begin{lemm}\label{HHMatPP}
We have the following:
\begin{itemize}
\item[{\rm (i)}]
$
e^{-\beta \mathbb{T}_{\mathrm{HH}}} \unrhd 0
$ w.r.t. $\Cone_{\mathrm{HH}}[M]$ for all $\beta \ge 0$.

\item[{\rm (ii)}] $
\mathbb{U}_{\mathrm{eff}} \unrhd 0
$ w.r.t. $\Cone_{\mathrm{HH}}[M]$.

\item[{\rm (iii)}] $
e^{-\beta E_{\mathrm{ph}}} \unrhd 0
$ w.r.t. $\Cone_{\mathrm{HH}}[M]$ for all $\beta \ge 0$.

\end{itemize}
\end{lemm}
{\it Proof.} 
(i) By (i) of Lemma \ref{TensorPP},  we see that 
\begin{align}
e^{-\beta \mathbb{T}_{\mathrm{HH}}}
=\int^{\oplus}_{\mathcal{Q}} \mathcal{L}\Big(e^{-\beta \mathbb{T}_{\mathrm{HH}}({\Bs q})}\Big)
\mathcal{R}\Big(e^{-\beta \mathbb{T}_{\mathrm{HH}}({\Bs q})}\Big) d{\Bs q}
\unrhd 0\ \ \ \mbox{w.r.t. $\Cone_{\mathrm{HH}}[M]$}.
\end{align}

Since $\mathbb{U}_{\mathrm{eff}} \unrhd 0$ w.r.t. $\Cone_{\mathrm{H}}[M]$ by Lemma \ref{Coulomb},
we obtain (ii) by (ii) of Lemma \ref{TensorPP}.
Finally, because 
$e^{-\beta E_{\mathrm{rad}} }\unrhd 0$ w.r.t. $L^2(\mathcal{Q})_+$ for all $\beta \ge 0$ by Example \ref{HarmonicOS},
 (iii) follows from  Proposition \ref{BasicDP22}. $\Box$

\begin{Prop} \label{HHPP}
 We have
 $
 \big( \mathbb{H}_{\mathrm{HH}}[M]+s\big)^{-1} \unrhd 0
 $ w.r.t. $\Cone_{\mathrm{HH}}[M]$ for all $s \ge -E\big(
 \mathbb{H}_{\mathrm{HH}}[M]
 \big)$.   
\end{Prop}
{\it Proof.} By Lemma \ref{HHMatPP} and Theorem \ref{StandPP}  with $A=\mathbb{T}_{\mathrm{HH}}+E_{\mathrm{rad}}$ and $B=\mathbb{U}_{\mathrm{eff}}$, we obtain 
that $e^{-\beta 
\mathbb{H}_{\mathrm{HH}}[M]
} \unrhd 0$ w.r.t.  $\Cone_{\mathrm{HH}}[M]$ for all $\beta\ge 0$.
Hence, by using Proposition \ref{PPEquiv}, we conclude the desired result in Proposition \ref{HHPP}.
 $\Box$
\medskip

Remark that, though subjects are different, there are some similarities between the  idea here and that of \cite{NM}.

In \cite{Miyao5}, the following much  stronger property was proven:

\begin{Thm}\label{HHPI}
We have
 $
 \mathscr{U} \big(
 \tilde{H}_{\mathrm{HH}, |\Lambda|}[M]
 +s\big)^{-1}\mathscr{U}^*
 =
 \big(\mathbb{H}_{\mathrm{HH}}[M]+s\big)^{-1} \rhd 0
 $ w.r.t. $\Cone_{\mathrm{HH}}[M]$ for all $s>-E\big(
 \tilde{H}_{\mathrm{HH}, |\Lambda|}[M]
 \big)$.
\end{Thm}

\subsection{Completion of proof of Theorem \ref{GeneHH}}
We will  check all conditions in Definition \ref{FirstSt} with 
\begin{align*}
&(H, \ H_*;\  \mathfrak{H},\  \mathfrak{H}_*;\  P;\  \mathfrak{P}, \ \mathfrak{P}_*; O; U)\no
=&
\Big(\tilde{H}_{\mathrm{HH}, |\Lambda|}[M], \, \tilde{H}_{\mathrm{H}, |\Lambda|}[M];\,  W\mathfrak{E}_{n=|\Lambda|}[M]\otimes L^2(\mathcal{Q}),\  
W\mathfrak{E}_{n=|\Lambda|}[M];\no 
& \ \ \ \ P_{\mathrm{ph}};\,
\mathfrak{P}_{\mathrm{HH}}[M],\,  \mathfrak{P}_{\mathrm{H}}[M];\ 
\tilde{S}_{\mathrm{tot}, |\Lambda|}^2[M]; \mathscr{U}\Big).
\end{align*}

By Proposition \ref{HHProjP}, (i) of Definition \ref{FirstSt} is satisfied.
By Proposition \ref{HHProjP2},  (ii) of Definition \ref{FirstSt} is fulfilled.
 (iii) and (iv) of  Definition \ref{FirstSt} follow from Theorems \ref{HHPI} and \ref{HPP}, respectively.
 By Proposition \ref{UniEQ} with $V=W$, one obtains Theorem \ref{GeneHH}
  $\Box$

\section{Proof of Theorem \ref{SpinHrad}}\label{PfLi4}
\setcounter{equation}{0}
\subsection{Preliminaries}

A path integral approach to  the quantized   radiation fields was initiated by Feynman \cite{Feynman}.
There is still a lack of  mathematical justification of it, 
but  its euclidean version (i.e.,  the imaginary time path integral of the  radiation fields)
has been studied  rigorously, see, e.g.,  \cite{BMM, JHB, Miyao3,Spohn}.
In these works, field operators are expressed  as linear operators acting on a certain $L^2$-space. 
In this subsection, we give such an  expression  of the radiation fields.
Our  construction  is based on Feynman\rq{}s  original work \cite{Feynman}, and has the benefit of usability.

\subsubsection{Second quantization }

Let $\mathfrak{X}$ be a complex Hilbert space. The bosonic Fock space over $\mathfrak{X}$
is given by $\Fock(\mathfrak{X})=\bigoplus_{n=0}^{\infty} \otimes^n_{\mathrm{s}} \mathfrak{X}$.
Let $A$ be a positive self-adjoint operator on $\mathfrak{X}$. The second quantization of $A$
 is defined by 
 \begin{align}
 \dG(A)=0\oplus \bigoplus_{n=1}^{\infty} \Bigg[\sum_{j=1}^n 1\otimes \cdots \otimes \underbrace{ A }_{j^{\mathrm{th}}}\otimes \cdots \otimes 1\Bigg].
 \end{align}
 We denote  by $a(f)$ the annihilation operator in $\Fock(\mathfrak{X})$ with test vector $f$ \cite[Section X. 7]{ReSi2}. By definition, $a(f)$ is densely defined, closed, and antilinear in $f$.

We first recall the following factorization  property of the bosonic Fock space:
\begin{align}
\Fock(\mathfrak{X}_1\oplus \mathfrak{X}_2 ) =\Fock(\mathfrak{X}_1) \otimes \Fock(\mathfrak{X}_2). \label{Fact}
\end{align}
Corresponding to (\ref{Fact}), we have
\begin{align}
\dG(A\oplus B)&=\dG(A) \otimes 1+1\otimes \dG(B), \label{Fact2}
\\ 
a(f\oplus g)&=a(f)\otimes 1+1\otimes a(g).\label{Fact3}
\end{align}

Let $E_{\mathrm{rad}}$ be the closure of $
\sum_{\lambda=1,2}\sum_{k\in V^*} \omega(k)a(k, \lambda)^*a(k, \lambda)
$.
We denote   by $[\omega]$ a multiplication operator $\omega\oplus \omega$
 on $\ell^2(V^*) \oplus \ell^2(V^*)$.
Remark that $E_{\mathrm{rad}}$ is positive,  self-adjoint,  and 
\begin{align}
E_{\mathrm{rad}}=\dG([\omega]),
\end{align}
where $\dG(A)$ is the second quantization of $A$ in $\Fock_{\mathrm{rad}}$.

\subsubsection{The ultraviolet cutoff decomposition}

Let  $V_{\le \kappa}^* =\{k\in V^*\, |\, |k| \le \kappa\}$, where $\kappa$ is the ultraviolet cutoff 
introduced  in Section \ref{RadHami}.
Since $\ell^2(V^*)=\ell^2(V_{\le \kappa}^*)\oplus \ell^2((V_{\le \kappa}^*)^c)$ with $
(V_{\le \kappa}^*)^c
$, the complement of $V_{\le \kappa}^*$, we have the identification
$
\ell^2(V^*)\oplus \ell^2(V^*)=
\Big(\ell^2(V_{\le \kappa}^*) \oplus \ell^2(V_{\le \kappa}^*)\Big)
\oplus
\Big(
\ell^2((V_{\le \kappa}^*)^c) \oplus \ell^2((V_{\le \kappa}^*)^c)
\Big)
$, which implies
\begin{align}
\Fock =\Fock_{ \le \kappa} \otimes \Fock_{>\kappa},
\end{align}
 by (\ref{Fact}), where $
 \Fock_{ \le \kappa}
 $ and $\Fock_{>\kappa}$ are  the bosonic Fock spaces  over $\ell^2(V_{\le \kappa}^*) \oplus \ell^2(V_{\le \kappa}^*)$ and  $\ell^2((V_{\le \kappa}^*)^c) \oplus \ell^2((V_{\le \kappa}^*)^c)$, respectively.

Let $\dG_{\le \kappa}(A)$ and $\dG_{>\kappa}(B)$ be the second quantized operators in $\Fock_{\le \kappa}$ and $\Fock_{>\kappa}$, respectively. 
 We have
\begin{align}
E_{\mathrm{rad}}=\dG_{\le \kappa}([\omega]) \otimes 1+1\otimes \dG_{>\kappa}([\omega])
\end{align}
by (\ref{Fact2}).

We introduce two closed subspaces of $\ell^2(V^*_{\le \kappa})$ as follows:
\begin{align}
\ell^2_{
{\boldsymbol \vepsilon}_1
}(V^*_{\le \kappa})=
\overline{\bigcup_{j=1, 2, 3} \ran(\vepsilon_{1j})
}, \ \ \
\ell^2_{
{\boldsymbol \vepsilon}_2
}(V^*_{\le \kappa})= \overline{
\bigcup_{j=1,2,3}  \ran(\vepsilon_{2j})
},
\end{align} 
where ${\boldsymbol \vepsilon}_{\lambda}=(\vepsilon_{\lambda 1}, \vepsilon_{\lambda2}, \vepsilon_{\lambda3})$
are the polarization vectors given by (\ref{Pola}).
Here, we identify  the  functions $\vepsilon_{1j}$ and $\vepsilon_{2j}$ with  the corresponding 
multiplication operators acting on $\ell^2(V_{\le \kappa}^*)$.

\begin{lemm}\label{VDec}
Let
$
 \mathcal{H}_{\vepsilon}=\ell^2_{
{\boldsymbol \vepsilon}_1
}(V^*_{\le \kappa})\oplus
\ell^2_{
{\boldsymbol \vepsilon}_2
}(V^*_{\le \kappa})
$. 
Let $\mathcal{I}=\{k=(k_1, k_2, k_3)\in V^*_{\le \kappa}\, |\, k_1=k_2=0\}$.
We have the following identifications:
\begin{align}
\mathcal{H}_{\vepsilon} &\cong \ell^2(V^*_{ \kappa}) \oplus \ell^2(V^*_{ \kappa}),
\ \ V_{\kappa}^*:=V_{\le \kappa}^*\backslash \mathcal{I},
\\
\mathcal{H}_{\vepsilon}^{\perp}&\cong\ell^2(\mathcal{I}) \oplus \ell^2(\mathcal{I}). \label{IdenC}
\end{align}
\end{lemm}
{\it Proof.} 
Since 
\begin{align}
\Bigg(\overline{\bigcup_{j=1, 2, 3} \ran(\vepsilon_{1j})
}\Bigg)^{\perp}
=\bigcap_{j=1,2,3}\Big (\R(\vepsilon_{1j})\Big)^{\perp}
=\bigcap_{j=1,2,3} \ker(\vepsilon_{1j}),
\end{align}
we have,  by (\ref{Pola})
\begin{align}
\mathcal{H}_{\vepsilon}^{\perp}&=
\{
f\in \ell^2(V^*_{\le \kappa}) \oplus \ell^2(V^*_{\le \kappa})\, |\, 
f(k)=0\ \ \mbox{for $\forall k\in \mathcal{I}^c$}
\} \cong \ell^2(\mathcal{I}) \oplus \ell^2(\mathcal{I}).
\end{align}
Thus, we obtain
\begin{align}
\mathcal{H}_{\vepsilon} & =\{
f\in \ell^2(V^*_{\le \kappa}) \oplus \ell^2(V^*_{\le \kappa})\, |\, 
f(k)=0 
\ \ \mbox{for $\forall k\in \mathcal{I}$}
\}\cong \ell^2(V_{\kappa}^*) \oplus \ell^2(V_{\kappa}^*).
\end{align}
This completes the proof. $\Box$
\medskip\\

By (\ref{Fact}) and Lemma \ref{VDec}, we obtain
\begin{align}
\Fock_{\le \kappa}=\Fock(\mathcal{H}_{\bep})\otimes
 \Fock(\mathcal{H}_{\bep}^{\perp}).\label{DirctD}
\end{align}

\begin{lemm}
Corresponding to the decomposition (\ref{DirctD}), we have
\begin{align}
\dG_{\le \kappa}([\omega])=\dG_{\bep}([\omega])\otimes 1+1\otimes  \dG_{\perp}([\omega])
,  \label{Dec1}
\end{align}
where
$
\dG_{\bep}([\omega])
$ is the second quantization of $[\omega]\restriction \mathcal{H}_{\bep}$ in $\Fock( \mathcal{H}_{\bep})$
and  $\dG_{\perp}([\omega])$ is the second quantization of $[\omega] \restriction \mathcal{H}_{\bep}^{\perp}$ in  $\Fock(\mathcal{H}_{\bep}^{\perp}).$

\end{lemm} 
{\it Proof.} 
Remark that  $[\omega]=\big([\omega] \restriction \mathcal{H}_{\bep}\big) \oplus \big([\omega] \restriction \mathcal{H}_{\bep}^{\perp}\big)$.
Using (\ref{Fact2}), we obtain
\begin{align}
\dG_{\le\kappa}([\omega])=\dG_{\le \kappa}\Big(\big([\omega]\restriction \mathcal{H}_{\vep} \big)
\oplus \big([\omega] \restriction \mathcal{H}_{\bep}^{\perp} \big)\Big)
= \mbox{RHS of (\ref{Dec1}).}
\end{align}
This completes the proof. $\Box$
\medskip\\

Let 
\begin{align}
\ell^2_{{\mathrm{even}}}(V^*_{\kappa}) &=\{f\in \ell^2(V^*_{\kappa})\, |\, f(-k)=f(k)\},\\
\ell^2_{{\mathrm{odd}}}(V^*_{\kappa})&=\{f\in \ell^2(V^*_{\kappa})\, |\, f(-k)=-f(k)\}.
\end{align}
We set 
\begin{align}
\mathfrak{h}_1=\mathfrak{h}_3=\ell^2_{\mathrm{even}}(V^*_{\kappa}),\ \ \ \
\mathfrak{h}_2=\mathfrak{h}_4=\ell^2_{\mathrm{odd}}(V^*_{\kappa}).
\end{align}
Since $\ell^2_{\bep_1}(V^*_{\kappa})=\mathfrak{h}_1\oplus \mathfrak{h}_2$ and 
 $\ell^2_{\bep_2}(V^*_{ \kappa})=\mathfrak{h}_3\oplus \mathfrak{h}_4$,
we obtain
\begin{align}
 \mathcal{H}_{\bep}=\mathfrak{h}_1\oplus \mathfrak{h}_2\oplus
 \mathfrak{h}_3\oplus \mathfrak{h}_4. \label{Decp}
\end{align}
Using the decomposition (\ref{Decp}), the annihilation operator  on $\Fock(\mathcal{H}_{\bep})$
can be expressed as $a(f_1, f_2, f_3, f_4)$. 
In what follows, we use the following notations:
\begin{align}
a(f, 0, 0, 0)&=a_1(f), \ a(0, f, 0, 0)=a_2(f),\\ a(0, 0, f, 0)&=a_3(f),\ a(0,0,0, f)=a_4(f).
\end{align}
 Thus,
$
a(f_1, f_2, f_3, f_4)=\sum_{r=1}^4 a_r(f_r)
$.

 By (\ref{Fact}) and (\ref{Decp}), 
we have the following:
\begin{align}
\Fock(\mathcal{H}_{\bep})=\bigotimes_{r=1}^4\Fock(\mathfrak{h}_r). \label{4Fact}
\end{align} 
We will often use the following identifications corresponding to (\ref{4Fact}):
\begin{align}
a_1(f)=a_1(f)\otimes 1\otimes 1\otimes 1,\ \ a_2(f)=1\otimes a_2(f)\otimes 1\otimes 1,\ \  \mathrm{etc.}
\end{align}
For notational convenience, we   express $a_r(f)$ as 
\begin{align}
a_r(f)=\sum_{k\in V_{ \kappa}^*} f(k)^* a_r(k),\ \ f\in \mathfrak{h}_r.
\end{align}
 
 The following lemma will be useful:
\begin{lemm}\label{PQA}
\begin{itemize}
\item[{\rm (i)}] Let $\dG_r(\omega)$ be the second quantization of $\omega$ in $\Fock(\mathfrak{h}_r)$.
We have
\begin{align}
\dG_{\bep}([\omega])=\dG_1(\omega)\otimes 1 \otimes 1 \otimes 1 +
1 \otimes\dG_2(\omega)\otimes  1 \otimes 1+\no
+ 1 \otimes 1 \otimes \dG_3(\omega)\otimes 1 
+
1\otimes 1 \otimes 1 \otimes \dG_4(\omega).
\end{align}
\item[{\rm (ii)}] We have
\begin{align}
A(x)=&A_1(x)\otimes 1 \otimes 1 \otimes 1+1 \otimes
 A_2(x)\otimes 1 \otimes 1+ \no
&+1 \otimes 1 \otimes  A_3(x)\otimes 1
+1 \otimes 1 \otimes 1 \otimes A_4(x),
\end{align}  
where
\begin{align}
A_1(x)&=\sum_{k\in V_{\kappa}^*}\bep_1(k) \chi_{\kappa}(k)  \cos (k\cdot x)\phi_1(k),\ 
A_2(x)=\sum_{k\in V_{\kappa}^*}\bep_1(k) \chi_{\kappa}(k)  \sin (k\cdot x)\pi_2(k),\no
A_3(x)&=\sum_{k\in V_{\kappa}^*}\bep_2(k) \chi_{\kappa}(k)  \cos (k\cdot x)\phi_3(k),\ 
A_4(x)=\sum_{k\in V_{\kappa}^*}\bep_2(k) \chi_{\kappa}(k)  \sin (k\cdot x)\pi_4(k),\no
\pi_r(k)&=\frac{i}{\sqrt{2\omega(k)}}(a_r(k)-a_r(k)^*),\ \ 
 \phi_r(k)=\frac{1}{\sqrt{2\omega(k)}}(a_r(k)+a_r(k)^*). \label{PIPHI}
\end{align}  
\end{itemize}
\end{lemm}

\subsubsection{The Feynman-Schr\"odinger representation}\label{FSRP}
We set $N=\# V_{ \kappa}^*$, the cardinality of $V_{\kappa}^*$. Let us consider a Hilbert space 
$L^2(\BbbR^N)$.
For each $\vphi\in L^2(\BbbR^N)$, we define a multiplication operator $q(k)$ by 
\begin{align}
\big(
q(k) \vphi
\big)({\boldsymbol q})=q(k) \vphi({\boldsymbol q}),\ \ \ {\boldsymbol q}=\{q(k)\}_{k\in V_{ \kappa}^*} \in \BbbR^N.
\end{align}
We also define $p(k)$ by $p(k)=-i \partial /\partial q(k)$.

Now, let us switch to a larger Hilbert space
$
L^2(\BbbR^N) \otimes L^2(\BbbR^N)
= L^2(\BbbR^{2N})
$. We set 
\begin{align}
q_1(k)=q(k) \otimes 1, \ \ q_2(k)=1\otimes q(k).
\end{align}
Similarly, we define $p_1(k)$ and  $p_2(k)$.
It is easy to see that $[q_r(k), p_{r\rq{}}(k\rq{})]
=i \delta_{rr\rq{}} \delta_{kk\rq{}}$.
The annihilation operator  is defined by $
b_r(k)=\sqrt{\frac{\omega(k)}{2}} q_r(k)+\frac{1}{\sqrt{2\omega(k)}} ip_r(k)
$. Remark that $
[b_r(k), b_{r\rq{}}(k\rq{})^*]=\delta_{rr\rq{}} \delta_{kk\rq{}}
$ holds.
Let 
\begin{align}
\Phi_0(\{{\boldsymbol q}_r\}_{r=1}^2)
=\pi^{-N/2} \exp\Bigg\{
-\sum_{r=1}^2 \sum_{k\in V_{\kappa}^*} q_r(k)^2
\Bigg\},\ \ {\boldsymbol q}_r=\{q_r(k)\}_{k\in V_{\kappa}^*}\in \BbbR^N.
\end{align}
Then, we have $b_r(k) \Phi_0=0$.

We define linear operators $b_1, b_2, b_3$ and $b_4$ by 
\begin{align}
b_{r}(g)=\sum_{k\in V_{\kappa}^*} g(k)^* \tilde{b}_r(k), \ \ r=1, 2, 3, 4
\end{align}
  for each $g\in \mathfrak{h}_r$, where $\tilde{b}_1(k)=\tilde{b}_2(k)=b_1(k)$ and 
  $\tilde{b}_2(k)=\tilde{b}_4(k)=b_2(k)$.
It is easily verified that $[b_r(g), b_{r\rq{}}(g\rq{})^*]=\delta_{rr\rq{}} \la g|g\rq{}\ra$.

The following lemma is easily checked:
\begin{lemm}
We have
\begin{align}
L^2(\BbbR^{2N})=
\overline{\mathrm{Lin}}\Big\{
b_{r_1}(g_1)^* \cdots b_{r_n}(g_n)^* \Phi_0, \Phi_0\, \Big|\,  g_j\in \mathfrak{h}_{r_j},\, r_j=1, \dots, 4,\, j=1, \dots,  n\in \BbbN
\Big\},
\end{align}
where $\overline{\mathrm{Lin}}(S)$ represents the closure of $\mathrm{Lin}(S)$.
\end{lemm}

Next, we define a unitary operator $U$ from $\bigotimes_{r=1}^4\Fock(\mathfrak{h}_r)$
onto $L^2(\BbbR^{2N})$ by 
\begin{align}
U \Omega_{\mathrm{rad}}&=\Phi_0,\\
Ua_{r_1}(g_1)^*\cdots a_{r_n}(g_n)^* \Omega_{\mathrm{rad}}&=b_{r_1}(g_1)^*\cdots b_{r_n}(g_n)^* \Phi_0.
\end{align}

Let $N_r=\dG_r(1), r=1,\dots, 4$ .
Let $\mathbf{U}$ be  a unitary operator  given by 
\begin{align}
\mathbf{U}=U\exp\{-i \pi (N_2+N_4)/2\}.\label{DEFU}
\end{align}

For each {\it real-valued} $g\in \mathfrak{h}_r\, (r=1,2,3,4)$, we set
\begin{align}
\phi_r(g)=\sum_{k\in V_{\kappa}^*} g(k) \phi_r(k), \ \ 
\pi_r(g)=\sum_{k\in V_{\kappa}^*} g(k) \pi_r(k),
\end{align}
where $\phi_r(k)$ and $\pi_r(k)$ are defined by (\ref{PIPHI}).
Similarly, we set $q_{\mu}(g)=\sum_{k\in V_{ \kappa}^*} g(k) q_{\mu}(k)$ for each $\mu=1,2$.
By the definition of $\mathbf{U}$, we have the following:

\begin{itemize}
\item $\mathbf{U} \phi_1(g)\mathbf{U}^{-1} =q_1(g)$ for each $g\in \mathfrak{h}_1$, real-valued; 
\item $\mathbf{U} \pi_2(g) \mathbf{U}^{-1} =q_1(g)$ for each $g\in \mathfrak{h}_2$, real-valued;
\item $\mathbf{U} \phi_3(g)\mathbf{U}^{-1} =q_2(g)$ for each $g\in \mathfrak{h}_3$, real-valued;
\item $\mathbf{U} \pi_4(g) \mathbf{U}^{-1} =q_2(g)$ for each $g\in \mathfrak{h}_4$, real-valued.
\end{itemize}
The point here is that $\phi_1(g),\, \pi_2(g), \, \phi_3(g)$ and $\pi_4(g)$ can be identified with  the {\it 
multiplication operators} $q_1(g)$ and $q_2(g)$.

\begin{Prop}\label{SFrep}
We have the following:
\begin{itemize}
\item[{\rm (i)}]
\begin{align}
\mathbf{U}\dG_{\bep}([\omega]) \mathbf{U}^{-1}
=\sum_{\mu=1,2} \sum_{k\in V_{ \kappa}^*} \frac{1}{2}\Big\{
p_{\mu}(k)^2+\omega(k)^2q_{\mu}(k)^2
\Big\}-\sum_{k\in V_{\kappa}^*} \omega(k).
\end{align}

\item[{\rm (ii)}] 
\begin{align}
\mathbf{U} A(x)\mathbf{U}^{-1}=&\sum_{k\in V^*_{\kappa}}
\chi_{\kappa}(k)\Big\{
\bep_1(k)
\Big(\cos(k\cdot x)+\sin(k\cdot x)\Big) q_{1}(k)\no
&+
\bep_2(k)
\Big(\cos(k\cdot x)+\sin(k\cdot x)\Big) q_2(k)
\Big\}. \label{VecP}
\end{align}
\end{itemize} 
\end{Prop}
{\it Proof.} (i) is easy to check.
(ii) immediately follows from (ii) of Lemma \ref{PQA} and properties just above Proposition \ref{SFrep}. $\Box$

\subsection{Natural self-dual cones}
By \cite[Theorem I.11]{Simon}, we can identify $\Fock_{>\kappa}$ with $L^2(\mathcal{M}, d\mu)$,
where $d\mu$ is some Gaussian probability measure.
Also, by (\ref{IdenC}), we can identify $\Fock(\mathcal{H}_{\bep}^{\perp})$ as 
$
\Fock({\mathcal{H}_{\bep}^{\perp}})=\Fock(\BbbC^{2|\mathcal{I}|})=L^2(\BbbR^{2|\mathcal{I}|}; d{\Bs \eta})=L^2(\BbbR^{2|\mathcal{I}|})$.
Combining these with the Feynman-Schr\"{o}dinger  representation in Section \ref{FSRP}, we obtain
\begin{align}
\Fock_{\mathrm{rad}}
=\Fock_{\le \kappa} \otimes \Fock_{>\kappa}
=\Fock(\mathcal{H}_{\bep})\otimes \Fock(\mathcal{H}_{\bep}^{\perp}) \otimes \Fock_{>\kappa}
=L^2(\BbbR^{2N}) \otimes L^2(\BbbR^{2|\mathcal{I}|}) \otimes L^2(\mathcal{M}, d\mu). \label{FockTenIdn}
\end{align}
Thus, denoting $\mathscr{R}=\BbbR^{2N}\times \BbbR^{2|\mathcal{I}|} \times \mathcal{M}$ and $d\nu= d{\boldsymbol q} d{\Bs \eta} d\mu $, we have $\Fock_{\mathrm{rad}}=L^2(\mathscr{R}, d\nu)$.
Moreover, 
$
\mathbf{U} E_{\mathrm{rad}}\mathbf{U}^{-1}=
L_0+L_1,
$
where 
\begin{align}
L_0
&=\sum_{\mu=1,2} \sum_{k\in V_{ \kappa}^*} \frac{1}{2}\Big\{
p_{\mu}(k)^2+\omega(k)^2q_{\mu}(k)^2
\Big\}-\sum_{k\in V_{\kappa}^*} \omega(k),\\
L_1&=\sum_{\mu=1,2} \sum_{k\in \mathcal{I}} \frac{1}{2}\bigg\{
-\frac{\partial^2}{\partial \eta_{\mu}^2(k)}+\omega(k)^2\eta_{\mu}(k)^2
\bigg\}-\sum_{k\in \mathcal{I}} \omega(k)
+\dG_{>\kappa}([\omega]).
\end{align}

In this representation, we have $
\mathfrak{E}_{n=|\Lambda|}[M]\otimes \Fock_{\mathrm{rad}}
=\mathfrak{E}_{n=|\Lambda|}[M]\otimes L^2(\mathscr{R}, d\nu)
$ and 
$
\mathbf{U} H_{\mathrm{rad}}\mathbf{U}^{-1}
=K_{\mathrm{rad}}+L_1,
$
where
\begin{align}
K_{\mathrm{rad}}=\sum_{x, y\in \Lambda}\sum_{\sigma=\uparrow, \downarrow} t_{xy}
\exp\bigg\{
i \int_{C_{xy}} dr\cdot \mathscr{A}(r)
\bigg\} c_{x\sigma}^*c_{y\sigma}
+\sum_{x, y\in \Lambda} \frac{U_{xy}}{2}(n_x-1)(n_y-1)+L_0.
\end{align}
 Here, $\mathscr{A}(x)=(\mathscr{A}_1(x), \mathscr{A}_2(x), \mathscr{A}_3(x))$ is
the triplet of the multiplication operator defined by the RHS of (\ref{VecP}).

Let $\Cone_{\mathrm{H}}[M]$ be the self-dual cone in $W\mathfrak{E}_{n=|\Lambda|}[M]$ defined by 
(\ref{SCH}).
The following self-dual cones are important in the remainder of this section:
\begin{itemize}
\item $
\displaystyle
\mathfrak{P}_{\mathrm{rad}}[M]=\int^{\oplus}_{\mathscr{R}}\mathfrak{P}_{\mathrm{H}}[M] d\nu$ in 
$\mathfrak{E}_{n=|\Lambda|}[M] \otimes \Fock_{\mathrm{rad}}$;
\item 
$\displaystyle
\mathfrak{P}_{\bep}[M]=\int^{\oplus}_{\mathscr{Q}} \mathfrak{P}_{\mathrm{H}}[M] d{\Bs q}
$ in $\mathfrak{E}_{n=|\Lambda|}[M] \otimes \Fock(\mathcal{H}_{\bep})$,
where $\mathscr{Q}=\BbbR^{2N}$.
\end{itemize}

\begin{Prop}\label{PropPrad}
Let $P_{\mathrm{rad}}=1\otimes |\Omega_{\mathrm{rad}}\ra\la\Omega_{\mathrm{rad}}|$.
We have the following:
\begin{itemize}
\item[{\rm (i)}] $P_{\mathrm{rad}} \unrhd 0$ w.r.t. $\Cone_{\mathrm{rad}}[M]$; 
\item[{\rm (ii)}]
Under the identification in Section \ref{StaLiIIExh}, we have $P_{\mathrm{rad}} \Cone_{\mathrm{rad}}[M]=\Cone_{\mathrm{H}}[M]$.
\end{itemize}
\end{Prop}
{\it Proof.} Proofs of (i) and (ii) are similar to those of Propositions \ref{HHProjP} and \ref{HHProjP2}. $\Box$
\medskip\\

We will study the hole-particle transformed Hamiltonians $\tilde{H}_{\mathrm{rad}, |\Lambda|}$ and 
$\tilde{K}_{\mathrm{rad}, |\Lambda|}$.

\begin{Prop}\label{EquivPP}
 If 
  $ \big( \tilde{K}_{\mathrm{rad}, |\Lambda|}[M]+s\big)^{-1} \rhd 0
 $ w.r.t. $\mathfrak{P}_{\bep}[M]$ for all $s>-
 E\big(
 \tilde{K}_{\mathrm{rad}, |\Lambda|}[M]
 \big)
 $,  then 
 $ \mathbf{U}\big( \tilde{H}_{\mathrm{rad}, |\Lambda|}[M]+s\big)^{-1}
 \mathbf{U}^{-1}\rhd 0
 $ w.r.t. $\mathfrak{P}_{\mathrm{rad}}[M]$ for all $s>
 -E\big(
 \tilde{H}_{\mathrm{rad}, |\Lambda|}[M]
 \big)$.
\end{Prop}
{\it Proof.} Using (\ref{FockTenIdn}), we rewrite
$\mathfrak{E}_{n=|\Lambda|}[M]\otimes \Fock_{\mathrm{rad}}$ as 
\begin{align}
\mathfrak{E}_{n=|\Lambda|}[M]\otimes \Fock_{\mathrm{rad}}
=\mathfrak{X} \otimes L^2(\BbbR^{2|\mathcal{I}|} \times \mathcal{M}, d{\Bs \eta}d\mu),\label{Factorize1}
\end{align}
where $\mathfrak{X}=\mathfrak{E}_{n=|\Lambda|}[M]\otimes L^2(\mathscr{Q})$.
We set \begin{align}
L_1^{(1)} &=\sum_{\mu=1,2} \sum_{k\in \mathcal{I}} \frac{1}{2}\bigg\{
-\frac{\partial^2}{\partial \eta_{\mu}^2(k)}+\omega(k)^2\eta_{\mu}(k)^2
\bigg\}-\sum_{k\in \mathcal{I}} \omega(k),
\\
L_1^{(2)} &= d\Gamma_{>\kappa}([\omega]).
\end{align}
By Example \ref{HarmonicOS} and \cite[Theorem I.16]{Simon}, we have $
e^{-\beta L_1^{(1)}} \rhd 0
$ w.r.t. $L^2(\BbbR^{2|\mathcal{I}|}, d{\Bs \eta})_+$ and $e^{-\beta L_1^{(2)}} \rhd 0$ w.r.t. $L^2(\mathcal{M}, d\mu)_+$ for all $\beta>0$.
Thus, the ground states of $L_1^{(1)}$ and $L_1^{(2)}$ are unique by Theorem \ref{PFF}. Let $\Omega^{(1)}$ (resp. $\Omega^{(2)}$)
 be the unique ground state of $L_1^{(1)}$ (resp. $L_1^{(2)}$). 
 Trivially, $\Omega_1=\Omega^{(1)} \otimes \Omega^{(2)}$ is the unique ground state of $L_1$. By Corollary \ref{TPV}, we know that $\Omega_1>0$ w.r.t. $\int_{\mathcal{M}}^{\oplus} L^2(\BbbR^{2|\mathcal{I}|}, d{\Bs \eta})_+d\mu=L^2(\BbbR^{2|\mathcal{I}|\times \mathcal{M}}, d{\Bs \eta}d\mu)_+$.

 By the assumption and Theorem \ref{PFF}, the ground state of $
 \tilde{K}_{\mathrm{rad}, |\Lambda|}[M]
 $
 is unique. Let $\psi$ be the unique ground state of $\tilde{K}_{\mathrm{rad}, |\Lambda|}[M]$.
 We also know that $\psi>0$ w.r.t. $\Cone_{{\Bs \varepsilon}}[M]$.
 Because $
 \mathbf{U}\tilde{H}_{\mathrm{rad}, |\Lambda|}[M]
 \mathbf{U}^{-1}
 =\tilde{K}_{\mathrm{rad}, |\Lambda|}[M]\otimes 1+1\otimes L_1
 $,  $\psi\otimes \Omega_1$ is the unique ground state of $
 \mathbf{U}\tilde{H}_{\mathrm{rad}, |\Lambda|}[M]
 \mathbf{U}^{-1}$.
 Because $\Cone_{\mathrm{rad}}[M]=
 \int^{\oplus}_{
 \BbbR^{2|\mathcal{I}|}
\times \mathcal{M}}
\Cone_{{\Bs \varepsilon}}[M]
d\mu d{\Bs \eta}
$,  we have
   $\psi\otimes \Omega_1>0$ w.r.t. $\Cone_{\mathrm{rad}}[M]$ by Corollary \ref{TPV}. 
 By applying Theorem \ref{PFF} again, we conclude that 
  $\mathbf{U}\big(\tilde{H}_{\mathrm{rad}, |\Lambda|}[M]+s\big)^{-1}
 \mathbf{U}^{-1}\rhd 0
 $ w.r.t. $\mathfrak{P}_{\mathrm{rad}}[M]$ for all $s>
 -E\big(
 \tilde{H}_{\mathrm{rad}, |\Lambda|}[M]
 \big)$. $\Box$
\medskip\\

We wish to prove $ \mathbf{U}\big( \tilde{H}_{\mathrm{rad}, |\Lambda|}[M]+s\big)^{-1}
 \mathbf{U}^{-1} \rhd 0
 $ w.r.t. $\mathfrak{P}_{\mathrm{rad}}[M]$ for all $s>-E\big(
 \tilde{H}_{\mathrm{rad}, |\Lambda|}[M]
 \big)$.
By Proposition \ref{EquivPP},  it suffices to show that  $ \big( \tilde{K}_{\mathrm{rad}, |\Lambda|}[M]
+s\big)^{-1}
 \rhd 0
 $ w.r.t. $\mathfrak{P}_{\bep}[M]$ for all $s>-E\big(
 \tilde{K}_{\mathrm{rad}, |\Lambda|}[M]
 \big)$.
To this end, we need some preparations.

\begin{lemm}
We have 
$
\tilde{K}_{\mathrm{rad}, |\Lambda|}[M] =S_{+, \uparrow}+S_{-, \downarrow}+\tilde{U}+L_0,
$
where
\begin{align}
S_{\pm, \sigma}&=\sum_{x, y\in \Lambda} t_{xy} \exp\Big\{\pm i\Phi_{x, y}^{\mathscr{A}}\Big\}
c_{x\sigma}^*c_{y\sigma},\label{SpmEx}\\
\Phi_{xy}^{\mathscr{A}}&=\int_{C_{xy}} dr\cdot \mathscr{A}(r),\\
\tilde{U}&= \sum_{x, y\in \Lambda}\frac{U_{xy}}{2}(n_{x\uparrow}-n_{x\downarrow})(n_{y\uparrow}-n_{y\downarrow}).
\end{align}
\end{lemm}
{\it Proof.} Observe that 
\begin{align}
&W \sum_{x, y\in \Lambda} t_{xy} \exp\big\{
+i \Phi_{xy}^{\mathscr{A}} \big\}c_{x\downarrow}^*c_{y\downarrow}
W^{-1}\no
=&\sum_{x, y\in \Lambda} t_{xy} \exp\big\{
+i \Phi_{xy}^{\mathscr{A}} \big\}
\gamma(x)\gamma(y)
c_{x\downarrow}c_{y\downarrow}^*. \label{HPT}
\end{align}
Since $\Lambda$ is bipartite in terms of  $\{t_{xy}\}$, we have $\gamma(x)\gamma(y)t_{xy}=-t_{xy}$.
Thus, because $\Phi_{xy}^{\mathscr{A}}=-\Phi_{yx}^{\mathscr{A}}$, we obtain
\begin{align}
\mbox{RHS of (\ref{HPT})}
=&\sum_{x, y\in \Lambda}t_{xy} \exp\big\{
+i \Phi_{xy}^{\mathscr{A}} \big\}
c_{y\downarrow}^*c_{x\downarrow}\no
=&\sum_{x, y\in \Lambda}t_{xy} \exp\big\{
-i \Phi_{yx}^{\mathscr{A}} \big\}
c_{y\downarrow}^*c_{x\downarrow}\no
=& S_{-, \downarrow}.
\end{align}
Similarly, we see that $
W \sum_{x, y\in \Lambda} t_{xy} \exp\big\{
+i \Phi_{xy}^{\mathscr{A}} \big\}c_{x\uparrow}^*c_{y\uparrow}
W^{-1}=S_{+, \uparrow}
$. $\Box$

\begin{coro}
Let $\mathbb{K}[M] = \tilde{K}_{\mathrm{rad}, |\Lambda|} [M]$.
We have 
$
\mathbb{K}[M]=\mathbb{S}-\mathbb{U}+L,
$
where
\begin{align}
\mathbb{S}&=\int^{\oplus}_{\mathscr{Q}} \mathcal{L}(\mathbf{S}_+({\Bs q})) d{\Bs q}
+\int^{\oplus}_{\mathscr{Q}} \mathcal{R}(\mathbf{S}_+({\Bs q}))d{\Bs q}, \label{SDirctR}\\
\mathbf{S}_+({\Bs q}) &= \sum_{x, y\in \Lambda}t_{xy} \exp\Big\{
+i \Phi_{xy}^{\mathscr{A}}({\Bs q})
\Big\} \mathsf{c}_x^*\mathsf{c}_y
+\sum_{x, y\in \Lambda} \frac{U_{xy}}{2} \mathsf{n}_x\mathsf{n}_y,\\
\Phi_{xy}^{\mathscr{A}}(\Bs q) &=\int_{C_{xy}} dr \cdot \mathscr{A}(r)[{\Bs q}],\\
\mathscr{A}(r)[{{\Bs q}}]&=\sum_{k\in V^*_{\kappa}}
\chi_{\kappa}(k)\Big\{
\bep_1(k)
\Big(\cos(k\cdot x)+\sin(k\cdot x)\Big) q_{1}(k)\no
&\ \ +
\bep_2(k)
\Big(\cos(k\cdot x)+\sin(k\cdot x)\Big) q_2(k)
\Big\},\ \ {\Bs q}=\{q_1(k),  q_2(k)\}_{k\in V_{\kappa}^*} \in \mathscr{Q},\\
\mathbb{U} &= \sum_{x, y\in \Lambda} U_{xy} \mathcal{L}(\mathsf{n}_x) \mathcal{R}(\mathsf{n}_y).
\end{align}
\end{coro}
{\it Proof.} The proof is similar to that of Corollary \ref{coroidenHH}. So we only explain a point to  which attention should be paid.
We observe that, by (\ref{SpmEx}),
\begin{align}
S_{-, \downarrow} &=\sum_{x, y\in \Lambda} t_{xy} \int^{\oplus}_{\mathscr{Q}} \exp
\big\{
-i \Phi_{xy}^{\mathscr{A}}({\Bs q})
\big\} \mathcal{R}(\mathsf{c}_y^* \mathsf{c}_x) d{\Bs q}\no
&=\sum_{x, y\in \Lambda} t_{yx} \int^{\oplus}_{\mathscr{Q}} \exp
\big\{
+i \Phi_{yx}^{\mathscr{A}}({\Bs q})
\big\} \mathcal{R}(\mathsf{c}_y^* \mathsf{c}_x) d{\Bs q}\no
&=\sum_{x, y\in \Lambda} t_{xy} \int^{\oplus}_{\mathscr{Q}} \exp
\big\{
+i \Phi_{xy}^{\mathscr{A}}({\Bs q})
\big\} \mathcal{R}(\mathsf{c}_x^* \mathsf{c}_y) d{\Bs q}, \label{SEqual}
\end{align}
where we used the fact $
\Phi_{xy}^{\mathscr{A}}({\Bs q})=-\Phi_{yx}^{\mathscr{A}}({\Bs q})
$, and the last equality in (\ref{SEqual}) comes from  relabeling the indecies.
Using this, we obtain the second term in (\ref{SDirctR}). $\Box$

\begin{lemm}\label{HRMatPP}
We have the following:
\begin{itemize}
\item[{\rm (i)}]
$
e^{-\beta \mathbb{S}} \unrhd 0
$ w.r.t. $\Cone_{\bep}[M]$ for all $\beta \ge 0$.

\item[{\rm (ii)}] $
\mathbb{U}\unrhd 0
$ w.r.t. $\Cone_{\bep}[M]$.

\item[{\rm (iii)}] $
e^{-\beta E_{\mathrm{rad}}} \unrhd 0
$ w.r.t. $\Cone_{\bep}[M]$ for all $\beta \ge 0$.

\end{itemize}
\end{lemm}
{\it Proof.} 
(i) By (i) of Lemma \ref{TensorPP} and (\ref{SDirctR}),  we see that 
\begin{align}
e^{-\beta \mathbb{S}}
=\int^{\oplus}_{\mathscr{Q}} \mathcal{L}\Big(e^{-\beta \mathbf{S}_+({\Bs q})}\Big)
\mathcal{R}\Big(e^{-\beta \mathbf{S}_+({\Bs q})}\Big) d{\Bs q}
\unrhd 0\ \ \ \mbox{w.r.t. $\Cone_{\bep}[M]$}.
\end{align}

Since $\mathbb{U} \unrhd 0$ w.r.t. $\Cone_{\mathrm{H}}[M]$ by Lemma \ref{Coulomb}, and $e^{-\beta E_{\mathrm{rad}}} \unrhd 0$ w.r.t. $L^2(\mathscr{Q})_+$ by Example \ref{HarmonicOS}, 
(ii) and (iii) immediately follow from (ii) 
of Lemma \ref{TensorPP}
and Proposition  \ref{BasicDP22}, respectively. $\Box$

\begin{Prop} \label{HRPP}
 We have
 $
 \big(\mathbb{K}[M]+s\big)^{-1}\unrhd 0
 $ w.r.t. $\Cone_{\bep}[M]$ for all $s>-E
 \big(
 \mathbb{K}[M]
 \big)
 $.   
\end{Prop}
{\it Proof.} By Proposition \ref{PPEquiv}, Theorem \ref{StandPP} and Lemma \ref{HRMatPP}, we obtain the desired result in Proposition \ref{HRPP}. $\Box$
\medskip\\

Applying the method developed in \cite{Miyao5}, we can prove the following:
\begin{Thm}\label{HRPI}
We have
 $
 \big(\mathbb{K}[M]+s\big)^{-1} \rhd 0
 $ w.r.t. $\Cone_{\bep}[M]$ for all $s>E
 \big(
 \mathbb{K}[M]
 \big)$.
\end{Thm}
{\it Proof.} Since the proof is similar to \cite[Theorem 4.1]{Miyao5}, we omit it. 
$\Box$
\medskip

By Proposition \ref{EquivPP} and Theorem \ref{HRPI}, we finally obtain the following: 
\begin{coro}\label{RadPI}
 $ \mathbf{U}\big( \tilde{H}_{\mathrm{rad}, |\Lambda|}[M]+s\big)^{-1}
 \mathbf{U}^{-1}\rhd 0
 $ w.r.t. $\mathfrak{P}_{\mathrm{rad}}[M]$ for all $
 s>
 E\big(
 \tilde{H}_{\mathrm{rad}, |\Lambda|}[M]
 \big)$.
\end{coro}

\subsection{Completion of proof of Theorem \ref{GeneHH}}
We will  check all conditions in Definition \ref{FirstSt} with 
\begin{align*}
&(H, \ H_*;\  \mathfrak{H},\  \mathfrak{H}_*;\  P;\  \mathfrak{P}, \ \mathfrak{P}_*; O; U)\no
=&
\Big(\tilde{H}_{\mathrm{rad}, |\Lambda|}[M], \, \tilde{H}_{\mathrm{H}, |\Lambda|}[M];\,  W\mathfrak{E}_{n=|\Lambda|}[M]\otimes L^2(\mathscr{R}, d\nu),\  
W\mathfrak{E}_{n=|\Lambda|}[M];\no 
& \ \ \ \ P_{\mathrm{rad}};\,
\mathfrak{P}_{\mathrm{rad}}[M],\,  \mathfrak{P}_{\mathrm{H}}[M];\ 
\tilde{S}_{\mathrm{tot}, |\Lambda|}^2[M];  \mathbf{U}\Big).
\end{align*}

By Proposition \ref{PropPrad}, (i) and (ii) of Definition \ref{FirstSt} are satisfied.
 (iii) and (iv) of  Definition \ref{FirstSt} follow from Theorem \ref{HPP} and Corollary  \ref{RadPI}, respectively.
 By applying Proposition \ref{UniEQ}  with $V=W$, we obtain Theorem \ref{GeneHH}.
  $\Box$

\section{Proof of Thereom \ref{Cl1}}\label{ProofCl1}
(i)
Let us consider the on-site Coulomb interaction $\{U\delta_{xy}\}$.
Clearly, $\{U\delta_{xy}\}$ is positive definite provided that $U>0$.
Thus, by Theorem \ref{HPP}, it holds that $\big( \tilde{H}^{(U)}_{\mathrm{H}, |\Lambda|
}[M]+s\big)^{-1} \rhd 0$ w.r.t. $\Cone_{\mathrm{H}}[M]$ for all $s>-E\big(
\tilde{H}^{(U)}_{\mathrm{H}, |\Lambda|
}[M]
\big)$.
Therefore, we can check all conditions in Definitions \ref{FirstSt} with 
\begin{align*}
&(H, \ H_*;\  \mathfrak{H},\  \mathfrak{H}_*;\  P;\  \mathfrak{P}, \ \mathfrak{P}_*; O; U)\no
=&
\Big(\tilde{H}_{\mathrm{H}, |\Lambda|}[M], \, \tilde{H}_{\mathrm{H}, |\Lambda|}^{(U)}[M];\,  W\mathfrak{E}_{n=|\Lambda|}[M],\  
W\mathfrak{E}_{n=|\Lambda|}[M]
;\,  1;\,
\mathfrak{P}_{\mathrm{H}}[M],\,  \mathfrak{P}_{\mathrm{H}}[M];\ 
\tilde{S}_{\mathrm{tot}, |\Lambda|}^2[M]; 1\Big),
\end{align*}
see Appendices \ref{PfLi1} and \ref{PfLi2} for definitions of the above operators. 
Thus, we have $
\tilde{H}_{\mathrm{H}, |\Lambda|}[M] \leadsto \tilde{H}_{\mathrm{H}, |\Lambda|}^{(U)}[M]
$.
By interchanging the roles of $
\tilde{H}_{\mathrm{H}, |\Lambda|}[M] $ and $\tilde{H}_{\mathrm{H}, |\Lambda|}^{(U)}[M]
$, we obtain $  \tilde{H}_{\mathrm{H}, |\Lambda|}^{(U)}[M] \leadsto
\tilde{H}_{\mathrm{H}, |\Lambda|}[M] $.

In a similar way, we can prove (ii)
 and (iii). $\Box$

\section{Proof of Theorems \ref{HHNT}, \ref{HHNT1} and \ref{HHNT2}}\label{PfStaNTThm}
\setcounter{equation}{0}
\subsection{Self-dual cones}

In this subsection, we construct some self-dual cones to prove the stabilities of the Nagaoka-Thouless theorem in Section \ref{StaNTThm}.
In the rest of this section, we continue to assume {\bf (B. 1)}, {\bf (B. 2)} and {\bf (B. 3)}.

\subsubsection{Self-dual cone in $\mathfrak{H}_{\mathrm{NT}}[M]$}

For each $(x, {\Bs \sigma})\in \mathcal{C}_M$, we set
${\Bs \sigma}\rq{}=\{\sigma_z\rq{} \}_{z\in \Lambda} \in \{\uparrow, \downarrow\}^{\Lambda}$
by 
\begin{align}
\sigma_z\rq{}=
\begin{cases}
\uparrow & \mbox{if $z=x$}\\
\sigma_z & \mbox{otherwise},
\end{cases}
\end{align}
where $\mathcal{C}_M$ is defined in Section \ref{NTBasicDef}.
With this notation, we define a complete orthonormal system (CONS) $
\{|x, {\Bs \sigma} \ra\, |\, (x, {\Bs \sigma}) \in \mathcal{C}_M\} \subseteq 
\mathfrak{H}_{\mathrm{NT}}[M]$ by 
$\displaystyle 
|x, {\Bs \sigma}\ra=c_{x\uparrow} \prod_{z\in \Lambda}\rq{} c_{z\sigma_z\rq{}}^*\Omega,
$
where $\displaystyle 
\prod_{z\in \Lambda}\rq{}$ indicates the ordered product according to an arbitrarily fixed order in $\Lambda$. Remark that this CONS was introduced by Tasaki \cite{Tasaki2, Tasaki22}.

\begin{define}
{\rm
For each $M\in \mathrm{spec}\big(
\mathcal{S}^{(3)}
\big)$,
a canonical self-dual cone in $\mathfrak{H}_{\mathrm{NT}}[M]$ is defined by 
$
\mathfrak{Q}_{\mathrm{H}}[M]=\mathrm{Coni} \{
|x, {\Bs \sigma}\ra\, \, |\, (x, {\Bs \sigma}) \in \mathcal{C}_M
\}. 
$
Recall that this type of self-dual cone is defined in Example \ref{Cone1} of Section \ref{GeneralThy}. $\diamondsuit$
}
\end{define}

\subsubsection{Self-dual cone in $\mathfrak{H}_{\mathrm{NT}}[M] \otimes \Fock_{\mathrm{ph}}$}
We switch to the Schr\"{o}dinger representation already discussed in Section \ref{SchRP}: $\Fock_{\mathrm{ph}}=L^2(\mathcal{Q})$.
Under the identification $
\mathfrak{H}_{\mathrm{NT}}[M] \otimes \Fock_{\mathrm{ph}}=
\mathfrak{H}_{\mathrm{NT}}[M] \otimes L^2(\mathcal{Q})$, we define 
a self-dual cone $\mathfrak{Q}_{\mathrm{HH}}[M]$ by 
\begin{align}
\mathfrak{Q}_{\mathrm{HH}}[M]
=\int^{\oplus}_{\mathcal{Q}} \mathfrak{Q}_{\mathrm{H}}[M] d{\Bs q}. \label{SDCH}
\end{align}
Remark that the right hand side of (\ref{SDCH}) is a direct integral of $\mathfrak{Q}_{\mathrm{H}}[M]$,
see Appendix \ref{App2} for details.

\subsubsection{Self-dual cone in $\mathfrak{H}_{\mathrm{NT}}[M] \otimes \Fock_{\mathrm{rad}}$}
We work in  the Feynman-Schr\"{o}dinger representation  introduced  in Section \ref{FSRP}: $\Fock_{\mathrm{rad}}=L^2(\mathscr{R}, d\nu)$.
Under the identification $
\mathfrak{H}_{\mathrm{NT}}[M] \otimes \Fock_{\mathrm{rad}}=
\mathfrak{H}_{\mathrm{NT}}[M] \otimes L^2(\mathscr{R}, d\nu)$, we define 
a self-dual cone $\mathfrak{Q}_{\mathrm{rad}}[M]$ by 
\begin{align}
\mathfrak{Q}_{\mathrm{rad}}[M]
=\int^{\oplus}_{\mathscr{R}} \mathfrak{Q}_{\mathrm{H}}[M] d\nu.
\end{align}

\subsection{The Nagaoka-Thouless theorem}
Here, we will give a brief review of  the Nagaoka-Thouless theorem.
Remark that a strategy below is mathematically equivalent to Tasaki\rq{}s work \cite{Tasaki2}.

In \cite{Miyao6}, Miyao proved the following:
\begin{Thm}\label{NagaPI}
 $
 e^{-\beta H^{\infty}_{\mathrm{H}}[M]} \rhd 0
 $ w.r.t. $\mathfrak{Q}_{\mathrm{H}}[M]$ for all $\beta>0$.
\end{Thm}

As a corollary of Theorem \ref{NagaPI}, we get the  Nagaoka-Thouless theorem:
\begin{coro}[Theorem \ref{NagaokaTThm1}]\label{NTThm2}
The ground state of $H_{\mathrm{H}}^{\infty}$
 has  total spin $S=\frac{1}{2}(|\Lambda|-1)$ and is unique apart from the trivial $(2S+1)$-degeneracy.
\end{coro}
{\it Proof.} For each $x\in \Lambda$, let us introduce a spin configuration by 
\begin{align}
(\Uparrow_x)_y=\begin{cases}
 \uparrow & y\neq x\\
 0 & y=x.
\end{cases}
\end{align}
Then $\{
|x, \Uparrow_x\ra\, |\, x\in \Lambda
\}$ is a CONS of $\mathfrak{H}_{\mathrm{NT}}[M=\frac{1}{2}(|\Lambda|-1)]$.
Clearly, each $|x, \Uparrow_x\ra$ has total spin $S=\frac{1}{2}(|\Lambda|-1)$.

By Theorems \ref{PIEquiv}, \ref{PFF} and  \ref{NagaPI}, the ground state of $H^{\infty}_{\mathrm{H}}[M]$ with $M=\frac{1}{2}(|\Lambda|-1)$ is unique and strictly positive w.r.t. $\mathfrak{Q}_{\mathrm{H}}[M=\frac{1}{2}(|\Lambda|-1)].$
The ground state of $
H^{\infty}_{\mathrm{H}}[\tfrac{1}{2}(|\Lambda|-1)]
$ is denoted by $\psi$.
Since $\mathfrak{Q}_{\mathrm{H}} [\frac{1}{2}(|\Lambda|-1)]=
\mathrm{Coni}\{
|x, \Uparrow_x\ra\, |\, x\in \Lambda
\}$, we have
$\psi=\sum_{x\in \Lambda} \psi_x |x, \Uparrow_x\ra$ 
with $\psi_x>0$. Thus, $\psi$ has total spin $S=\frac{1}{2}(|\Lambda|-1)$.

We set $\mathcal{S}^{(\pm)}=\mathcal{S}^{(1)}\pm i \mathcal{S}^{(2)}$ as usual.
Let $\psi_{\ell}=(S^{(-)})^{\ell} \psi,\ \ell=0, 1, \dots, |\Lambda|-2$. Then
$\psi_{\ell}$ is the unique ground state of $H^{\infty}_{\mathrm{H}}[\frac{1}{2}(|\Lambda|-\ell-1)]$
 for each $\ell=0, 1,\dots, |\Lambda|-2$,
 and has total spin $S=\frac{1}{2}(|\Lambda|-1)$. $\Box$

\begin{flushleft}
{\it Completion of proof of Theorem \ref{HHNT} }
\end{flushleft}
By Theorem \ref{ElUniTh} and Corollary \ref{NTThm2}, one obtains Theorem \ref{HHNT}.
$\Box$

\subsection{Proof of Theorem \ref{HHNT1}}\label{PfHHNT1}
Corollaries \ref{StNaga1} and \ref{StNaga2} were already proved in \cite{Miyao6}.
Here, we prove these results in the context of   the  stability class introduced in Section \ref{GeneralThy}.

\begin{Prop}\label{NTP}
For each $M\in \mathrm{spec}\big(
\mathcal{S}^{(3)}
\big)$, 
we have $P_{\mathrm{ph}} \unrhd 0$ w.r.t. $\mathfrak{Q}_{\mathrm{HH}}[M]$.
\end{Prop}
{\it Proof.} In  the Schr\"{o}dinger representation, we have 
$
\Omega_{\mathrm{ph}}({\Bs q}) =\pi^{-|\Lambda|/4} \exp(-{\Bs q}^2/2)
$. Thus, $\Omega_{\mathrm{ph}} >0$ w.r.t. $L^2(\mathcal{Q})_+$,
 which implies that $
 |\Omega_{\mathrm{ph}} \ra\la \Omega_{\mathrm{ph}}| \unrhd 0
 $ w.r.t. $L^2(\mathcal{Q})_+$. Hence, by Proposition  \ref{BasicDP22}, we obtain the desired assertion in the lemma. $\Box$

\begin{Prop}\label{NTP2}
For each $M\in \mathrm{spec}\big(
\mathcal{S}^{(3)}
\big)$, 
we have $
P_{\mathrm{ph}}
\mathfrak{Q}_{\mathrm{HH}}[M]= \mathfrak{
Q}_{\mathrm{H}}[M]$.
\end{Prop}
{\it Proof.}
It is not hard to check that 
$P_{\mathrm{ph}}\mathfrak{Q}_{\mathrm{HH}}[M]=\mathfrak{Q}_{\mathrm{H}}[M]\otimes \Omega_{\mathrm{ph}}$. 
Thus, using the identification $\mathfrak{H}_{\mathrm{NT}}[M]= \mathfrak{H}_{\mathrm{NT}}[M] \otimes \Omega_{\mathrm{ph}}$  mentioned  in Section \ref{SectionStaNT1},  
we have $P_{\mathrm{ph}}\mathfrak{Q}_{\mathrm{HH}}[M]=\mathfrak{Q}_{\mathrm{H}}[M]$. $\Box$
\medskip\\

Recall the definition of the Lang-Firsov transformation $e^L$ in Section \ref{PropHHH}.
In \cite{Miyao6}, we proved the following:

\begin{Thm}\label{NTHHIP}
For each $M\in \mathrm{spec}\big(
\mathcal{S}^{(3)}
\big)$, 
we have
$
e^L \exp\Big\{-\beta H_{\mathrm{HH}}^{\infty}[M]\Big\} e^{-L} \rhd 0
$  w.r.t. $\mathfrak{Q}_{\mathrm{HH}}[M]$ for all $\beta >0$.
\end{Thm}
{\it Proof.} See \cite[Theorem 5.10]{Miyao6}. $\Box$
\begin{flushleft}
{\it
Completion of  proof of Theorem \ref{HHNT}
}
\end{flushleft}
We will check all conditions in Definition \ref{FirstSt}
with 
\begin{align}
&(H, \ H_*;\  \mathfrak{H},\  \mathfrak{H}_*;\  P;\  \mathfrak{P}, \ \mathfrak{P}_*; O; U)\no
=&\Big(
H_{\mathrm{HH}}^{\infty}[M],  H_{\mathrm{H}}^{\infty}[M];\ 
\mathfrak{H}_{\mathrm{NT}}[M] \otimes L^2(\mathcal{Q}), \mathfrak{H}_{\mathrm{NT}}[M];\ \no
&P_{\mathrm{ph}};\ \mathfrak{Q}_{\mathrm{HH}}[M], \mathfrak{Q}_{\mathrm{H}}[M];\ 
\mathcal{S}^2;\ e^L
\Big).
\end{align}
By Propositions \ref{NTP} and  \ref{NTP2}, we can confirm (i) and (ii) of Definition \ref{FirstSt}.
By Theorems \ref{NagaPI}, \ref{NTHHIP} and  \ref{PIEquiv}, (iii) and (iv) of Definition \ref{FirstSt} are satisfied.
 Hence, by Proposition \ref{UniEQ}, we obtain Theorem \ref{HHNT}.
$\Box$

\subsection{Proof of Theorem \ref{HHNT2}}
Because the idea of the proof is similar to that of Theorem \ref{HHNT1}, we provide a  sketch only.

Let $\Omega_{\mathrm{rad}}$ be the Fock vacuum in $\Fock_{\mathrm{rad}}$.
We set $P_{\mathrm{rad}}=1\otimes |\Omega_{\mathrm{rad}}\ra\la \Omega_{\mathrm{rad}}|$.

In a similar way as  in Section \ref{PfHHNT1}, we can prove the following two  propositions.
\begin{Prop}\label{NTPR}
For each $M\in \mathrm{spec}\big(
\mathcal{S}^{(3)}
\big)$, 
we have $P_{\mathrm{rad}} \unrhd 0$ w.r.t. $\mathfrak{Q}_{\mathrm{rad}}[M]$.
\end{Prop}

\begin{Prop}\label{NTPR2}
For each $M\in \mathrm{spec}\big(
\mathcal{S}^{(3)}
\big)$, 
we have
$P_{\mathrm{rad}}\mathfrak{Q}_{\mathrm{rad}}[M]=\mathfrak{Q}_{\mathrm{H}}[M]$. 

\end{Prop}

In \cite{Miyao6}, we proved the following:

\begin{Thm}\label{NTHRIP}
For each $M\in \mathrm{spec}\big(
\mathcal{S}^{(3)}
\big)$, 
we have
$
\mathsf{F} \exp\Big\{-\beta H_{\mathrm{rad}}^{\infty}[M]\Big\}  \mathsf{F}^{-1}\rhd 0
$
  w.r.t. $\mathfrak{Q}_{\mathrm{rad}}[M]$ for all $\beta >0$, where $\mathsf{F}=e^{i\pi N_{\mathrm{rad}}/2} \mathbf{U}$
   with $N_{\mathrm{rad}}=\dG(1)$. Here, $\mathbf{U}$ is defined by (\ref{DEFU}).
\end{Thm}

\begin{flushleft}
{\it
Completion of  proof of Theorem \ref{HHNT1}
}
\end{flushleft}
We will check all conditions in Definition \ref{FirstSt}
with 
\begin{align}
&(H, \ H_*;\  \mathfrak{H},\  \mathfrak{H}_*;\  P;\  \mathfrak{P}, \ \mathfrak{P}_*; O; U)\no
=&\Big(
H_{\mathrm{rad}}^{\infty}[M],  H_{\mathrm{H}}^{\infty}[M];\ 
\mathfrak{H}_{\mathrm{NT}}[M] \otimes L^2(\mathscr{R}, d\nu), \mathfrak{H}_{\mathrm{NT}}[M];\ \no
&P_{\mathrm{rad}};\ \mathfrak{Q_{\mathrm{rad}}}[M], \mathfrak{Q}_{\mathrm{H}}[M]; \ 
\mathcal{S}^2;\ \mathsf{F}
\Big).
\end{align}
By Propositions \ref{NTPR} and \ref{NTPR2}, we can confirm (i) and (ii) of Definition \ref{FirstSt}.
(iv) of Definition \ref{FirstSt} is satisfied by Theorems \ref{NTHRIP} and \ref{PIEquiv}.
By Theorems \ref{NagaPI} and \ref{PIEquiv}, we know that (iii) of Definition \ref{FirstSt} is fulfilled. 
Therefore,   by Proposition \ref{UniEQ}, we obtain Theorem \ref{HHNT1}.
$\Box$

\section{Basic properties of operator theoretic correlation inequalities}\label{App1}

\setcounter{equation}{0}

We begin with the following theorem.
\begin{Thm}\label{SAH}
Let $\Cone$ be a convex cone in $\h$.
$\Cone$ is self-dual if and only if:
\begin{itemize}
\item[{\rm (i)}] $ \la \xi| \eta\ra\ge 0$ for all $\xi, \eta\in \Cone$.
\item[{\rm (ii)}] Let $\h_{\BbbR}$
 be a real closed subspace of $\h$ generated by $\Cone$ . Then
	     for all $\xi\in \h_{\BbbR}$, there exist $\xi_+,
	     \xi_-\in \Cone$ such that $\xi=\xi_+-\xi_-$ and $\la \xi_+|
	     \xi_-\ra=0$.
\item[{\rm (iii)}] $\h=\h_{\BbbR}+i
	    \h_{\mathbb{R}}= \{\xi+ i \eta\, |\, \xi, \eta\in
	     \h_{\BbbR}\}$.
\end{itemize}
\end{Thm} 
{\it Proof.}  
For the  reader\rq{}s convenience, we provide a sketch of the  proof.
  
Assume that $\Cone$ is self-dual.  Then, by \cite{Bos} or \cite[Proof of Proposition 2.5.28]{BR1},
we easily check that  the conditions  (i)--(iii) are fulfilled.

Conversely, suppose that  $\Cone$ satisfies (i)--(iii).  We   see that 
$\Cone\subseteq \Cone^{\dagger}$  by  (i).
We will show the inverse. Let $\xi\in \Cone^{\dagger}$.
By (ii) and (iii), we can write $\xi$ as 
$\xi=(\xi_{R, +}-\xi_{R, -})+i(\xi_{I, +}-\xi_{I, -})$ with 
$\xi_{R, \pm}, \xi_{I, \pm} \in \Cone$, $\la \xi_{R, +}|\xi_{R, -}\ra=0$
and $\la \xi_{I, +}|\xi_{I, -}\ra=0$. 
Assume that $\xi_{I, +}\neq 0$. Then,
$\la \xi|\xi_{I, +}\ra$ is a
complex number, which contradicts with the fact that $\la \xi|\eta\ra\ge
0$ for all $\eta\in \Cone$. Thus, $\xi_{I, +}$ must be $0$.  Similarly, we have $\xi_{I, -}=0$.
Next, assume that $\xi_{R, -}\neq 0$. Because $\xi_{R, -}\in \Cone$, we
have
$
0\le \la \xi|\xi_{R, -}\ra=-\|\xi_{R, -}\|^2<0,
$
which is a contradiction. Hence, we conclude that $\xi=\xi_{R, +}\in
\Cone$. $\Box$

\begin{coro}\label{DecRI}
Let $\Cone$ be a self-dual cone in $\h$. For each $\xi\in \h$, 
we have the following decomposition:
\begin{align}
\xi=(\xi_1-\xi_2)+i(\xi_3-\xi_4), \label{ReIm}
\end{align}
where $\xi_1, \xi_2, \xi_3$ and $\xi_4$ satisfy $\xi_1, \xi_2, \xi_3,
 \xi_4\in \Cone$, $\la \xi_1|\xi_2\ra=0$ and $\la \xi_3|\xi_4\ra=0$. 
\end{coro} 

\begin{define}{\rm 
  Suppose that $A\h_{\BbbR}\subseteq
 \h_{\BbbR}$ and $B\h_{\BbbR} \subseteq
	     \h_{\BbbR}$. If $(A-B) \Cone\subseteq
	     \Cone$, then we write this as $A \unrhd B$ w.r.t. $\Cone$.  $\diamondsuit$
	     }
\end{define}

The following proposition is fundamental in the present paper

\begin{Prop}
Let $A, B, C, D\in \mathscr{B}(\h)$ and let $a, b\in
 \BbbR$. 
\begin{itemize}
\item[{\rm (i)}] If $A\unrhd 0, B\unrhd 0$ w.r.t. $\Cone$ and
	     $a, b\ge 0$, then $aA +bB \unrhd 0$
	     w.r.t. $\Cone$.
\item[{\rm (ii)}] If $A \unrhd B \unrhd 0$ and $C\unrhd D \unrhd 0$
	     w.r.t. $\Cone$,
	     then
$AC\unrhd BD \unrhd 0$ w.r.t. $\Cone$.
\item[{\rm (iii)}] If $A \unrhd 0 $ w.r.t. $\Cone$, then $A^*\unrhd 0$ w.r.t. $\Cone$.
\end{itemize} 
\end{Prop} 
{\it Proof.} See, e.g., \cite[Lemmas 2.6 and  2.7]{Miyao9}. $\Box$

\begin{Prop}\label{PPEquiv}
Let $A$ be a positive self-adjoint operator.
The following statements are mutually  equivalent:
\begin{itemize}
\item[{\rm (i)}] $e^{-tA} \unrhd 0$ w.r.t. $\Cone$ for all $t\ge 0$.
\item[{\rm (ii)}] $(A+s)^{-1} \unrhd 0$ w.r.t. $\Cone$ for all $s>-E(A)$.
\end{itemize}
\end{Prop}
{\it Proof.} The proposition  immediately follows from
the following formulas:
\begin{align}
(A+s)^{-1} =\int_0^{\infty} e^{-t(A+s)} dt,\ \ \ e^{-tA}=\slim\Big(1+\frac{s}{n}A\Big)^{-n},
\end{align}
where $\slim$ indicates the strong limit. $\Box$

\begin{Prop}\label{PIEquiv}
Let $A$ be a positive self-adjoint operator.
The following statements are mutually  equivalent:
\begin{itemize}
\item[{\rm (i)}] The semigroup $e^{-tA} $ is ergodic, that is, 
 for every  $\xi, \eta\in \Cone\backslash \{0\}$, there exists a  $t_0\ge 0$ such that 
 $\la \xi|e^{-t_0A} \eta\ra>0$. Note that $t_0$ could depend on $\xi$ and $\eta$.
\item[{\rm (ii)}] $(A+s)^{-1} \rhd 0$ w.r.t. $\Cone$ for all $s>-E(A)$.
\end{itemize}
In particular, if $e^{-t A} \rhd 0$ w.r.t. $\Cone$ for all $t>0$, then $(A+s)^{-1} \rhd 0$ w.r.t. $\Cone$
for all $s>-E(A)$. 
\end{Prop}
{\it Proof.}
 Use the elementary formula: $
(A+s)^{-1} =\int_0^{\infty} e^{-t(A+s)} dt
$. $\Box$

\begin{Prop}\label{PSum}
Assume that $A \unrhd 0$ w.r.t. $\Cone$. Then $\ex^{\beta A} \unrhd 0$
w.r.t. $\Cone$ for all $\beta \ge 0$.
\end{Prop} 
{\it Proof.} See, e.g., \cite[Proposition A. 3]{Miyao4}. $\Box$

\begin{Thm}\label{StandPP}
Let $A$ be a  self-adjoint positive operator on $\h$ and $B\in \mathscr{B}(\h)$. 
Suppose that 
\begin{itemize}
\item[{\rm (i)}] $e^{-\beta A} \unrhd 0$ w.r.t. $\Cone$ for all $\beta
	     \ge 0$;
\item[{\rm (ii)}] $B\unrhd 0$ w.r.t. $\Cone$.
\end{itemize} 
Then we have $e^{-\beta (A-B)}\unrhd 0$ w.r.t. $\Cone$ for all $\beta
 \ge 0$.
\end{Thm}
{\it Proof.} See, e.g., \cite[Proposition A. 5]{Miyao4}. $\Box$

\begin{Prop}\label{TK}
Assume that $\ex^{\beta A} \unrhd 0$ and $\ex^{\beta B} \unrhd 0$
 w.r.t. $\Cone$ for all $\beta \ge 0$. Then $\ex^{\beta (A+B)} \unrhd 0$
 w.r.t. $\Cone$ for all $\beta \ge 0$.
\end{Prop} 
{\it Proof.} See, e.g., \cite[Proposition A. 4]{Miyao4}. $\Box$
\medskip

The following theorem  plays  an important role in the present study.
\begin{Thm}\label{PFF}{\rm (Perron--Frobenius--Faris)}
Let $A$ be a  self-adjoint operator,  bounded from below. 
Let $E(A)=\inf \mathrm{spec}(A)$, where $\mathrm{spec}(A)$ is  spectrum of $A$.
Suppose that 
 $0\unlhd e^{-tA}$ w.r.t. $\Cone$ for all $t\ge 0$,  and that  $E(A)$ is an eigenvalue.
Let $P_A$ be the orthogonal projection onto the closed subspace spanned
 by  eigenvectors associated with   $E(A)$.
 Then  the following statements
 are equivalent:
\begin{itemize}
\item[{\rm (i)}] 
$\dim \ran P_A=1$ and $P_A\rhd 0$ w.r.t. $\Cone$.
\item[{\rm (ii)}] $ (A+s)^{-1}\rhd 0$ w.r.t. $\Cone$ for all
	     $s>-E(A)$.
\end{itemize}
\end{Thm} 
{\it Proof.} See, e.g.,    \cite{Faris, Miyao1, ReSi4}. $\Box$

\begin{rem}\label{HarEx}
{\rm 
By (i), there exists a unique $\xi\in \h$ such that $\xi>0$
 w.r.t. $\Cone$
and $P_A=|\xi\ra \la \xi|$. 
Of course, $\xi$ satisfies $A\xi=E(A)\xi$. Hence, (i) implies  that  the lowest eigenvalue of $A$ is nondegenerate, and the corresponding eigenvector is  strictly positive.
$\diamondsuit$
}
\end{rem}

\begin{Thm}\label{AbstRP}
Let us consider the case where $\Cone=\mathscr{L}^2(\mathfrak{H})_+$ given in Definition \ref{L2Define}.
Let $A, C_j\in \mathscr{B}(\mathfrak{H})$. Suppose that $A$ is self-adjoint. We set 
$
H=\mathcal{L}(A)+\mathcal{R}(A)-\sum_{j=1}^n \mathcal{L}(C_j) \mathcal{R}(C_j^*).
$
Then 
$e^{-\beta H} \unrhd 0$ w.r.t. $\Cone$ for all $\beta \ge 0$.
\end{Thm}
{\it Proof.}
We set
$
H_0=\mathcal{L}(A)+\mathcal{R}(A)$
and 
$ V=\sum_{j=1}^n \mathcal{L}(C_j) \mathcal{R}(C_j^*).
$
Trivially, $H=H_0-V$.
By Proposition \ref{GeneralPP}, we have
$
e^{-\beta H_0}=\mathcal{L}(e^{-\beta A}) \mathcal{R}(e^{-\beta A}) \unrhd 0
$
w.r.t. $\Cone$ for all $\beta \ge 0$. 
Because $V\unrhd 0$ w.r.t. $\Cone$, we can apply Theorem \ref{StandPP}.
  $\Box$

\begin{rem}
{\rm 
Jaffe and Pedrocchi give an alternative proof within an algebraic setting \cite{JP}. $\diamondsuit$
}
\end{rem}

\begin{Prop}\label{PIEquiv*}
Let $A\in \mathscr{B}(\mathfrak{H})$. The following statements are equivalent:
\begin{itemize}
\item[{\rm (i)}] $A\rhd 0$ w.r.t. $\Cone$.
\item[{\rm (ii)}] $A^*\rhd 0$ w.r.t. $\Cone$.
\end{itemize}
\end{Prop}
{\it Proof.} (i) $\Rightarrow $ (ii):
Let $\xi, \eta\in \Cone\backslash \{0\}$.
Because $A\xi >0$ w.r.t. $\Cone$, we have
$
\la \xi|A^*\eta\ra=\la A\xi|\eta\ra>0.
$
Since $\xi$ is arbitrary, 
we have $A^*\eta>0$ w.r.t. $\Cone$, which implies that  $A^*\rhd 0$ w.r.t. $\Cone$.
Simiraly, we can prove that  (ii) $\Rightarrow $ (i). $\Box$

\begin{Thm}\label{PEq}
Let $A$ and $B$ be self-adjoint operators,  bounded from below.
Assume the following conditions:
\begin{itemize}
\item[{\rm (a)}] There exists a sequence of bounded self-adjoint
	     operator $C_n$ such that $
A+C_n
$ converges to $B$ in the strong resolvent sense and $B-C_n$
converges to $A$ in the strong resolvent sense as $n\to \infty$;
\item[{\rm (b)}] $ e^{-tC_n}\unrhd 0$ w.r.t. $\Cone$ for all $t\in \BbbR$
	     and $n\in \BbbN$;
\item[{\rm (c)}] 
For all $\xi, \eta\in \Cone$ such that $\la \xi|\eta\ra=0$, it holds
	     that $\la \xi|e^{-tC_n}\eta\ra=0$ for all $n\in \BbbN$ and
	     $t\ge 0$.
\end{itemize} 
The following {\rm (i)} and {\rm (ii)} are mutually equivalent:
\begin{itemize}
\item[{\rm (i)}] $(A+s)^{-1} \rhd 0$ w.r.t. $\Cone$ for all $s>-E(A)$;
\item[{\rm (ii)}] $(B+s)^{-1} \rhd 0$ w.r.t. $\Cone$ for all $s>-E(B)$.
\end{itemize} 

\end{Thm} 
{\it Proof.} See, e.g., \cite[Theorem 3]{Faris}  and \cite[Theorem A.1]{Miyao6}. $\Box$

\section{Direct integrals of self-dual cones}\label{App2}

\setcounter{equation}{0}

Let $\mathfrak{H}$ be a complex Hilbert space.
Let $(M, \mathfrak{M}, \mu)$ be a $\sigma$-finite measure space.
The Hilbert space of  $L^2(M, d\mu; \mathfrak{H})$ of square integrable $\mathfrak{H}$-valued functions 
\cite[Section XIII.16]{ReSi4} is called a {\it constant fiber direct integral}, and is written as 
$\displaystyle 
\int^{\oplus}_M\mathfrak{H} d\mu.
$
The inner product on $\displaystyle  \int^{\oplus}_M\mathfrak{H} d\mu$ is  given by 
$\displaystyle 
\la \Phi|\Psi\ra=\int_M\la \Phi(m)|\Psi(m)\ra_{\mathfrak{H} }d\mu,
$
where $\la \cdot |\cdot \ra_{\mathfrak{H}}$ is the inner product on $\mathfrak{H}$.
Remark that  $L^2(M, d\mu; \mathfrak{H})$ can be naturally identified with $\mathfrak{H} \otimes L^2(M, d\mu)$:
\begin{align}
\mathfrak{H}\otimes L^2(M, d\mu)=\int_M^{\oplus} \mathfrak{H} d\mu. \label{TensorIdn}
\end{align}

$L^{\infty}(M, d\mu; \mathscr{B}(\h))$ denotes the space of measurable functions from $M$ to $\mathscr{B}(\h)$ with 
\begin{align}
\|A\|_{\infty}=\mathrm{ess.sup} \|A(m)\|_{\mathscr{B}(\h)}<\infty.
\end{align}
A bounded operator $A$ on $\int^{\oplus}_M\mathfrak{H} d\mu$ is said to be decomposed by the direct integral decomposition, if and only if there is a function $A(\cdot)\in L^{\infty}(M, d\mu; \mathscr{B}(\mathfrak{H}))$ such that 
\begin{align}
(A\Psi)(m)=A(m)\Psi(m) \label{Decomp1}
\end{align}
for all $\Psi\in \int^{\oplus}_M\mathfrak{H} d\mu$.
In this case, we call $A$ {\it decomposable} and write 
\begin{align}
A=\int^{\oplus}_MA(m)d\mu.
\end{align}

\begin{example}
{\rm
Let $B\in \mathscr{B}(\mathfrak{H})$. Under the identification (\ref{TensorIdn}), we have
\begin{align}
B\otimes 1=\int^{\oplus}_MBd\mu.\ \ \ \ \diamondsuit \label{BOI}
\end{align}
}
\end{example}
\begin{lemm}\label{ThmXIII83}
If $A(\cdot) \in L^{\infty}(M, d\mu; \mathscr{B}(\mathfrak{H}))$, then there is a unique decomposable operator $A \in \mathscr{B}(\int^{\oplus}_M\mathfrak{H} d\mu)$  such that (\ref{Decomp1}) holds.
\end{lemm}
{\it Proof.} See \cite[Theorem XIII. 83]{ReSi4}. $\Box$
\medskip\\

Let $\Cone$ be a self-dual cone in $\mathfrak{H}$. We set 
\begin{align}
\int^{\oplus}_M \Cone d\mu=\bigg\{
\Psi\in \int^{\oplus}_M \mathfrak{H} d\mu\, \Big|\, \mbox{$\Psi(m) \ge 0$ w.r.t. $\Cone$ for $\mu$-a.e.} 
\bigg\}.
\end{align}
It is not hard to check that $\int^{\oplus}_M\Cone d\mu$ is a self-dual cone in $\int^{\oplus}_M\mathfrak{H} d\mu$.
We call $\int^{\oplus}_M\Cone d\mu$ a {\it direct integral of $\Cone$}.

The following lemma is useful:
\begin{lemm}\label{PPEquivH}
Let $\Psi\in \int^{\oplus}_M\mathfrak{H}d\mu$. The following statements are equivalent:
\begin{itemize}
\item[{\rm (i)}] $\Psi\ge 0$ w.r.t. $\int^{\oplus}_M \Cone d\mu$.
\item[{\rm (ii)}] $\la \xi\otimes f|\Psi\ra\ge 0$ for all $\xi\in \Cone,\ f\in L^2(M, d\mu)_+$.
\end{itemize}
\end{lemm}
{\it Proof.}
To show (i) $\Rightarrow$ (ii) is easy. Let us show the inverse.
We set  $G_{\xi}(m):=\la \xi|\Psi(m)\ra$. By (ii),  we have 
$
\la \xi\otimes f|\Psi\ra=\int_MG_{\xi}(m) f(m) d\mu\ge 0 $
 for each 
$f\in L^2(M, d\mu)_+.
$
Since $f$ is arbitrary, we conclude that $G_{\xi}(m) \ge 0$.
Because $\xi$ is arbitrary, we finally arrive at $\Psi(m) \ge 0$ w.r.t. $\Cone$. $\Box$

\begin{Prop}\label{BasicDP}
Let $A=\int^{\oplus}_MA(m) d\mu$ be a decomposable operator on $\int^{\oplus}_M\mathfrak{H} d\mu$.
If $A(m) \unrhd 0$ w.r.t. $\Cone$ for $\mu$-a.e., then
$A\unrhd 0$ w.r.t. $\int^{\oplus}_M\Cone d\mu$.
\end{Prop}
{\it Proof.} For each $\Psi\in \int^{\oplus} _M\Cone d\mu$, we have
$
(A\Psi)(m)=A(m) \Psi(m)\ge 0 $ w.r.t. $\Cone$ for $\mu$-a.e..
Hence, $A\Psi \ge 0$ w.r.t. $\int^{\oplus}_M\Cone d\mu$. $\Box$

\begin{coro}\label{BasicDP2}
Let $B\in \mathscr{B}(\mathfrak{H})$. Under the identification (\ref{TensorIdn}), if $B\unrhd 0$ w.r.t. $\Cone$, then
$B\otimes 1 \unrhd 0$ w.r.t. $\int^{\oplus}_M\Cone d\mu$. 
\end{coro}
{\it Proof.} Use (\ref{BOI}). $\Box$

\begin{Prop}\label{BasicDP22}
Let $C$ be a bounded linear operator on $L^2(M, d\mu)$. Under the identification (\ref{TensorIdn}), 
if $C\unrhd 0$ w.r.t. $L^2(M, d\mu)_+$, then $1\otimes C \unrhd 0$ w.r.t. $\int^{\oplus}_M\Cone d\mu$.
\end{Prop}
{\it Proof.} Let $\Psi\in \int^{\oplus}_M\Cone d\mu$.
For each $\xi\in \Cone$ and $ f\in L^2(M, d\mu)_+$, we have 
\begin{align}
 \la 1\otimes C\Psi|\xi\otimes f\ra=\la \Psi|\xi\otimes C^*f\ra. \label{EqC1}
\end{align}
Because $C^*\unrhd 0 $ w.r.t. $L^2(M, d\mu)_+$, $C^*f\ge 0$ w.r.t. $L^2(M, d\mu)_+$,
 which implies that $\xi\otimes C^*f \ge 0$. Thus, the right hand side of (\ref{EqC1}) is positive.
By Lemma \ref{PPEquivH}, we conclude that $1\otimes C  \Psi\ge 0$ w.r.t. $\int^{\oplus}_M \Cone d\mu$. $\Box$
\medskip\\

The following lemma is useful in the present  study:
\begin{lemm}\label{TensorPP}
Let  $\mathcal{Z}=\BbbR^n$. Let $\Cone=\mathscr{L}^2(\mathfrak{X})_+$ be a natural self-dual cone 
 in $\mathfrak{X}\otimes \mathfrak{X}=\mathscr{L}^2(\mathfrak{X})$, see Section \ref{L2Define} for notations.
\begin{itemize}

\item[{\rm (i)}]
Let $B: \mathcal{Z}\to \mathscr{B}(\mathfrak{X});\ {\Bs q}\mapsto B({\Bs q})$ be continuous with $
\sup_{\Bs q} \|B({\Bs q})\| <\infty
$.
We have
\begin{align}
\int_{\mathcal{Z}}^{\oplus} \mathcal{L}(B({\Bs q})^*)\mathcal{R}(B({\Bs q}))d{\Bs q} \unrhd 0
\ \ \ \mbox{w.r.t. $\displaystyle \int_{\mathcal{Z}} ^{\oplus }\Cone d{\Bs q}$}.
\end{align}
\item[{\rm (ii)}] Let $C\in \mathscr{B}(\mathfrak{X})$. We have 
$\mathcal{L}(C^*) \mathcal{R}(C) \otimes 1 \unrhd 0$ w.r.t. $\int^{\oplus}_{\mathcal{Z}} \Cone d{\Bs q}$.

\end{itemize}
\end{lemm}
{\it Proof.} 
Set  $A({\Bs q}):= \mathcal{L}(B({\Bs q})^*)\mathcal{R}(B({\Bs q})) \in \mathscr{B}(\mathscr{L}^2(\mathfrak{X}))$.
We have 
$
\mathrm{ess.sup}_{\Bs q} \|A({\Bs q})\|\le (
\sup_{\Bs q} \|B({\Bs q})\| 
)^2<\infty.
$ 
Thus, we can define a bounded linear operator $
\int_{\mathcal{Z}}^{\oplus} \mathcal{L}(B({\Bs q})^*)\mathcal{R}(B({\Bs q}))d{\Bs q} 
$
on $\int^{\oplus}_{\mathcal{Z}} \mathscr{L}^2(\mathfrak{X}) d{\Bs q}$
  by Lemma \ref{ThmXIII83}.
Since $A({\Bs q}) \unrhd 0$ w.r.t. $\Cone$ by Proposition \ref{GeneralPP}, 
we can apply Proposition \ref{BasicDP} and obtain (i).

(ii) immediately follows from  Propositions \ref{GeneralPP} 
 and Corollary \ref{BasicDP2}. $\Box$

\begin{lemm}\label{PIEquivT}
Let $\Psi\in \int_M \mathfrak{H} d\mu$. The following statements are equivalent:
\begin{itemize}
\item[{\rm (i)}] $\Psi>0$ w.r.t. $\int_M^{\oplus} \Cone d\mu$.
\item[{\rm (ii)}] $\la\xi \otimes f|\Psi\ra>0$ for all $\xi\in \Cone\backslash \{0\}$
and $f\in L^2(M, d\mu)_+ \backslash \{0\}$.
\end{itemize}
\end{lemm}
{\it Proof.} To prove that  (i) $\Rightarrow$ (ii) is easy.
Let us prove the converse.
We set $G_{\xi}(m)
=\la \xi|\Psi(m)\ra_{\mathfrak{H}}
$. We  have
$
\la \xi\otimes f|\Psi\ra=\int_MG_{\xi}(m) f(m) d\mu>0$ for every $ f\in L^2(M, d\mu)_+\backslash \{0\}. 
$
Thus, we have $G_{\xi}(m)>0$ for $\mu$-a.e..
Because $\xi$ is arbitrary, we conclude that $\Psi(m)>0$ w.r.t. $\Cone$ for $\mu$-a.e.. $\Box$

\begin{coro}\label{TPV}
Let $\xi\in \Cone$ and let $f\in L^2(M, d\mu)_+$.
If $\xi>0$ w.r.t. $\Cone$ and $f>0 $ w.r.t. $L^2(M, d\mu)_+$, then it holds that $\xi\otimes f>0$
w.r.t. $\int_M^{\oplus}
\Cone d\mu$.
\end{coro}

t

\end{document}